\newif\ifpngfig
\pngfigtrue 

\ifpngfig
  \documentclass[useAMS,usenatbib,pdftex]{mn2e/mn2e}
\else
  \documentclass[useAMS,usenatbib]{mn2e/mn2e}
\fi
\usepackage{graphicx}
\usepackage{amsmath}
\usepackage{amssymb}
\usepackage{mathrsfs}
\usepackage[dvips,usenames]{color}
\usepackage{hyperref}
\usepackage[normalem]{ulem} 
\usepackage{lineno}
\usepackage[section]{placeins}

\newcommand{\jks}{\mbox{$J\!-\!K_{\rm s}$}}

\newcommand{\av}{\mbox{$A_V$}}
\newcommand{\muo}{\mbox{$\mu_0$}}

\newcommand{\feh}{\mbox{\rm [{\rm Fe}/{\rm H}]}}
\newcommand{\mh}{\mbox{\rm [{\rm M}/{\rm H}]}}
\newcommand{\Mact}{\mbox{${\mathcal M}$}}
\newcommand{\Ract}{\mbox{${\mathcal R}$}}
\newcommand{\Mini}{\mbox{${\mathcal M}_{\rm i}$}}
\newcommand{\Msun}{\mbox{$M_{\odot}$}}

\newcommand{\Teff}{\mbox{$T_{\rm eff}$}}

\newcommand{\logg}{\mbox{$\log g$}}

\newcommand{\beq}{\begin{equation}}
\newcommand{\eeq}{\end{equation}}
\newcommand{\beqa}{\begin{eqnarray}}
\newcommand{\eeqa}{\end{eqnarray}}
\newcommand{\bite}{\begin{itemize}}
\newcommand{\eite}{\end{itemize}}
\newcommand{\deltanu}{\mbox{$\Delta\nu$}}
\newcommand{\numax}{\mbox{$\nu_{\rm max}$}}
\newcommand{\deltaP}{\mbox{$\Delta P$}}

\newcommand{\comment}[1]{}

\title[Distances and extinctions from APOKASC]{Bayesian distances and extinctions for giants observed by {\it Kepler} and APOGEE}
\author[Tha\'ise S. Rodrigues et al.]{
Tha\'ise S. Rodrigues$^{1,2,3}$\thanks{E-mail: thaise.rodrigues@oapd.inaf.it}, 
L\'eo Girardi$^{1,3}$, 
Andrea Miglio$^{4,5}$, 
Diego Bossini$^{4,5}$, 
\newauthor
Jo Bovy$^{6,7}$, 
Courtney Epstein$^{8}$,
Marc H. Pinsonneault$^{8}$,
Dennis Stello$^{5,9}$, 
\newauthor 
Gail Zasowski$^{10}$, 
Carlos Allende Prieto$^{11,12}$, 
William J. Chaplin$^{4,5}$, 
Saskia Hekker$^{13}$,
\newauthor
Jennifer A. Johnson$^{8,14}$,
Szabolcs M{\'e}sz{\'a}ros$^{15}$,
Beno\^{\i}t Mosser$^{16}$, 
Friedrich Anders$^{17}$, 
\newauthor
Sarbani Basu$^{18}$, 
Timothy C. Beers$^{19,20}$,
Cristina Chiappini$^{17}$,
Luiz A.N. da Costa$^{3,21}$,
\newauthor
Yvonne Elsworth$^{4,5}$,
Rafael A. Garc\'{\i}a$^{22}$,
Ana E. Garc\'{\i}a P\'erez$^{23}$,
Fred R. Hearty$^{24}$,
\newauthor
Marcio A.G. Maia$^{3,21}$,
Steven R. Majewski$^{23}$,
Savita Mathur$^{25}$,
Josefina Montalb\'an$^{26}$,
\newauthor 
David L. Nidever$^{27}$,
Basilio Santiago$^{3,28}$,
Mathias Schultheis$^{29}$, 
Aldo Serenelli$^{30}$, 
\newauthor
Matthew Shetrone$^{31}$
\newauthor 
\\
$^{1}$Osservatorio Astronomico di Padova -- INAF, Vicolo dell'Osservatorio 5, I-35122 Padova, Italy\\
$^{2}$Dipartimento di Fisica e Astronomia, Universit\`a di Padova, Vicolo dell'Osservatorio 2, I-35122 Padova, Italy\\
$^{3}$Laborat\'orio Interinstitucional de e-Astronomia -- LIneA, Rua Gal.\ Jos\'e Cristino 77, Rio de Janeiro, RJ -- 20921-400, Brazil\\
$^{4}$School of Physics and Astronomy, University of Birmingham, Edgbaston, Birmingham, B15 2TT, UK\\
$^{5}$Stellar Astrophysics Centre (SAC), Department of Physics and Astronomy, Aarhus University, Ny Munkegade 120, \\ DK-8000 Aarhus C, Denmark \\
$^{6}$Institute for Advanced Study, Einstein Drive, Princeton, NJ 08540, USA\\
$^{7}$Hubble Fellow\\
$^{8}$Department of Astronomy, The Ohio State University, Columbus, OH 43210, USA\\
$^{9}$Sydney Institute for Astronomy (SIfA), School of Physics, University of Sydney, NSW 2006, Australia\\
$^{10}$Department of Physics and Astronomy, Johns Hopkins University, Baltimore, MD 21218, USA\\
$^{11}$Instituto de Astrof{\'{\i}}sica de Canarias (IAC), E-38200 La Laguna, Tenerife, Spain\\
$^{12}$Departamento de Astrof{\'{\i}}sica, Universidad de La Laguna (ULL), E-38206 La Laguna, Tenerife, Spain\\
$^{13}$Max-Planck-Institut f\"ur Sonnensystemforschung, Justus-von-Liebig-Weg 3, D-37077 G\"ottingen, Germany\\
$^{14}$Center for Cosmology and Astro-Particle Physics, Ohio State University, Columbus, OH 43210, USA\\
$^{15}$Department of Astronomy, Indiana University, Bloomington, IN 47405-7105, USA\\
$^{16}$Observatoire de Paris/CNRS, UMR 8109 F-92195 Meudon, France\\
$^{17}$Leibniz-Institut f\"ur Astrophysik Potsdam (AIP), An der Sternwarte 16, D-14482 Potsdam, Germany\\
$^{18}$Department of Astronomy, Yale University, PO Box 208101, New Haven, CT 06520-8101, USA\\
$^{19}$Department of Physics, University of Notre Dame, 225 Nieuwland Science 
Hall, Notre Dame, IN 46656, USA\\
$^{20}$JINA: Joint Institute for Nuclear Astrophysics, University of Notre 
Dame, Notre Dame, IN 46556, USA\\
$^{21}$Observat\'orio Nacional, Rua Gal. Jos\'e Cristino 77, Rio de Janeiro, RJ - 20921-400, Brazil\\
$^{22}$Laboratoire AIM, CEA/DSM -- CNRS - Univ. Paris Diderot -- IRFU/SAp, Centre de Saclay, 91191 Gif-sur-Yvette Cedex, France\\
$^{23}$Department of Astronomy, University of Virginia, Charlottesville, VA 22904-4325, USA \\
$^{24}$Department of Astronomy \& Astrophysics, The Pennsylvania State University, 525 Davey Laboratory, University Park PA  16802, USA\\
$^{25}$Space Science Institute, 4750 Walnut street Suite 205, Boulder, CO 80301, USA\\
$^{26}$Institut d'Astrophysique et de Geophysique de l'Universit\'e de Li\`ege, 17 all\'ee du 6 Ao\^ut, B-4000 Li\`ege, Belgium\\
$^{27}$Department of Astronomy, University of Michigan, Ann Arbor, MI, 48104, USA\\
$^{28}$Instituto de F\'{\i}sica, Universidade Federal do Rio Grande do Sul, Porto Alegre, Brazil\\
$^{29}$Laboratoire Lagrange (UMR7293), Universit\'{e} de Nice Sophia Antipolis, CNRS, Observatoire de la C\^{o}te d'Azur, BP 4229, 06304 \\ Nice Cedex 4, France\\
$^{30}$Institute of Space Sciences (CSIC-IEEC), Campus UAB, E-08193 Bellaterra, Spain\\
$^{31}$University of Texas at Austin, McDonald Observatory, 32 Fowlkes Rd., McDonald Observatory, TX 79734-3005, USA
} 

\date{Accepted\vspace{2cm}}

\pagerange{\pageref{firstpage}--\pageref{lastpage}} \pubyear{2014}

\begin{document}
\maketitle

\label{firstpage}

\begin{abstract}
We present a first determination of distances and extinctions for individual stars in the first release of the APOKASC catalogue, built from the joint efforts of the Apache Point Observatory Galactic Evolution Experiment (APOGEE) and the {\it Kepler} Asteroseismic Science Consortium (KASC). Our method takes into account the spectroscopic constraints derived from the APOGEE Stellar Parameters and Chemical Abundances Pipeline, together with the asteroseismic parameters from KASC. These parameters are then employed to estimate intrinsic stellar properties, including absolute magnitudes, using the Bayesian tool PARAM. We then find the distance and extinction that best fit the observed photometry in SDSS, 2MASS, and WISE passbands. 
The first 1989 giants targeted by APOKASC are found at typical distances between 0.5 and 5~kpc, with individual uncertainties of just $\sim\!1.8$ per cent. Our extinction estimates are systematically smaller than provided in the {\it Kepler} Input Catalogue and by the \citeauthor{schlegel98} maps. Distances to individual stars in the NGC~6791 and NGC~6819 star clusters agree to within their credible intervals. Comparison with the APOGEE red clump and SAGA catalogues provide another useful check, exhibiting agreement with our measurements to within a few percent. Overall, present methods seem to provide excellent distance and extinction determinations for the bulk of the APOKASC sample. Approximately one third of the stars present broad or multiple-peaked probability density functions and hence increased uncertainties. Uncertainties are expected to be reduced in future releases of the catalogue, when a larger fraction of the stars will have seismically-determined evolutionary status classifications. 
\end{abstract}


\section{Introduction}
\label{sec:intro}

A number of massive high-resolution spectroscopy surveys \citep[e.g., APOGEE, Gaia-ESO, ARGOS, GALAH; Majewski et al., in preparation;][]{ges, argos, galah} are presently being conducted as part of a major community effort to reveal the evolution and present structure of our Milky Way (MW) galaxy. These surveys promise to greatly expand the available data base of spectroscopic properties such as radial velocities, effective temperatures, surface gravities and chemical abundances. As demonstrated by several authors \citep{allendeprieto06, burnett10, binney14, hayden13, santiago14}, spectroscopic parameters coupled with photometry can provide distance estimates for all of the observed stars, especially when the surface gravity, \logg, is well-constrained. This is preferentially done via Bayesian methods that naturally take into account the many sources of measurement uncertainties and biases. However, it is also clear that a major effort is needed to calibrate such distance determinations and reduce their uncertainties below the $\sim$20 per cent level.
 
Future astrometry from Gaia will obviously provide distance calibrators over a wide range of apparent magnitudes and distances -- except for the very red and optically-obscured stars, hidden by dust lanes across the Galactic mid-plane. In the meantime, distance determinations for field giants in spectroscopic surveys must rely essentially on just two kinds of calibrators: stars in clusters, and stars with well-determined asteroseismic parameters. In this paper we concentrate on the latter, discussing the accuracies in distance determinations that are attainable via Bayesian methods (see Sec.~\ref{sec:clusters}).

We utilize a very special sample of stars -- the APOKASC sample. This unique data set results from a collaboration between {\it Kepler} Asteroseismic Science Consortium \citep[KASC\footnote{\url{http://astro.phys.au.dk/KASC}},][]{kjeldsen10} and Apache Point Observatory Galactic Evolution Experiment (APOGEE; Majewski et al., in preparation), which itself is part of the third phase of the Sloan Digital Sky Survey \citep[SDSS-III;][]{eisenstein11}. 
Almost 2,000 red giants targeted by the \emph{Kepler} satellite mission \citep{borucki10} have been observed by APOGEE during the first year and included in the SDSS-III Data Release 10 \citep[DR10,][]{dr10}. They correspond to the sample presented in \citet{pinsonneault14} and discussed in this work. The APOKASC sample will include over 8,000 giants by the end of the APOGEE survey, and will be further expanded during the upcoming SDSS-IV/APOGEE-2 campaign.

Solar-like oscillations are excited in cool stars, and the natural periods for low density red giants (of the order of days to weeks) are sufficiently long for them to be easily detected with the \emph{Kepler} cadence \citep{hekker10}. \emph{Kepler} asteroseismic data, as outlined in Sec.~\ref{sec:sample}, can be used to infer mean density, \logg, masses (\Mact), radii ($R$) -- when combined with an effective temperature estimate -- and diagnostics of evolutionary state. The APOGEE spectra provide accurate determinations of effective temperatures (\Teff) and chemical abundances of several elements (Garc\'ia P\'erez et al., in preparation)\footnote{APOGEE also provides estimates of \logg, which are less reliable than those derived from the asteroseismic constraints \citep[see e.g.,][]{meszaros13}.}. There is also an extensive data base of photometry for these stars, which can provide additional constraints on stellar properties.

This data set provides a powerful set of tools for estimating stellar distances. Precise asteroseismic surface gravities, combined with mass constraints, can be used to infer stellar radii. $T_{\rm eff}$ and extinction can be measured using spectroscopy and photometry. This information is straightforwardly converted into intrinsic luminosities from the standard relation
\beq
L=4\pi R^2\sigma\Teff^4,
\label{eq:lum}
\eeq
where $\sigma$ is the Stefan-Boltzmann constant. When $L$ is combined with a bolometric correction, an extinction, and the observed apparent magnitude in a given passband, a so-called `direct measurement' of the distance is possible \citep[see e.g.,][]{miglio13}. Bayesian methods combine the likelihood of all possible solutions to provide a better weighted -- and possibly more accurate -- solution that includes prior information about the data set. Essentially, Bayesian methods incorporate information from stellar models that allows us (1) to require a consistent stellar parameter measurement, as opposed to permitting unphysical combinations of mass, radius, temperature, and metallicity; (2) to account for population effects, such as lifetime, the star formation rate, and the initial mass function, which bias the true stellar parameters in a manner inconsistent with a purely Gaussian distribution; and (3) to reconcile independent methods for inferring properties such as the effective temperature.

Once distances to the asteroseismic targets are determined, they can be used for a series of applications related to Galactic archaeology \citep{miglio13}. Especially useful are the red giants, which can be measured at large distances due to their intrinsic brightness, hence probing regions of the MW far from the well-studied Solar Neighbourhood. In addition, the asteroseismic distances will help to obtain a better distance calibration for stars observed in broad-band photometry and high-resolution spectroscopy alone. Indeed, one of our main long-term goals is to derive the best possible distances for the over 100,000 stars being observed by APOGEE across the Galaxy, and for the additional $>$200,000 stars that will be observed by APOGEE-2. 

The structure of this paper is as follows. Sec.~\ref{sec:sample} describes the APOKASC sample. Sec.~\ref{sec:method} presents the Bayesian method that we apply to determine distances to APOKASC stars, taking a few stars as examples. Sec.~\ref{sec:discussion} discusses the results for the entire sample, comparing values derived from different assumptions and different priors, so that systematic uncertainties can be estimated. Sec.~\ref{sec:conclu} draws a few conclusions.

\section{Sample}
\label{sec:sample}

\subsection{Spectroscopic data from APOGEE}

APOGEE uses a high-resolution infrared spectrograph \citep{wilson12}, mounted at the Apache Point Observatory 2.5~m telescope \citep{gunn06}, with a mean resolution of $\sim$22,500 in the $H$-band (spectral coverage: $1.51 - 1.70$ $\mu$m). APOGEE has already observed more than 100,000 stars selected from 2MASS photometry, at typical signal-to-noise ratios of $\sim\!140$ per resolution element. The targeted stars are mostly red giant branch (RGB), red clump (RC), and asymptotic giant branch stars \citep{zasowski13}, and are spread over all regions of the MW, including the bulge, disk, and halo. The scientific exploitation of this enormous data base is facilitated by the APOGEE Stellar Parameter and Chemical Abundances Pipeline \citep[ASPCAP;][Garc\'ia P\'erez et al., in preparation]{meszaros13}. For each APOGEE target, ASPCAP returns basic stellar parameters (effective temperature, surface gravity, metallicity), and individual chemical abundances for a number of elements.
The raw ASPCAP stellar parameters were then compared with independent external measurements from star cluster members and asteroseismic targets for the key stellar parameters, namely overall metallicity (\mh), surface gravity, and effective temperature. The final DR10 results included a recommended set of corrections intended to make the ASPCAP results consistent with the values from these external checks \citep[see][]{meszaros13}.
In this work, we use the \Teff\ and \mh\ `corrected ASPCAP values' provided in DR10, instead of the raw ones; they include corrections that improve the agreement with other independent scales based on the infrared-flux method (IRFM) and on cluster data. The \mh\ are calibrated with the literature values of \feh\ in 20 star clusters.

\subsection{Asteroseismic data from {\it Kepler}}

The {\it Kepler} space telescope has observed $\sim$196,400 stars \citep{huber14} in a field of 105 deg$^2$ towards the constellations of Cygnus and Lyra \citep{borucki10}. Apart from the discovery of exoplanets and multiple stellar systems, the high temporal and photometric quality of the data provides the possibility to study red giants by detection of solar-like oscillations \citep[e.g.,][]{huber10, chaplin11}. For solar-like oscillators with pulsations excited in the turbulent outer layers, two global asteroseismic parameters can be extracted: the average large frequency separation, \deltanu, and the frequency of maximum oscillation power, \numax. The former is the dominant frequency separation of the near-regular pattern of high overtones, and depends to good approximation on the mean density $\overline{\rho}$ of the star \citep{vandakurov68}:
\beq
 \Delta\nu \propto \overline{\rho}^{\,1/2} \propto \Mact^{1/2} R^{-3/2}.
 \label{eq:deltanu}
\eeq
The latter is the frequency of maximum power of the Gaussian-like modulation of the mode amplitudes, which is related to the acoustic cut-off frequency of the star, and therefore to its fundamental parameters \citep{brown91, belkacem11}:
\beq
 \nu_{\rm max} \propto g T_{\rm eff}^{-1/2} \propto (\Mact/R^2)  T_{\rm eff}^{-1/2}.
 \label{eq:numax}
\eeq
Adopting homology relations and considering reference values of \deltanu\ and \numax\ derived from the Sun, these equations determine the mass and radius of a star independently of evolutionary stellar models, if a value for the effective temperature is available. This is the so-called `direct method' of parameter determination.
Asteroseismic radii agree to within 5 per cent of those inferred from interferometry \citep{huber12} and from stars with {\it Hipparcos} parallaxes \citep{silvaaguirre12}. Masses are more difficult to directly constrain, but eclipsing binaries in NGC~6791 with well-measured masses \citep{brogaard12} can be used to infer the expected masses for red giants. Asteroseismic mass estimates for cluster members \citep{miglio12_2} are close to, but greater than, these mass estimates; for these stars, systematic uncertainties in the asteroseismic masses are at the 10 per cent level. A larger systematic trend for metal-poor stars is found \citep{epstein14}, but such stars are rare in our sample. For our purposes, the primary impact of mass uncertainties is their impact on radius measurements, as overestimated masses at fixed surface gravity will require overestimated radii to compensate.

\begin{figure}
\begin{center}
  \ifpngfig
    \resizebox{0.85\hsize}{!}{\includegraphics{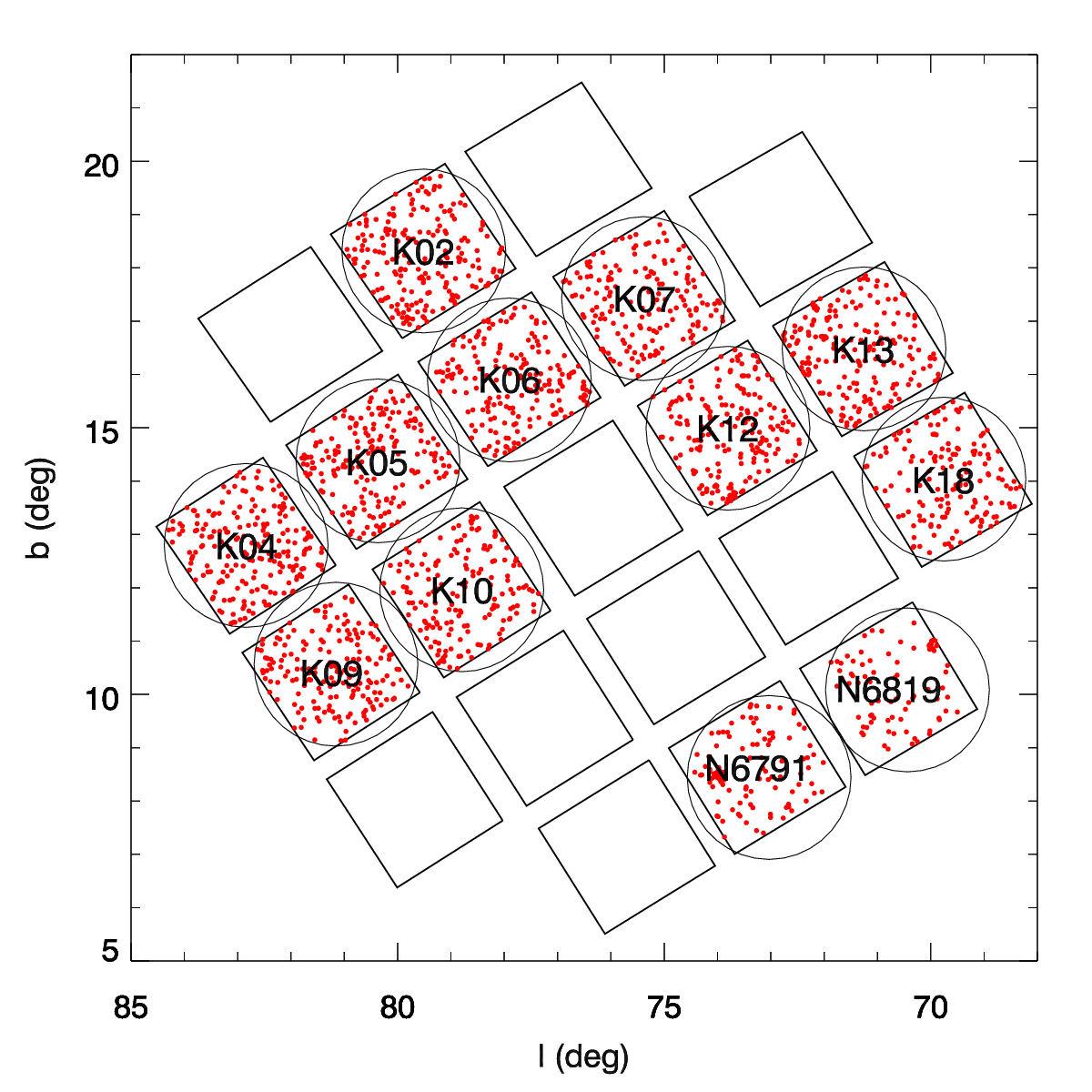}}
  \else
    \resizebox{0.85\hsize}{!}{\includegraphics{figs/l_b_av/l_b_kfov.eps}}
  \fi
\caption{Position of APOKASC fields (circles) in Galactic coordinates relative to the {\it Kepler} field (squares). Red dots represent the stars observed during the first year and analysed in this paper. The final APOKASC sample will include a significantly larger sample across the entire {\it Kepler} field.}
 \label{fig:l_b_kfov}
\end{center}
\end{figure}

This excellent and accurate alternative to derive stellar properties encouraged the APOGEE team to include {\it Kepler} stars on their target list, giving rise to the APOGEE-KASC collaboration (APOKASC). Approximately $\sim$10,000 stars in the magnitude range $7 \leq H \leq 11$, including giants from the open clusters NGC~6791 and NGC~6819, were already observed; out of 2,000 stars are part of the first APOKASC public release \citep{pinsonneault14} and are distributed in the sky as in Fig.~\ref{fig:l_b_kfov}. The squares show the {\it Kepler} field of view, and the circles indicate the APOGEE plates observed during the first year with the stars in the sample (red dots). The target selection and first release of the APOKASC catalogue are described in \citet{pinsonneault14}. A total of 1989 stars having both seismic and spectroscopic data are analysed in this work. 

\subsection{Photometry}

In addition to the spectroscopic and asteroseismic parameters, stars in the APOKASC catalogue have measured apparent magnitudes in 
\begin{itemize}
\item SDSS $griz$ and $DDO51$, as measured by the KIC team \citep{brown11}, and corrected by \citet{pinsonneault12};
\item $JHK_{\rm s}$ from 2MASS \citep{cutri03, skrutskie06};
\item the {\it Kepler} magnitude, $Kp$, as derived from a combination of the $griz$ magnitudes \citep{brown11};
\item WISE photometry (at 3.35, 4.6, 11.6 and 22.1$\mu$m, or W1 to W4) from the Preliminary Release Source Catalog \citep{wright10}.
\end{itemize}
For this work, we discard the $Kp$ magnitude because it does not represent an independent photometric measurement, and $DDO51$ because it is a relatively narrow (and non-standard) passband, which causes problems in our synthetic photometry \citep[see][]{girardi02}. The WISE photometry in the filters $W3$ and $W4$ are disregarded because of their larger measurement uncertainties and possible contamination by warm interstellar dust \citep[][and references therein]{davenport14}. Thus, we make use of a set of nine photometric measurements covering the entire wavelength range from the blue to the mid-infrared, using standard filter transmission curves and well-defined zero-points, which are all easily reproducible by stellar models, as illustrated below.

\section{Description of the Bayesian method}
\label{sec:method}

In principle, one could simply derive independent observational estimates for the stellar observables, using the direct method. However, there are important effects which are neglected by treating all stellar parameters as uncorrelated and all errors as strictly Gaussian. For example, stars are much more likely to be observed in long-lived evolutionary phases than in short-lived ones; less massive stars are more common than higher mass ones; and stellar theory makes strong predictions about the allowed combinations of mass, radius, \Teff, and abundance. Bayesian methods provide a natural way of taking these effects into account.

In this work, we adopt a Bayesian method implemented as an extension to the PARAM code\footnote{\url{http://stev.oapd.inaf.it/param}} \citep{dasilva06}, which estimates stellar properties by comparing observational data with the values derived from stellar models, in this case a data set of theoretical isochrones. It is similar to the methods developed by \citet{hernandez99} and \citet{jorgensen05}, in the sense that it (1) provides the likelihood of all stellar parameters, after computing every possible solution, it (2) provides an easy and reliable way to estimate uncertainties, since it considers the observational ones and weights the contribution of each component according to its observational uncertainties, and it (3) applies Bayesian inference, i.e., it takes into account prior information on the data set. PARAM was extended to build a well-sampled grid of stellar models including seismic properties. Similar grid-based methods are described in \citet{stello09} and \citet{basu10}.

Our method works as follows: first, it determines the intrinsic stellar properties, and in a second step it estimates the distances and extinctions. These two steps are explained in Sec.~\ref{sec:step1} and \ref{sec:step2}, respectively. Sec.~\ref{sec:whytwo} explains why the method is separated in these two steps.

\subsection{Step 1: determining intrinsic stellar properties}
\label{sec:step1}

\begin{figure*}
\begin{center}
  \ifpngfig
    \resizebox{0.7\hsize}{!}{\includegraphics{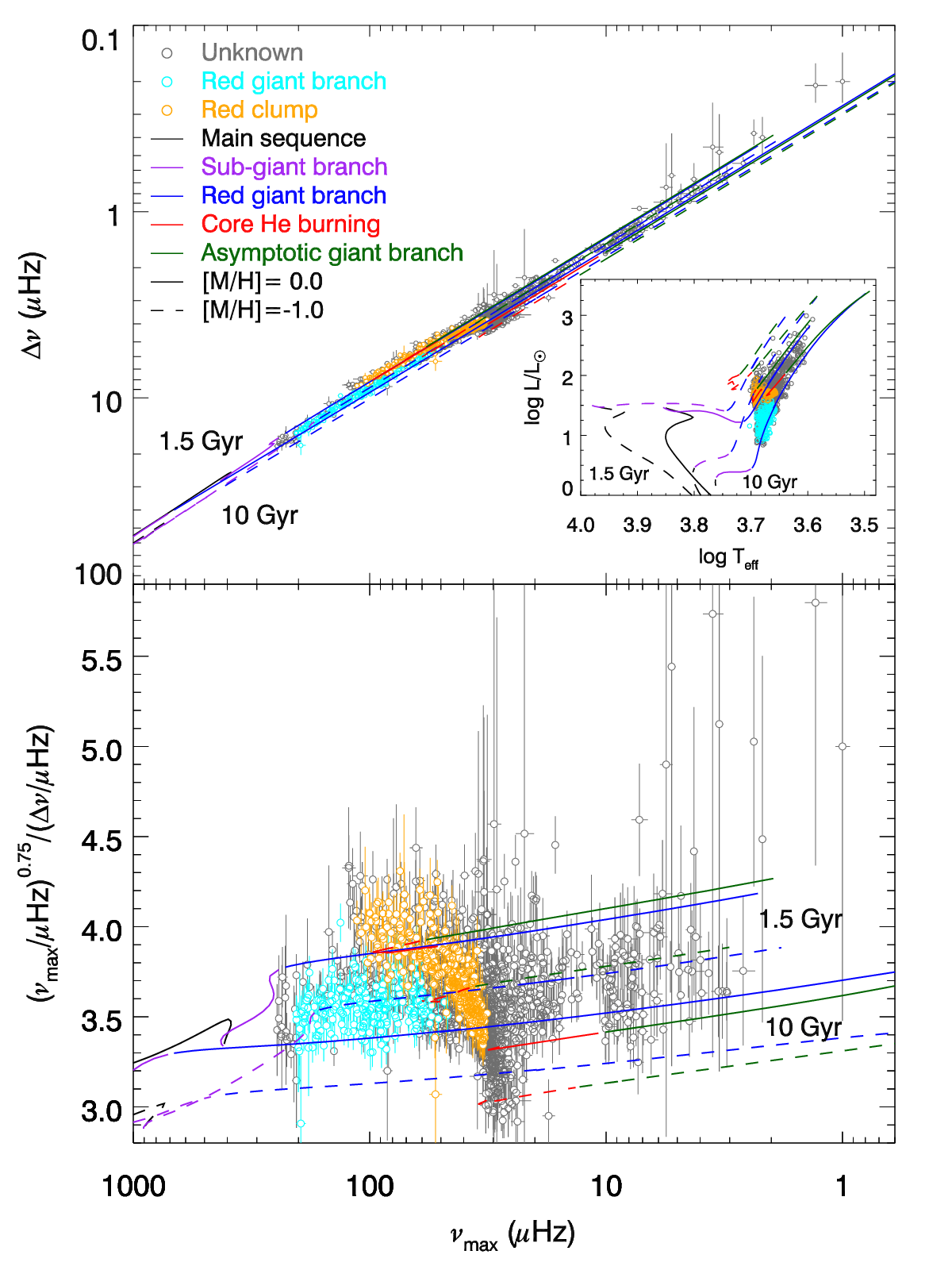}}
  \else
   \resizebox{0.7\hsize}{!}{\includegraphics{figs/isoc/HR_dnu_numax_inv.eps}}
  \fi
 \caption{Top panel: an illustration of stellar isochrones in the \deltanu\ versus \numax\ plane, covering the same range as the APOKASC giants. The isochrones are shown for two different ages (1.5 and 10 Gyr) and metallicities (0.0 and $-1.0$). Different evolutionary stages along the isochrones are marked with different colors. The grey, cyan, and orange dots are stars with asteroseismic evolutionary stage classification as unknown, RGB, and RC. The inset shows the same models and data in the more familiar H--R diagram. Bottom panel: same as in the top panel, but plotting the ratio between $\numax^{0.75}$ and \deltanu, which removes the radius dependence.}
 \label{fig:hr}
\end{center}
\end{figure*} 

The adopted set of isochrones is PARSEC v1.1\footnote{\url{http://stev.oapd.inaf.it/cmd}} \citep{bressan12}, from the Padova-Trieste stellar evolution group, which rely on updated input physics, and includes a solar model that reproduces tight constraints from helioseismology. For this work, isochrones were re-generated with a very small stepsize in both age and metallicity -- namely 0.02~dex in $\log\tau$ and 0.01~dex in \mh. At every point on the isochrones \deltanu\ and \numax\ are computed from the scaling relations
\begin{align}
\frac{\deltanu}{\deltanu_\odot} &= \left( \frac{\Mact}{\Msun} \right)^{1/2} 
  \left(\frac{R}{R_\odot}\right)^{-3/2}, \nonumber \\
\frac{\numax}{\numax_\odot}  &= \frac{\Mact}{\Msun} 
  \left(\frac{R}{R_\odot}\right)^{-2} 
  \left(\frac{\Teff}{\Teff_{\odot}}\right)^{-1/2},
\label{eq:dnu_numax}
\end{align}
where the solar values of $\deltanu_\odot=135.03$~$\mu$Hz and $\numax_\odot=3140.0$~$\mu$Hz have been used \citep[see][]{pinsonneault14}. The top panel of Fig.~\ref{fig:hr} illustrates how the isochrones appear in a $\deltanu$ versus $\numax$ diagram, in comparison with the more familiar Hertszprung--Russell (H--R) diagram (inset). In order to clarify, the bottom panel shows the ratio $\nu_{\rm max}^{0.75}/\Delta \nu \propto M^{0.25} T_{\rm eff}^{-0.375} $ \citep[cf.][]{huber11}, which removes the radius dependence, consequently the luminosity.

In addition, we have stored information about the evolutionary stage along the isochrones, which allows us to separate isochrone sections into two broad groups of `core-He burners' and `non-core He burners'. Many stars in the APOKASC catalogue can be safely classified into these two groups via the so-called period spacing of mixed modes, \deltaP\ \citep{bedding11, beck11, mosser11, stello13}. Mixed modes result from gravity waves propagating in the radiative interior which couple to pressure waves in the envelope, so that they become observable at the stellar surface, providing direct information about the stellar deep interior.

From the \Teff, \mh, \deltanu\ and \numax\ measurements, PARAM derives a probability density function (PDF) for the following stellar parameters: \Mact, $R$, $\log{g}$, age ($\tau$), mean density, and absolute magnitudes in several passbands, $M_\lambda$. First, the code computes the posterior probability, which is the probability of a chosen set of models given the prior probability on the models and the measured data, expressed as
\beq
p(\mathbf{x}|\mathbf{y}) = \frac{p(\mathbf{y}|\mathbf{x}) p(\mathbf{x})}{p(\mathbf{y})},
\label{eq_ppdf}
\eeq
where $p(\mathbf{x})$ represents the prior function, $p(\mathbf{y}|\mathbf{x})$  the likelihood function, and $p(\mathbf{y})$ is a normalization factor (which does not depend on $\mathbf{x}$, and can be ignored); $\mathbf{x}$ and $\mathbf{y}$ are the set of parameters to be derived and of measured data, respectively,
\begin{align}
\mathbf{x} & =(\Mact,R,\logg,\tau,M_\lambda), \nonumber \\
\mathbf{y} & =(\mh,\Teff,\deltanu,\numax). \nonumber
\end{align}
 The theoretical isochrones  make the connection between $\mathbf{x}$ and $\mathbf{y}$, $\mathbf{y}=\mathcal{I}(\mathbf{x})$. Assuming that the uncertainties of the measured data can be described as a normal distribution with a mean $y^\prime$ and standard deviation $\sigma_{y^\prime}$, the likelihood function can be written as
\begin{align}
p(\mathbf{y^\prime}|\mathbf{x}) &= L(\mathbf{y^\prime}, \mathcal{I}(\mathbf{x})) \nonumber \\
&= \prod_i \frac{1}{\sqrt{2\pi}\sigma_{y_i}}  \times \exp{\left(\frac{-(y_i^\prime-y_i)^2}{2\sigma_{y_i}^2} \right)}.
\label{eq_gdist}
\end{align}
The prior function is given by
\beq
 p(\mathbf{x}) = p(\Mact) \times p(\tau) \times p(\mh),
\eeq
where we adopted a flat prior for metallicity and age, $p(\tau)=p(\mh)=1$, i.e., that all metallicities and ages are equally probable, inside the interval $[10^6,10^{10}]$~yr. 
The prior in mass, $p(\Mact)$, is given by the \citet{chabrier01} initial mass function, $p(\Mini)$, but corrected for the small amount of mass lost close to the tip of the RGB, by adopting a relation $\Mact=\Mini-\Delta\Mini$. This correction $\Delta\Mini$ is computed from a \citet{reimers75} law with efficiency parameter $\eta=0.2$ \citep{miglio12_2}, and turns out to be close to null for the bulk of RGB stars, and smaller than 0.1~\Msun\ for all RC stars. No additional prior was adopted for the other parameters.

Finally, the marginal distribution $p(x_i|\mathbf{y^\prime})$ (hereafter $p(x_i)$) can be calculated, which is the PDF of each parameter $x_i$ obtained by integrating the posterior PDF given in Eq.~\ref{eq_ppdf} over all parameters, except $x_i$.

\begin{figure*}
  \ifpngfig
    \begin{minipage}{0.7\textwidth}
      \resizebox{\hsize}{!}{\includegraphics{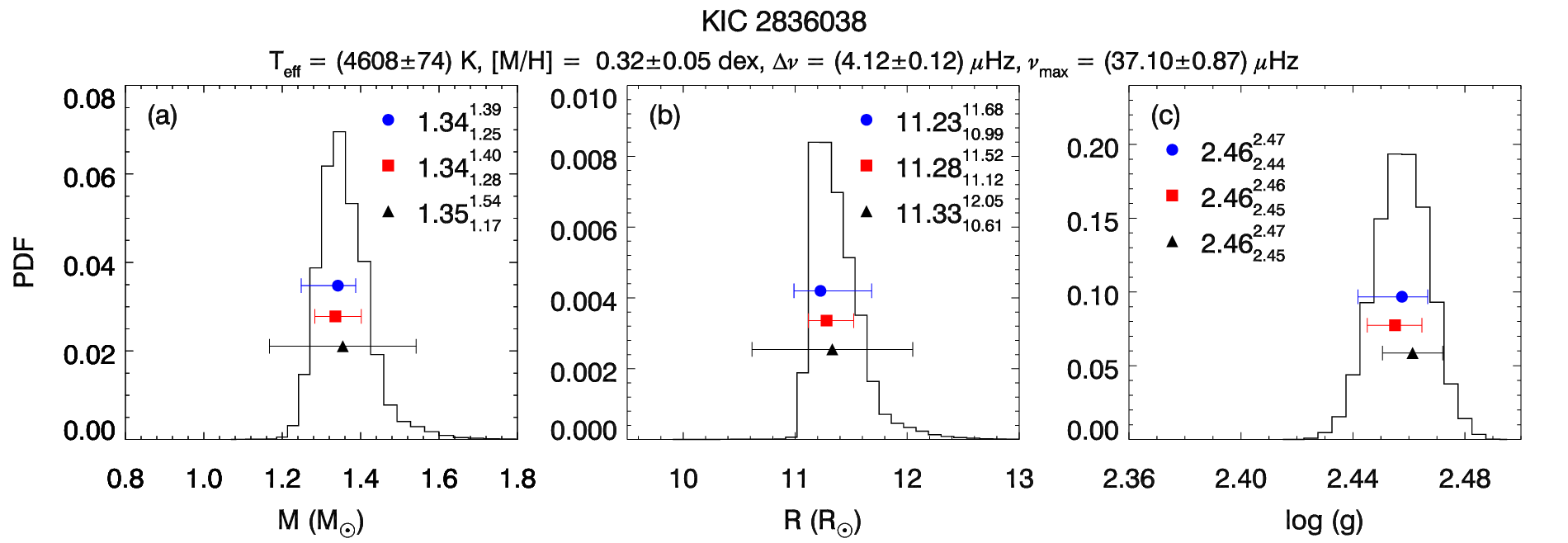}}
    \end{minipage}
    \begin{minipage}{0.7\textwidth}
      \resizebox{\hsize}{!}{\includegraphics{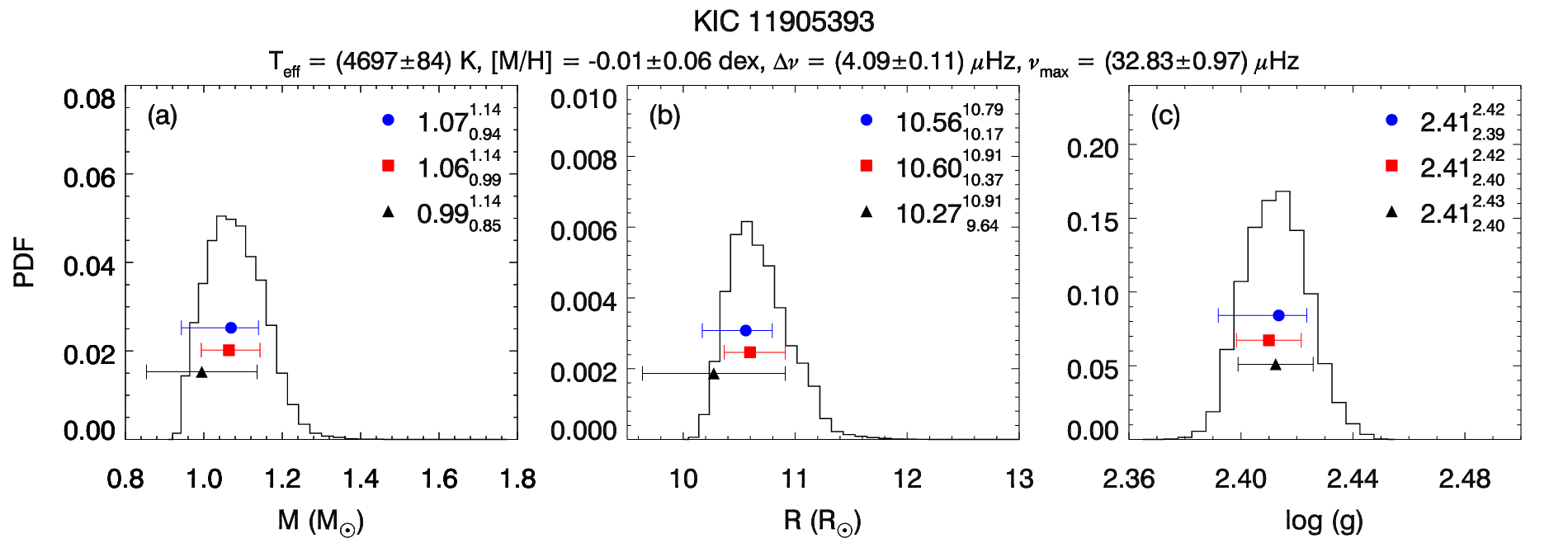}}
    \end{minipage}
    \begin{minipage}{0.7\textwidth}
       \resizebox{\hsize}{!}{\includegraphics{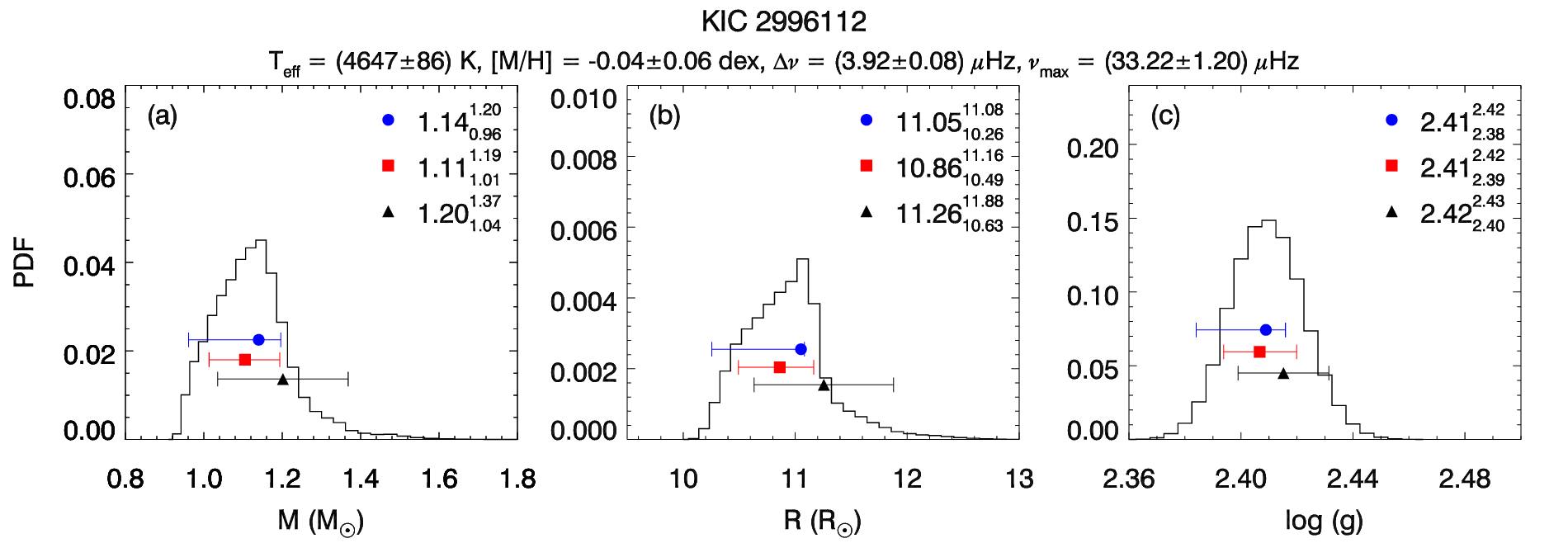}}  
    \end{minipage}
    \begin{minipage}{0.7\textwidth}
       \resizebox{\hsize}{!}{\includegraphics{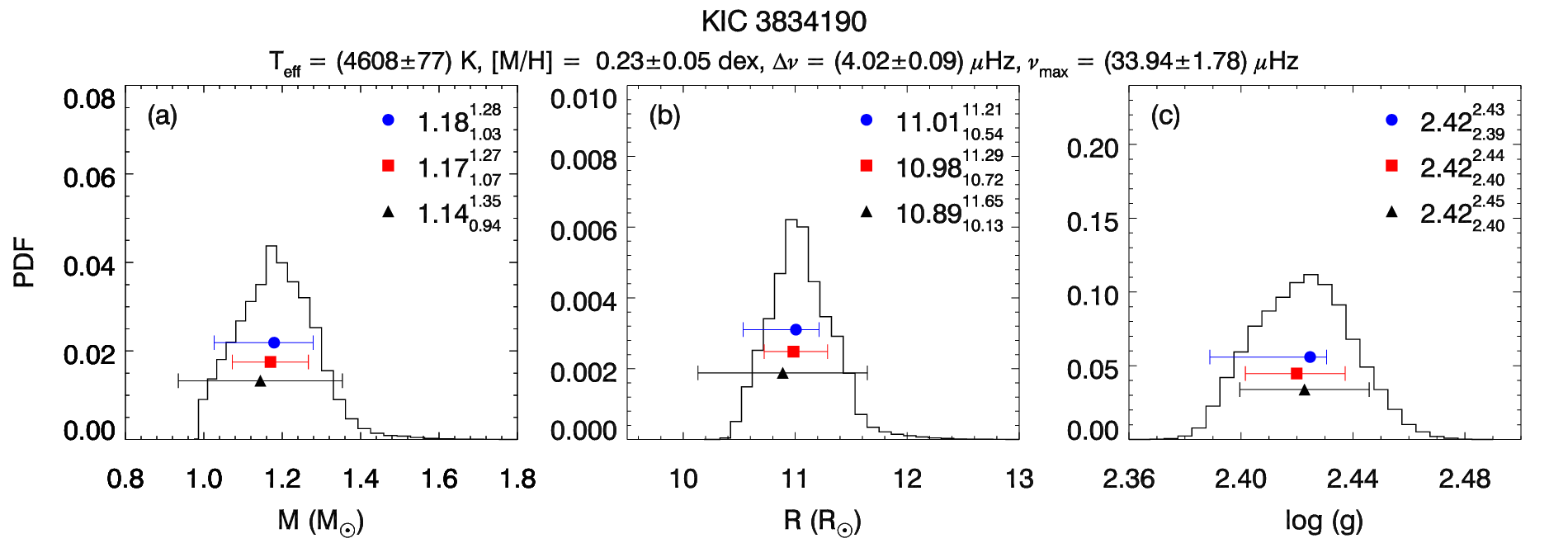}}  
    \end{minipage}
  \else
    \begin{minipage}{0.7\textwidth}
  \resizebox{\hsize}{!}{\includegraphics{figs/m_r_g/pdf_m_r_g_2836038.eps}}
 \end{minipage}
 \begin{minipage}{0.7\textwidth}
  \resizebox{\hsize}{!}{\includegraphics{figs/m_r_g/pdf_m_r_g_11905393.eps}}
 \end{minipage}
 \begin{minipage}{0.7\textwidth}
   \resizebox{\hsize}{!}{\includegraphics{figs/m_r_g/pdf_m_r_g_2996112.eps}}  
 \end{minipage}
 \begin{minipage}{0.7\textwidth}
  \resizebox{\hsize}{!}{\includegraphics{figs/m_r_g/pdf_m_r_g_3834190.eps}}  
 \end{minipage}
\fi
 \caption{(a) PDF of the mass, (b) radius, and (c) logarithm of surface gravity, for a set of typical APOKASC targets presenting single-peaked PDFs, and in a sequence of increasing uncertainties in the derived parameters. The adopted values of \Teff, \mh, \deltanu\ and \numax\ are indicated in the plots. The median and its 68 per cent CI, and the mode and its 68 per cent CI are represented by red and blue symbols, respectively. The black triangles are the results of the direct method.}
 \label{fig:m_r_a_g}
\end{figure*}

As an example, Fig.~\ref{fig:m_r_a_g} presents the resulting marginal PDFs of \Mact, $R$, and \logg\ for a series of four stars with well-behaved, single-peaked results. The adopted values of \Teff, \mh, \deltanu\ and \numax\ are indicated in the plots. For each PDF we have computed the median and 68 per cent credible intervals (CI; red symbols) by simply determining the points along the cumulative distribution function where suitable values were reached. In addition, we also indicate the mode and the 68 per cent CI (blue symbols), which are more suitable to represent the parameters inferred via the Bayesian method. The 68 per cent CI of the mode is determined by looking at the shortest interval that contains 68 per cent of the PDF area \citep{box73}. For simplicity, in what follows, the half-widths of these 68 per cent CI will be referred to as $\sigma(x)$, and used as an estimate of the uncertainties for each parameter $x$.

\begin{figure*}
  \ifpngfig
    \begin{minipage}{0.65\columnwidth}
      \resizebox{\hsize}{!}{\includegraphics{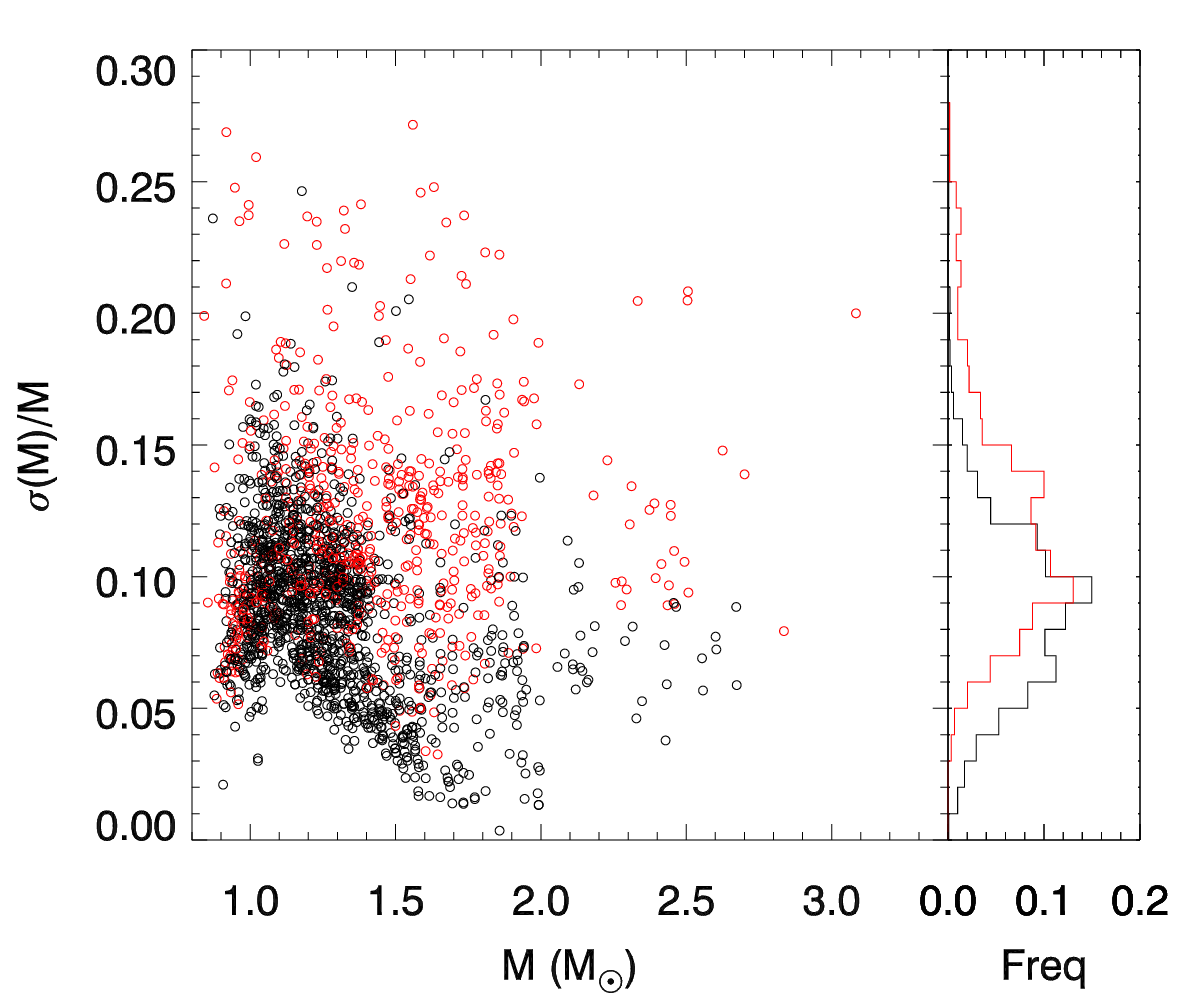}}
    \end{minipage}
    \begin{minipage}{0.65\columnwidth}
      \resizebox{\hsize}{!}{\includegraphics{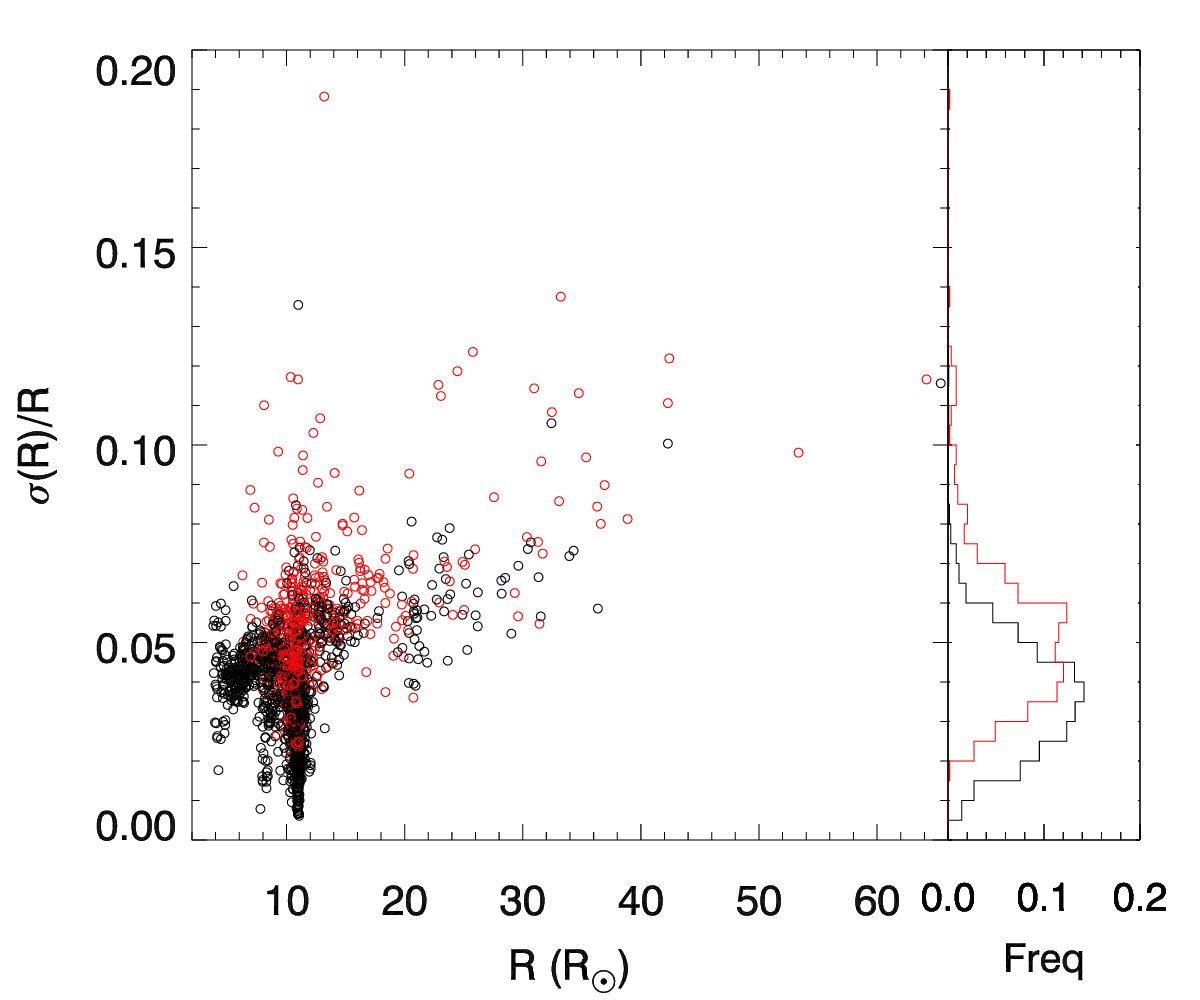}}
    \end{minipage}
    \begin{minipage}{0.65\columnwidth}
      \resizebox{\hsize}{!}{\includegraphics{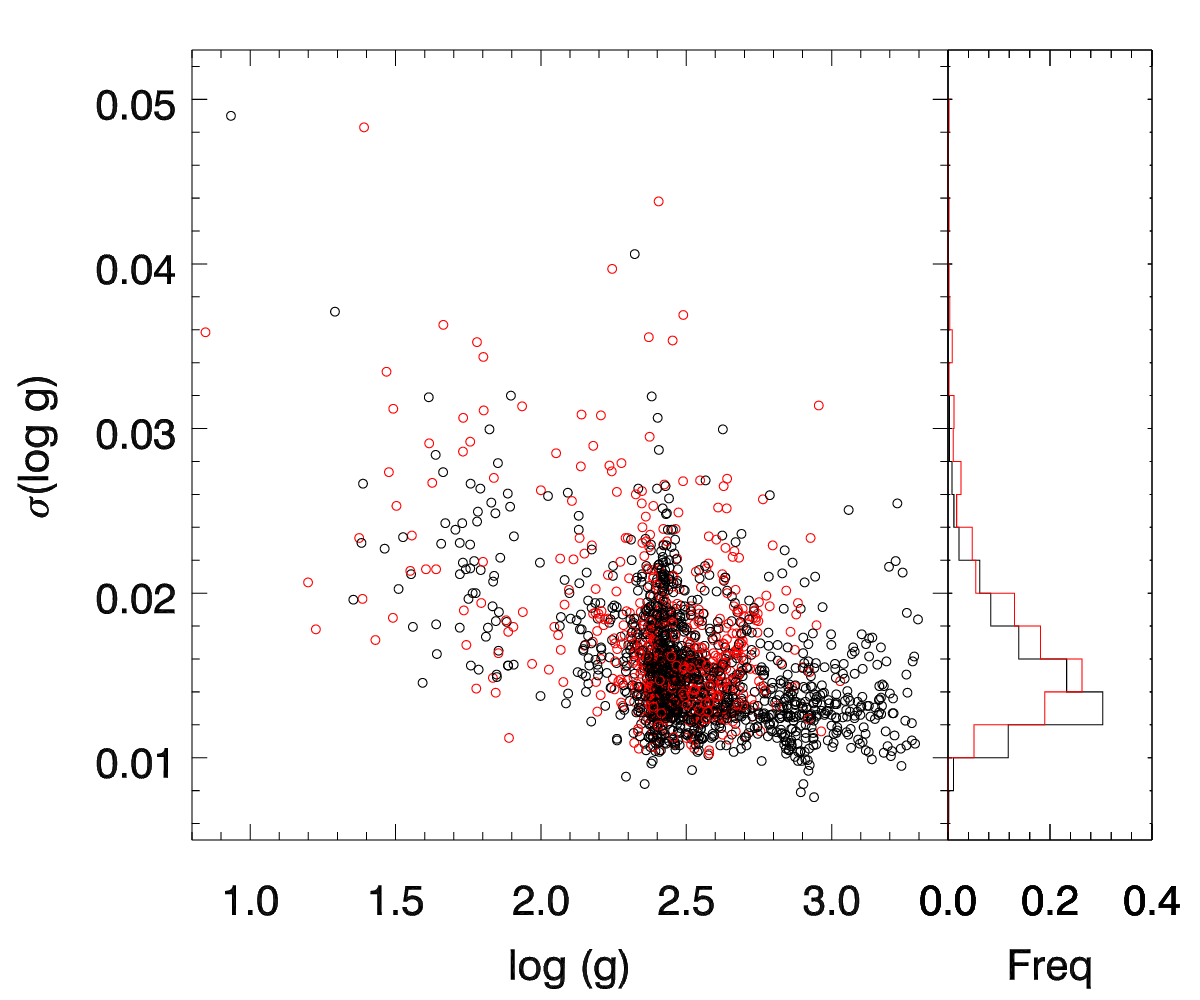}}
    \end{minipage}
  \else
    \begin{minipage}{0.65\columnwidth}
      \resizebox{\hsize}{!}{\includegraphics{figs/uncert/m_er_hist.eps}}
    \end{minipage}
    \begin{minipage}{0.65\columnwidth}
      \resizebox{\hsize}{!}{\includegraphics{figs/uncert/r_er_hist.eps}}
    \end{minipage}
    \begin{minipage}{0.65\columnwidth}
      \resizebox{\hsize}{!}{\includegraphics{figs/uncert/logg_er_hist.eps}}
    \end{minipage}
  \fi
 \caption{Distributions of relative (left and middle panels) and absolute (right) uncertainties for the stars in the sample, for the quantities derived in Step~1 (Sec.~\ref{sec:step1}) -- namely mass, radius and \logg. Black dots are stars with single-peaked PDFs, red dots are with broad/multiple-peaked ones. The right sub-panels show histograms of these uncertainty distributions.}
 \label{fig:relaterrors1}
\end{figure*}

It is worth noting that the PDFs for \Mact\ and $R$ are usually asymmetric, although they are derived from parameters with assumed Gaussian-distributed uncertainties. Uncertainties in mass have a median of $\sigma(\Mact)/\Mact=0.09$. More important in the context of this work is that the PDFs for $R$ and \logg\ are usually quite well-constrained, with median values of $\sigma(R)/R=0.040$ and $\sigma(\logg)=0.015$~dex, respectively. The full distribution of relative and absolute uncertainties is presented in Fig.~\ref{fig:relaterrors1}.

\begin{figure*}
  \ifpngfig
    \begin{minipage}{0.7\textwidth}
      \resizebox{\hsize}{!}{\includegraphics{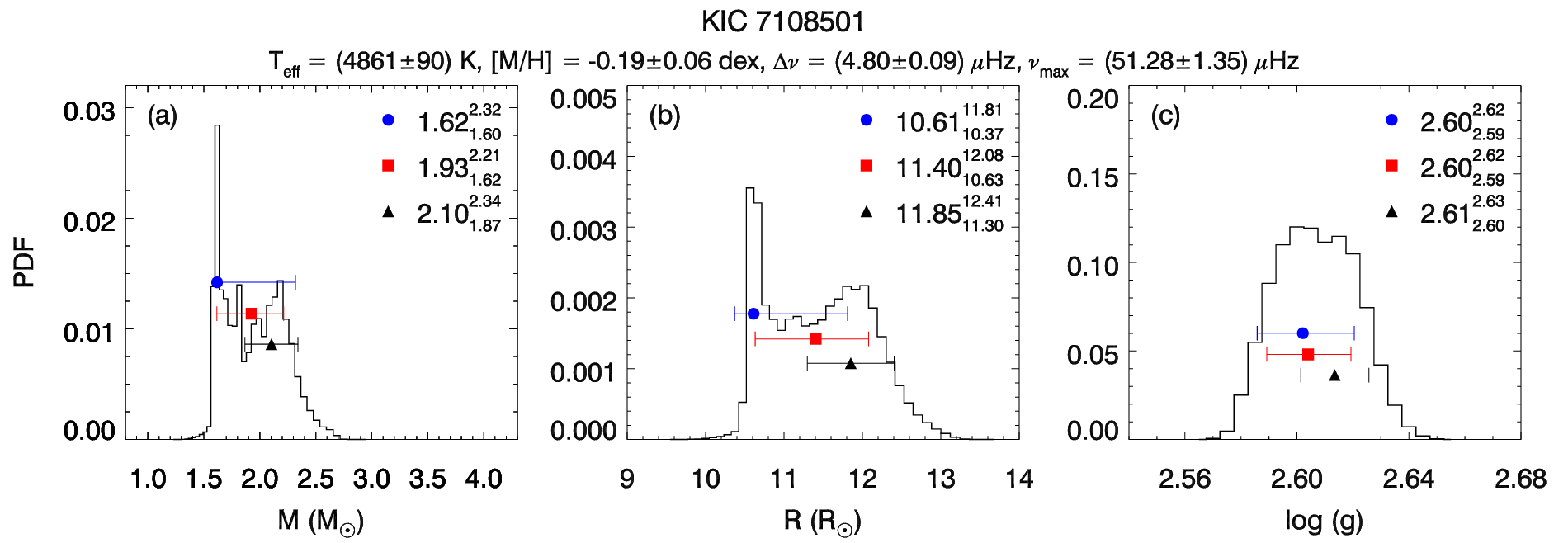}}
    \end{minipage}
    \begin{minipage}{0.7\textwidth}
       \resizebox{\hsize}{!}{\includegraphics{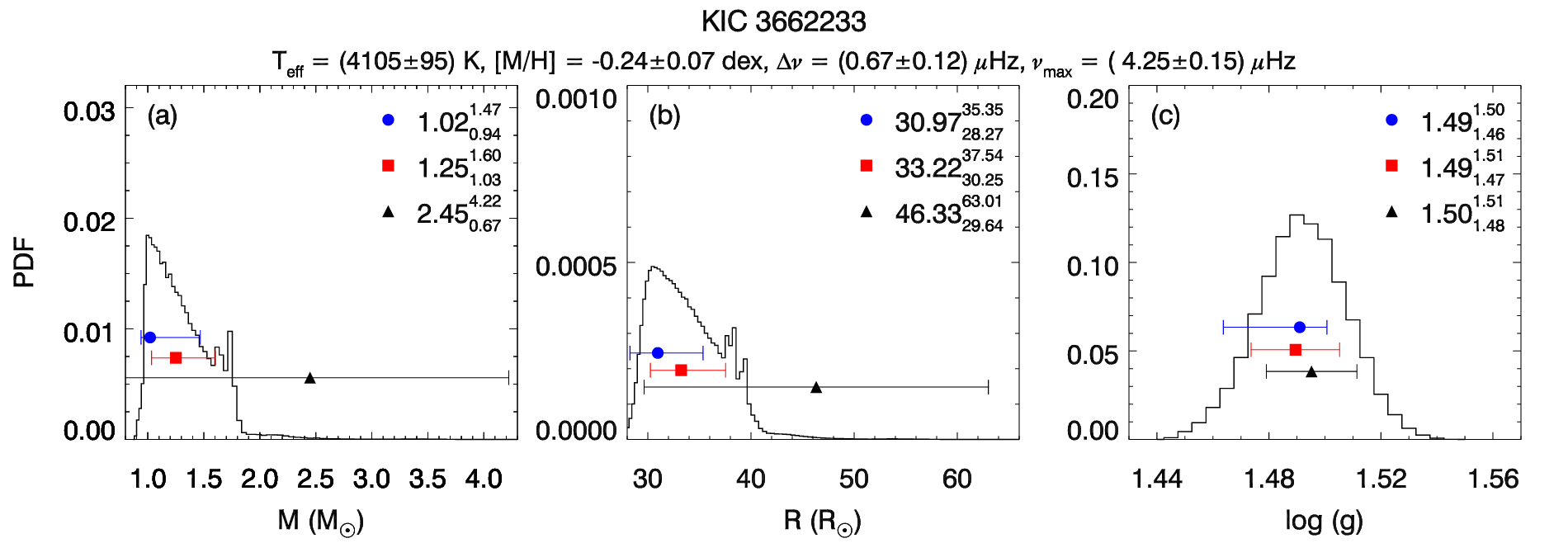}}
    \end{minipage}
    \begin{minipage}{0.7\textwidth}
      \resizebox{\hsize}{!}{\includegraphics{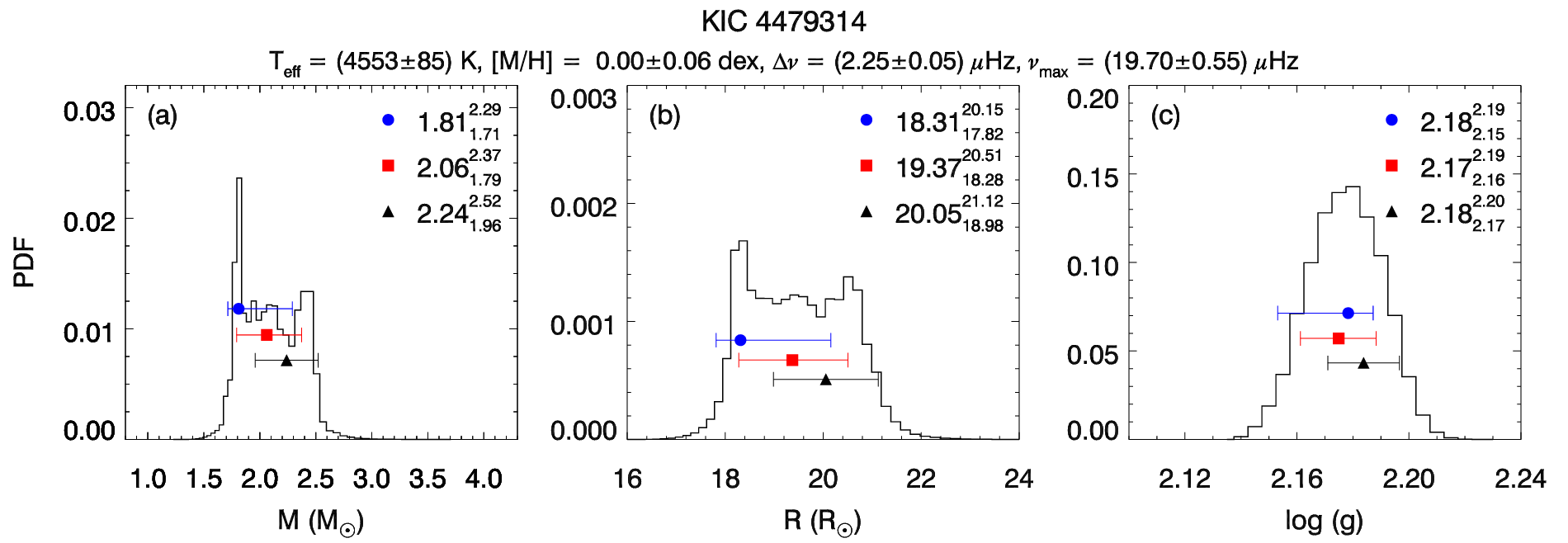}}
    \end{minipage}
    \begin{minipage}{0.7\textwidth}
      \resizebox{\hsize}{!}{\includegraphics{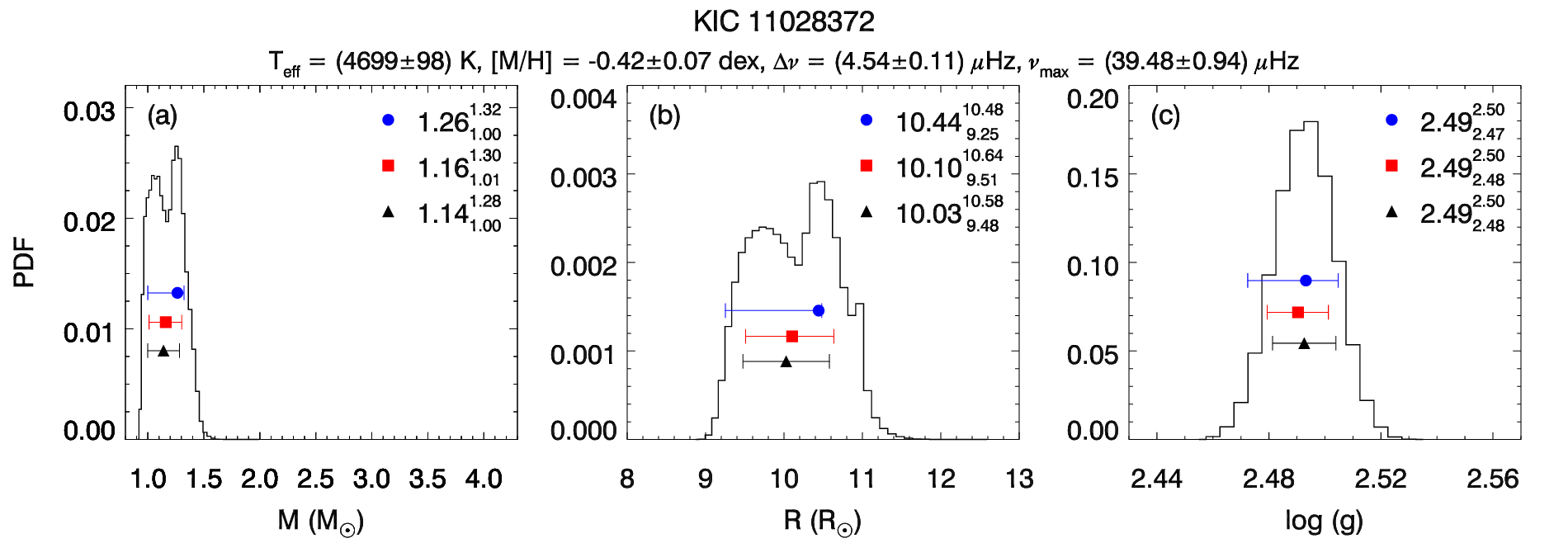}}
    \end{minipage}
  \else
    \begin{minipage}{0.7\textwidth}
      \resizebox{\hsize}{!}{\includegraphics{figs/m_r_g/pdf_m_r_g_7108501.eps}}
    \end{minipage}
    \begin{minipage}{0.7\textwidth}
       \resizebox{\hsize}{!}{\includegraphics{figs/m_r_g/pdf_m_r_g_3662233.eps}}
    \end{minipage}
    \begin{minipage}{0.7\textwidth}
      \resizebox{\hsize}{!}{\includegraphics{figs/m_r_g/pdf_m_r_g_4479314.eps}}
    \end{minipage}
    \begin{minipage}{0.7\textwidth}
      \resizebox{\hsize}{!}{\includegraphics{figs/m_r_g/pdf_m_r_g_11028372.eps}}
    \end{minipage}
  \fi
 \caption{The same as in Fig.~\ref{fig:m_r_a_g}, but for some stars with broad and/or multiple-peaked PDFs. 
}
 \label{fig:m_r_a_g_weird}
\end{figure*}

Fig.~\ref{fig:m_r_a_g_weird} shows a small set of stars for which the PDFs are extremely broad, and present multiple peaks. A large number of such situations appear in our results; a total of $\sim\!600$ out of 1989 stars. These stars are indicated as red dots in Fig.~\ref{fig:relaterrors1}. These cases happen more frequently in the upper part of the color-magnitude diagram, where confusion between stars in different long-lived evolutionary stages is possible, for instance: confusion between stars in the RGB and RC, between the RC and the RGB-bump, and between the asymptotic giant branch bump and upper RGB. Such confusion happens simply because the typical distance between such evolutionary stages, both in the \numax\ versus \deltanu\ plane and in the H--R diagram, is small and comparable to the error bars, as can be seen in Fig.~\ref{fig:hr}.

As a rule, multiple peaks are often present in \Mact\ and $R$ PDFs, but rarely in \logg. This happens because \logg\ is a direct output of \numax\ \citep[see also][]{gai11}. The most likely values for \Mact\ and $R$ may turn out to be poorly defined in these cases, which will reflect on the results of the next section. 

The situation is much improved for stars in which \deltaP\ -- and hence the evolutionary stage -- is measured. These stars often present single-peaked PDFs, although the compact nature of the RC in the \numax\ versus \deltanu\ diagram, and the presence of a slight halt in the RGB evolution at the RGB-bump, may still cause the presence of broad and multiple-peaked PDFs.

\begin{figure*}
  \ifpngfig
    \begin{minipage}{0.65\columnwidth}
      \resizebox{\hsize}{!}{\includegraphics{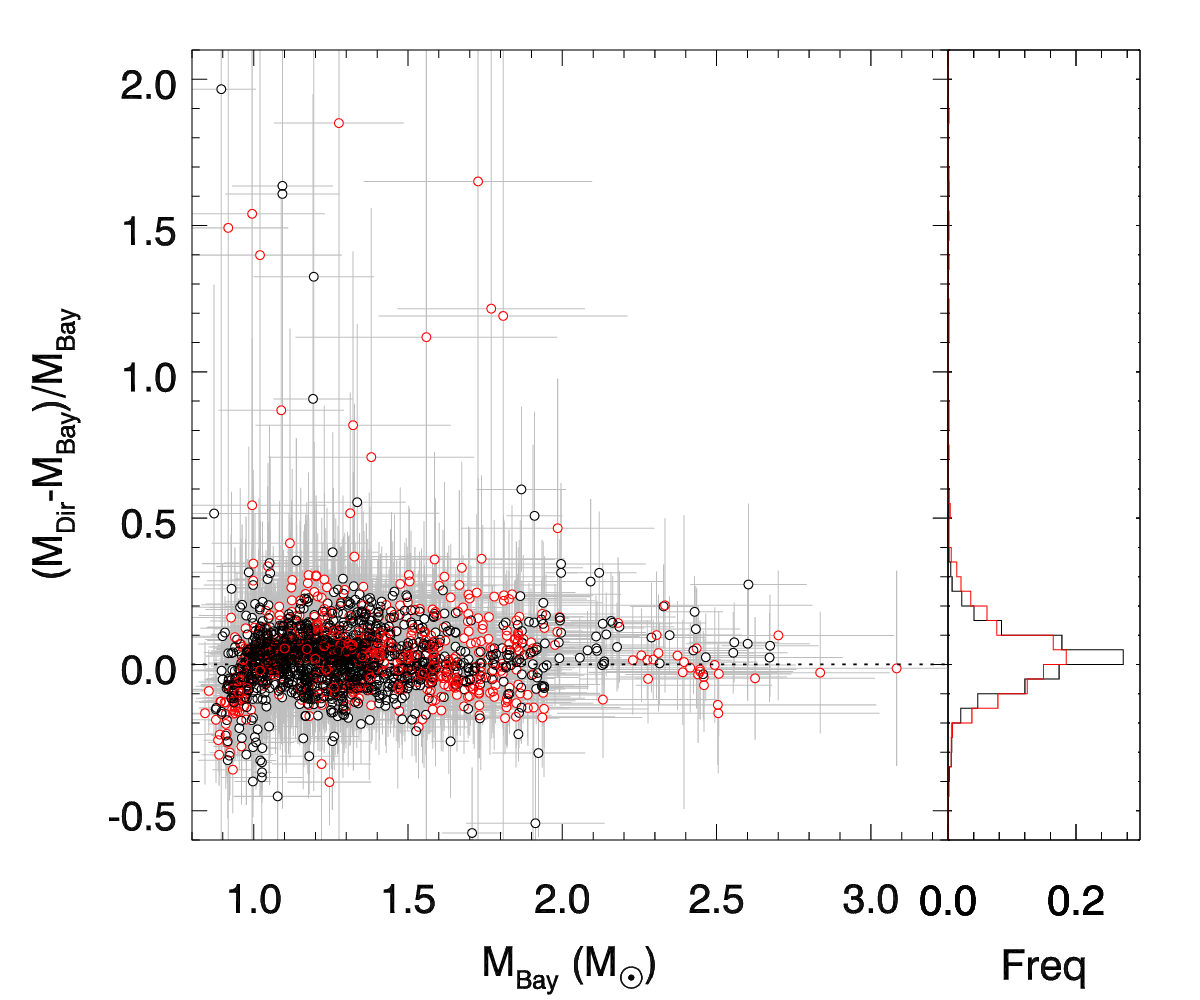}}
    \end{minipage}
    \begin{minipage}{0.65\columnwidth}
      \resizebox{\hsize}{!}{\includegraphics{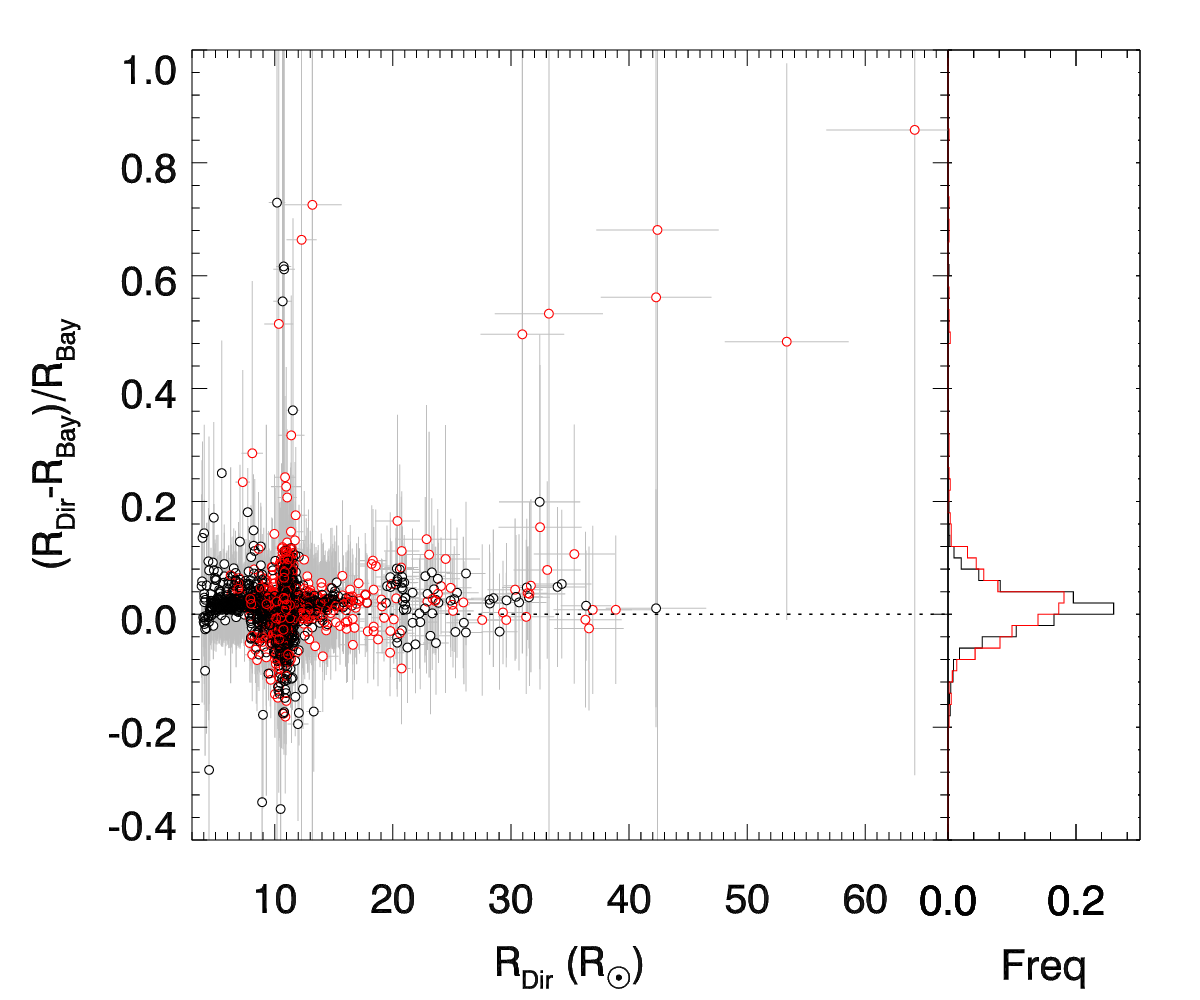}}
    \end{minipage}
    \begin{minipage}{0.65\columnwidth}
      \resizebox{\hsize}{!}{\includegraphics{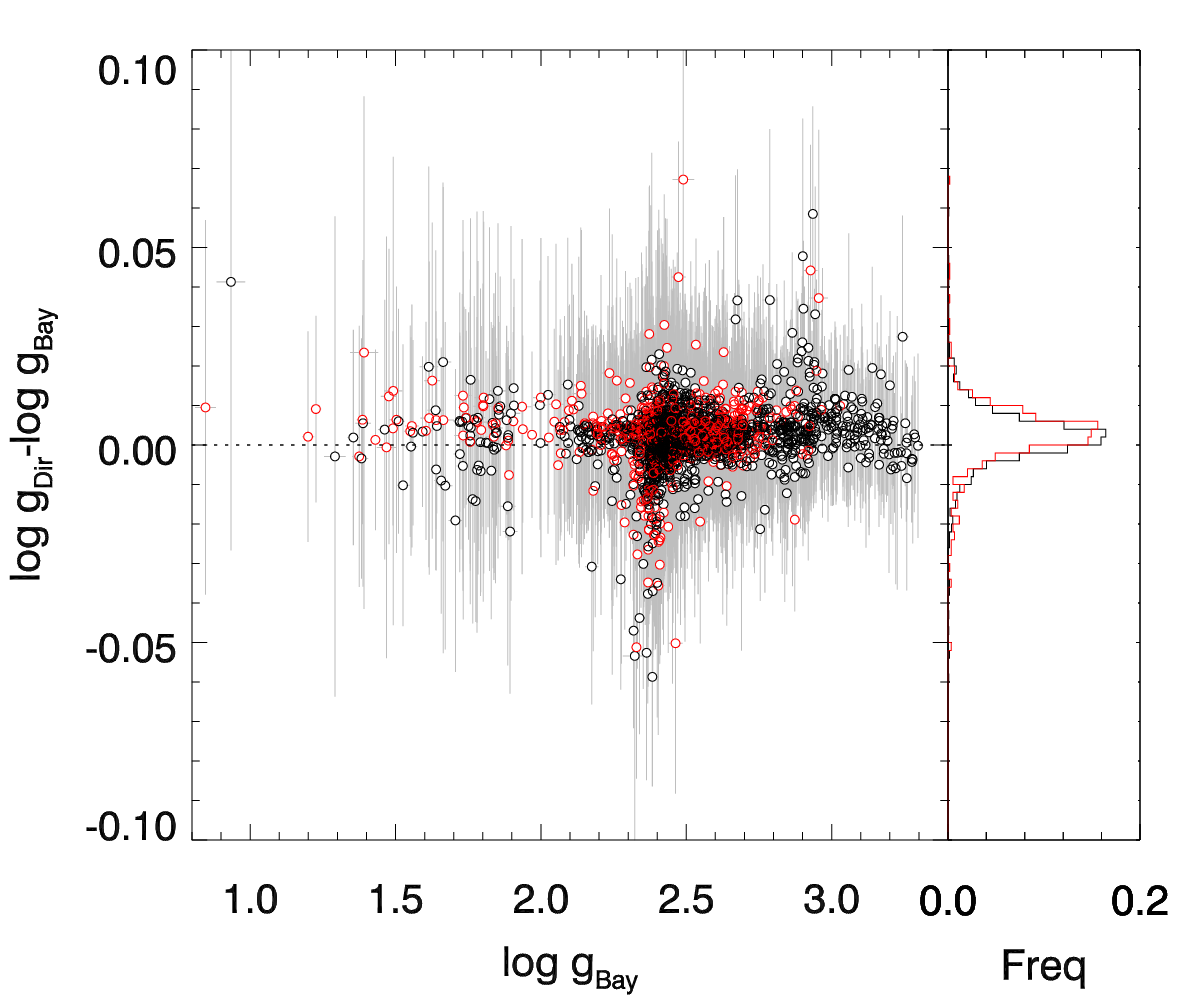}}
    \end{minipage}
  \else
    \begin{minipage}{0.65\columnwidth}
      \resizebox{\hsize}{!}{\includegraphics{figs/metdir/m_bay_dir.eps}}
    \end{minipage}
    \begin{minipage}{0.65\columnwidth}
      \resizebox{\hsize}{!}{\includegraphics{figs/metdir/r_bay_dir.eps}}
    \end{minipage}
    \begin{minipage}{0.65\columnwidth}
      \resizebox{\hsize}{!}{\includegraphics{figs/metdir/logg_bay_dir.eps}}
    \end{minipage}
  \fi
\caption{Relative (left and middle panels) and absolute (right) differences between masses, radii and \logg\ for the stars in the sample, derived with the Bayesian and the direct methods. The dashed line is the identity line. Black dots are stars with single-peaked PDFs, red dots are with broad/multiple-peaked ones. The right sub-panels show histograms of the distribution of these differences.}
 \label{fig:compdir1}
\end{figure*}

We recall that our results compare very well with the results of the direct method, as shown in Fig.~\ref{fig:compdir1}. In the direct method, the mass, radius, and \logg~are calculated directly by the scaling relations in Eq.~\ref{eq:dnu_numax}, and their uncertainties by a simple propagation of the involved uncertainties. For comparison, these values are also plotted with the Bayesian ones in Figs.~\ref{fig:m_r_a_g} and \ref{fig:m_r_a_g_weird} (black triangles). Their relative (median) uncertainties are $\sigma(\Mact_\text{Dir})/\Mact_\text{Dir}=0.13$, $\sigma(R_\text{Dir})/R_\text{Dir}=0.055$ and $\sigma(\logg_\text{Dir})=0.012$~dex. Since the Bayesian method constrains the derived parameters to be within the grid provided by the stellar models, its uncertainties are, in general, smaller than those provided by the direct method \citep[by a factor of $\sim$1.4 in radius, see also][]{gai11}. It is also interesting to note the much smaller spread in the parameters of stars at $\logg\simeq2.4$, $R\simeq10~R_\odot$, which correspond to the RC. Just a handful of outliers are observed in Fig.~\ref{fig:compdir1}; they correspond to stars with high relative uncertainties ($\gtrsim$0.2) in their seismic parameters. The mean differences of the parameters between both methods are $(\Mact_\text{Dir}\!-\!\Mact_\text{Bay})/\Mact_\text{Bay} = -0.003 \pm 0.004$, $(R_\text{Dir}\!-\!R_\text{Bay})/R_\text{Bay} = 0.003 \pm 0.002$ and $(\logg_\text{Dir}\!-\!\logg_\text{Bay}) = 0.0020 \pm 0.0004$~dex.

Our results also compare very well with the final result of the grid-based models presented in the APOKASC catalogue \citep{pinsonneault14}. These authors employed the Bellaterra Stellar Properties Pipeline \citep{serenelli13}, the grid of the BaSTI models of \citet{pietrinferni04}, and the corrected spectroscopic parameters, referred as Scale 2 in the catalogue paper. The mean of the relative differences are $(\Mact_\text{Scale2}\!-\!\Mact_\text{Bay})/\Mact_\text{Bay}=0.0003 \pm 0.0034$, $(R_\text{Scale2}\!-\!R_\text{Bay})/R_\text{Bay}=0.003 \pm 0.001$ and $(\logg_\text{Scale2}\!-\!\logg_\text{Bay})=0.0017 \pm 0.0004$~dex.


\subsection{Step 2: Determining distances and extinctions}
\label{sec:step2}

In the second step, we {\em assume} that the spectroscopically derived \Teff\ and \mh, as well as the asteroseismic \logg, are of superior quality with respect to the photometrically-derived values. This assumption is based on the fact that the results of spectroscopy and asteroseismology are essentially not affected by the stellar distances and extinctions. This allows us to derive the PDFs of the absolute magnitudes, $p(M_\lambda)$, exactly in the same way as the other stellar parameters discussed in the previous section, by properly weighting the absolute magnitudes of different isochrone sections. Details about the tables of bolometric corrections used inside the isochrones are provided in \citet{girardi02, girardi04, bonatto04}; and, \citet{marigo08}. For the ranges of \Teff\ and \logg\ relevant to our work, the bolometric corrections are based on the library of ATLAS9 synthetic spectra from \citet{castelli03}.

Based on the above, we have the PDFs of the stellar absolute magnitudes in several passbands from Step~1. These can be used to derive distances $d$ (in parsecs) via the distance modulus $\mu_0$,
\beq
 d = 10^{0.2\mu_0+1} = 10^{0.2(\mu_\lambda-A_\lambda)+1} = 10^{0.2(m_\lambda - M_\lambda - A_\lambda)+1}, 
\label{eq:distance}
\eeq
where $\mu_\lambda$, $m_\lambda$, $M_\lambda$, and $A_\lambda$ are the apparent distance modulus, apparent magnitude, absolute magnitude, and extinction in a passband denoted by $\lambda$, respectively. Assuming further that all $A_\lambda$ are related by a single interstellar extinction curve expressed in terms of its $V$-band value (that is, $A_\lambda(A_V)$), this equation can be used to derive the total extinction, $A_V$, and $d$ simultaneously. More specifically, we can derive the joint PDF: $p(d,\av)$ or $p(\muo,\av)$. We choose the second form for computational convenience, since $p(\muo)$ can be easily converted into a $p(d)$,
\beq
p(d) = \frac{5}{ln 10} \frac{p(\muo)}{d}.
\eeq

Since we have now obtained the PDFs of $M_\lambda$, it is easy to show that the PDF of the apparent distance modulus, $p(\mu_{\lambda})$, is given by the cross-correlation between the PDF of $m_\lambda$ and $M_\lambda$, assuming a normal distribution for the apparent magnitude:
\beq
p(\mu_{\lambda})  =  p(m_\lambda) \star p(M_\lambda).
\eeq
The $p(\mu_{\lambda})$ is then translated by a given value of $A_\lambda$, resulting in the joint PDF of the apparent distance modulus 
\beq
p(\mu_{0\lambda},A_\lambda) = p(\mu_{\lambda} - A_\lambda),
\eeq
which is more conveniently written as a function of the $V$-band extinction only:
\beq
p(\mu_{0\lambda},A_V) = p\left[\mu_{\lambda} - A_\lambda(A_V)\right].
\eeq
Finally, when all the passbands are combined, the result is a joint PDF for the distance modulus and extinction, 
\beq
p(\mu_0,A_V) = \prod_{i} p(\mu_{0{\lambda_i}},A_V). 
\eeq

The best agreement between $p(\mu_{0\lambda})$ will occur for a particular value of extinction that maximizes $p(\mu_0,A_V)$. This allows us to estimate the extinction simultaneously with the distance modulus. The implicit assumption that we have to make is that the extinction curve -- i.e., the coefficients $A_\lambda/A_V$ -- is well known for every star.

For the filters considered in this work, extinction coefficients are computed as described in \citet{girardi02, girardi08}, after adopting the \citet{cardelli89} and \citet{ODonnell94} extinction curve with $R_V=A_V/E({B-V})=3.1$. We adopt, for each star, the extinction coefficients derived for the measured $\Teff$ and $\logg$, and for the solar metallicity. This is a fairly good approximation, indeed. For instance, changes of $250$~K in \Teff, 0.2~dex in \logg, and 0.5~dex in \mh, cause changes in the $A_g/A_V$ coefficient of just $\sim\!0.003$, and even smaller changes for redder passbands.

Following this procedure, we computed $p(\mu_{0},A_V)$ for a range of \av\ varying from $-0.5$ to $1.0$~mag, in steps of $0.01$~mag, and covering a sufficiently large range of $\mu_{0}$, hence mapping the joint PDF of both parameters. The range of extinction includes negative values, which are obviously unphysical. Statistically, one should consider an infinite range for the parameters when building the PDFs, but in practice, one allows a very large range around the expected values to cover all the possible solutions with a significant probability. In this case, a small dispersion around $A_V=0.0$ is expected, since this is a statistical method.

\begin{figure*}
  \ifpngfig
    \begin{minipage}{0.49\textwidth}
      \resizebox{\hsize}{!}{\includegraphics{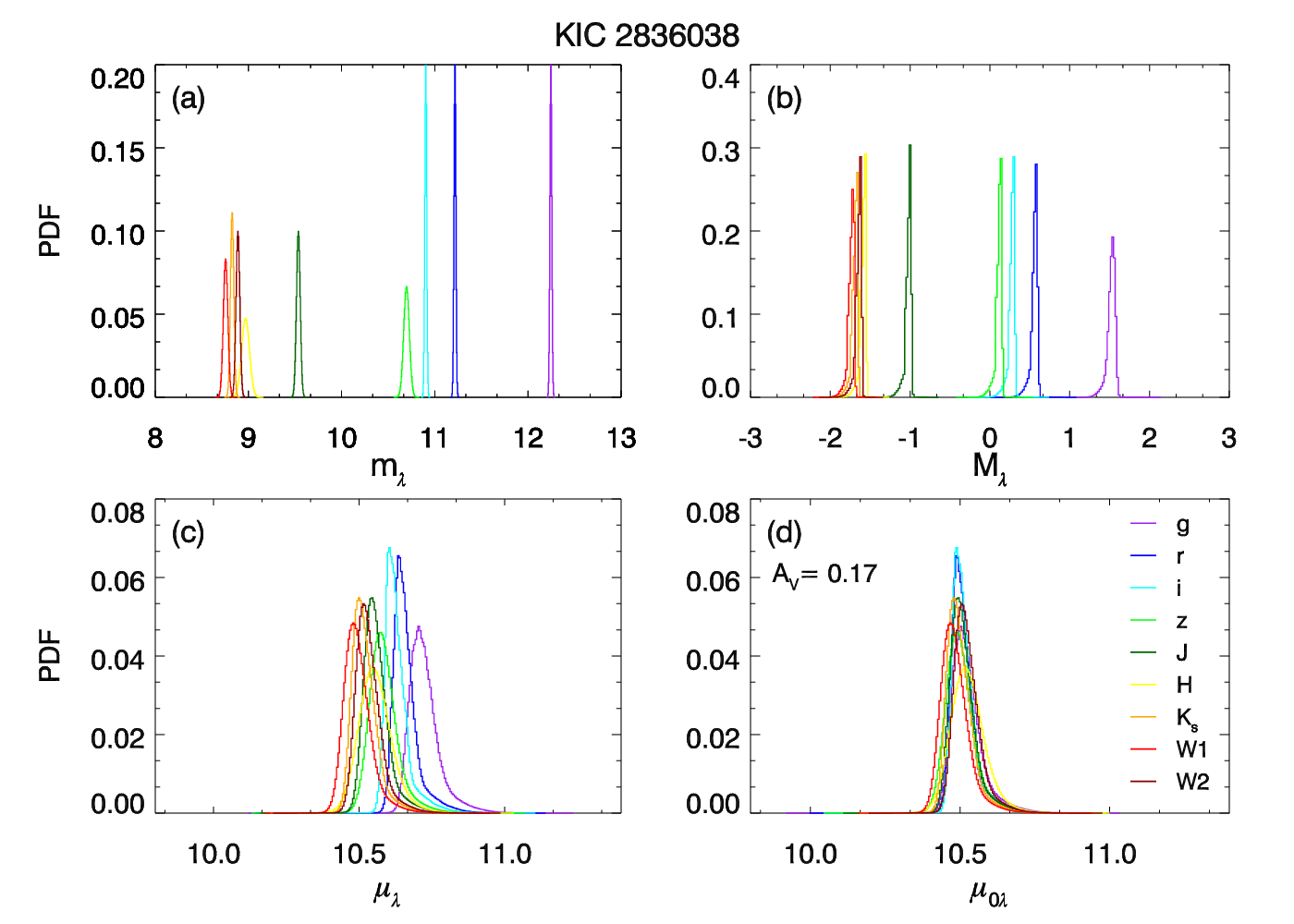}}
      \resizebox{\hsize}{!}{\includegraphics{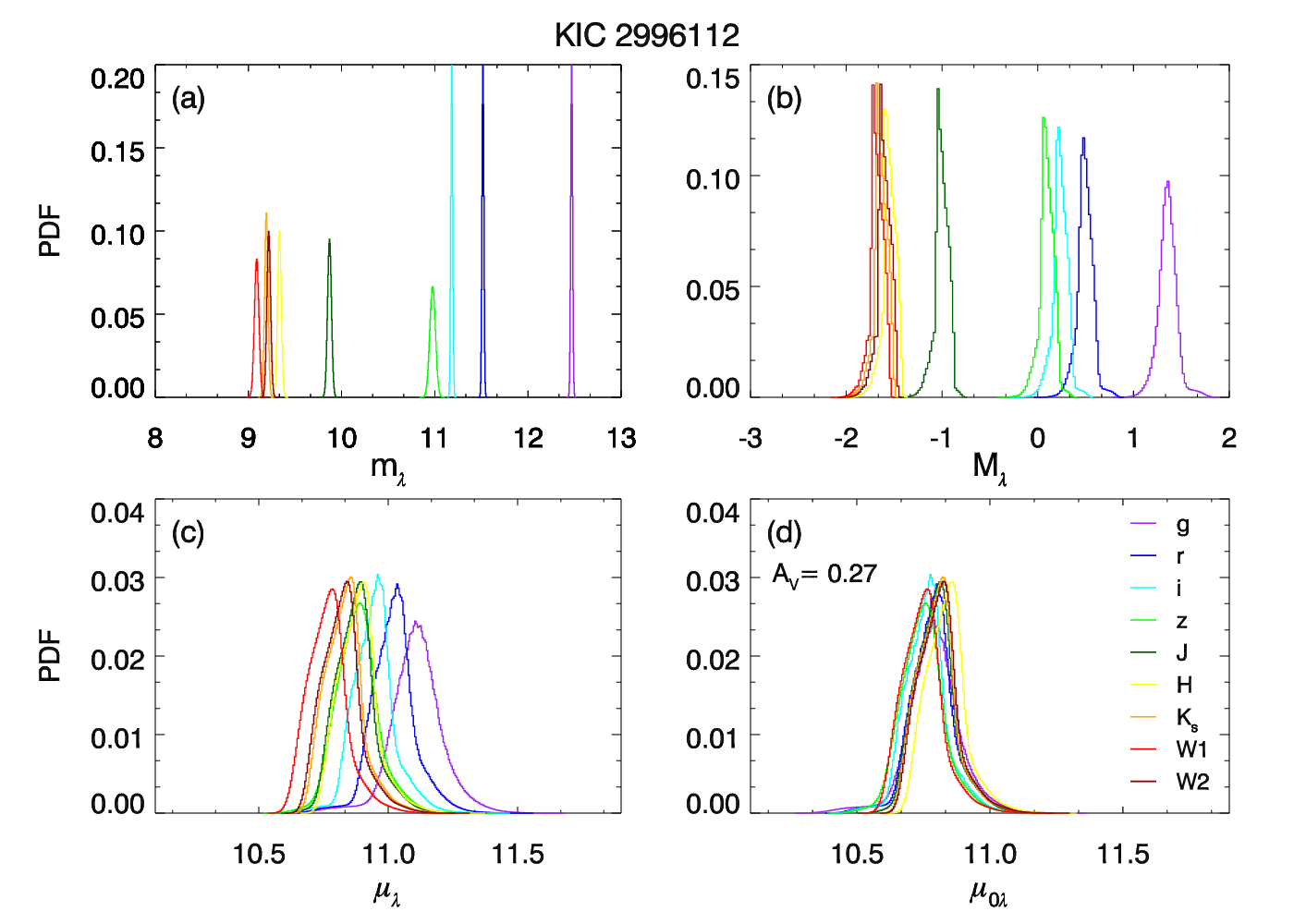}}
    \end{minipage}
    \begin{minipage}{0.49\textwidth}
      \resizebox{\hsize}{!}{\includegraphics{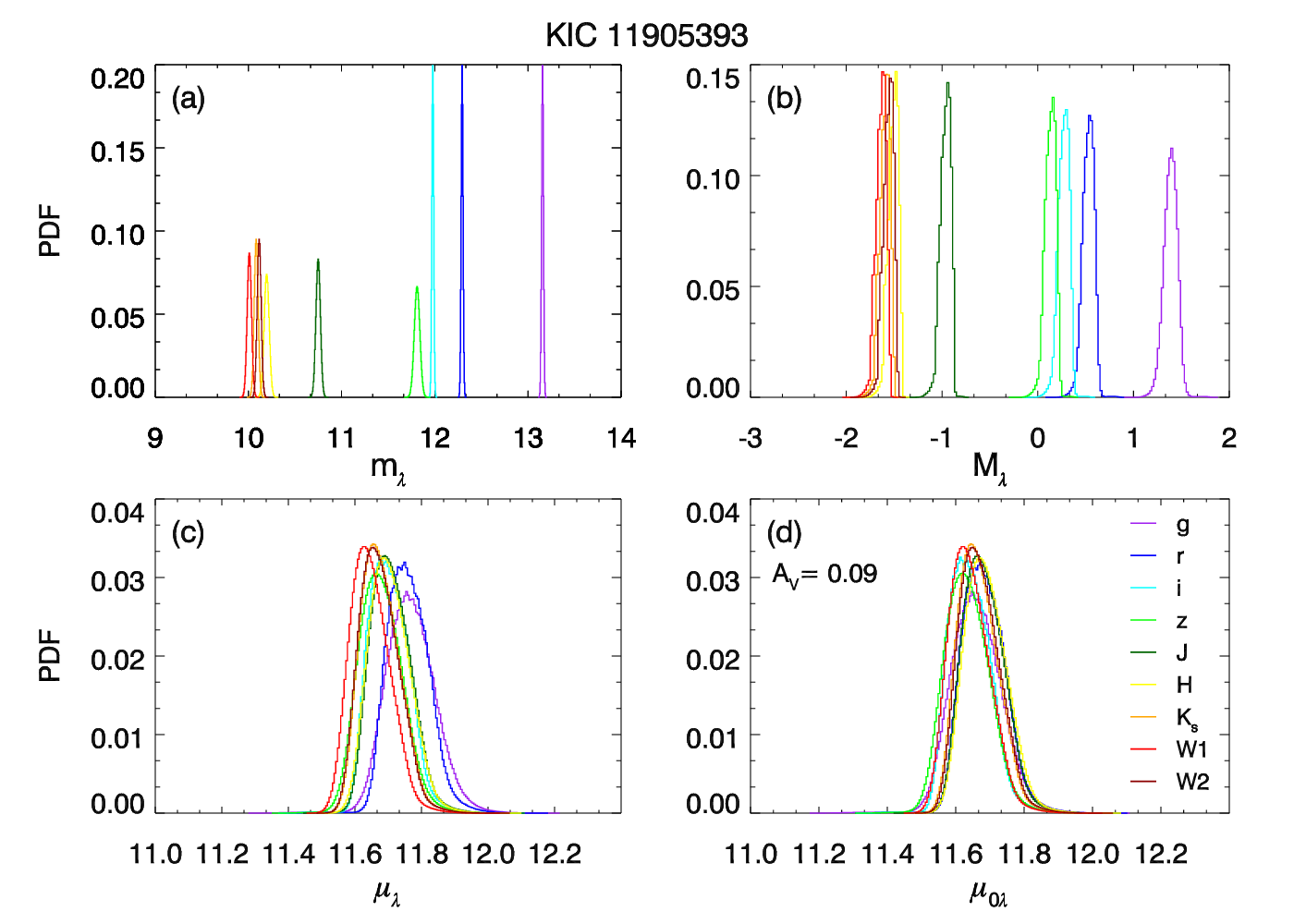}}
      \resizebox{\hsize}{!}{\includegraphics{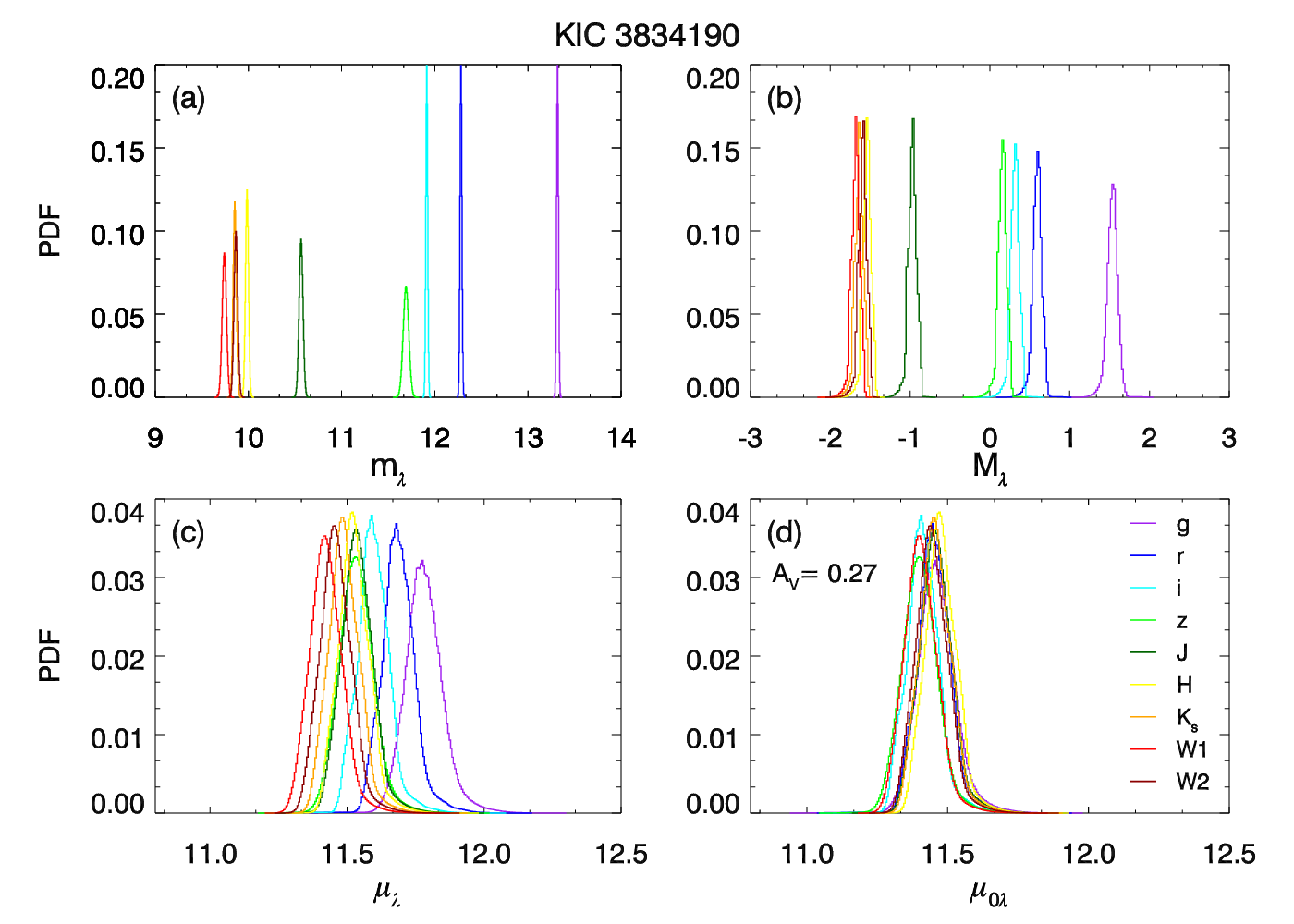}}
    \end{minipage}
  \else
    \begin{minipage}{0.49\textwidth}
      \resizebox{\hsize}{!}{\includegraphics{figs/m_M_mu_mu0/pdf_m_M_mu_mu0_2836038.eps}}
      \resizebox{\hsize}{!}{\includegraphics{figs/m_M_mu_mu0/pdf_m_M_mu_mu0_2996112.eps}}
    \end{minipage}
    \begin{minipage}{0.49\textwidth}
      \resizebox{\hsize}{!}{\includegraphics{figs/m_M_mu_mu0/pdf_m_M_mu_mu0_11905393.eps}}
      \resizebox{\hsize}{!}{\includegraphics{figs/m_M_mu_mu0/pdf_m_M_mu_mu0_3834190.eps}}
    \end{minipage}
  \fi
 \caption{(a) PDF of the apparent magnitude, (b) absolute magnitude, (c) apparent distance modulus, and (d) distance modulus for the same stars as in Fig.~\ref{fig:m_r_a_g}, and for all available passbands as detailed in the legend. The best fitting extinction is indicated in panel (d).}
 \label{fig:m_M_mu_mu0}
\end{figure*}

\begin{figure*}
  \ifpngfig
    \begin{minipage}{0.49\textwidth}
      \resizebox{\hsize}{!}{\includegraphics{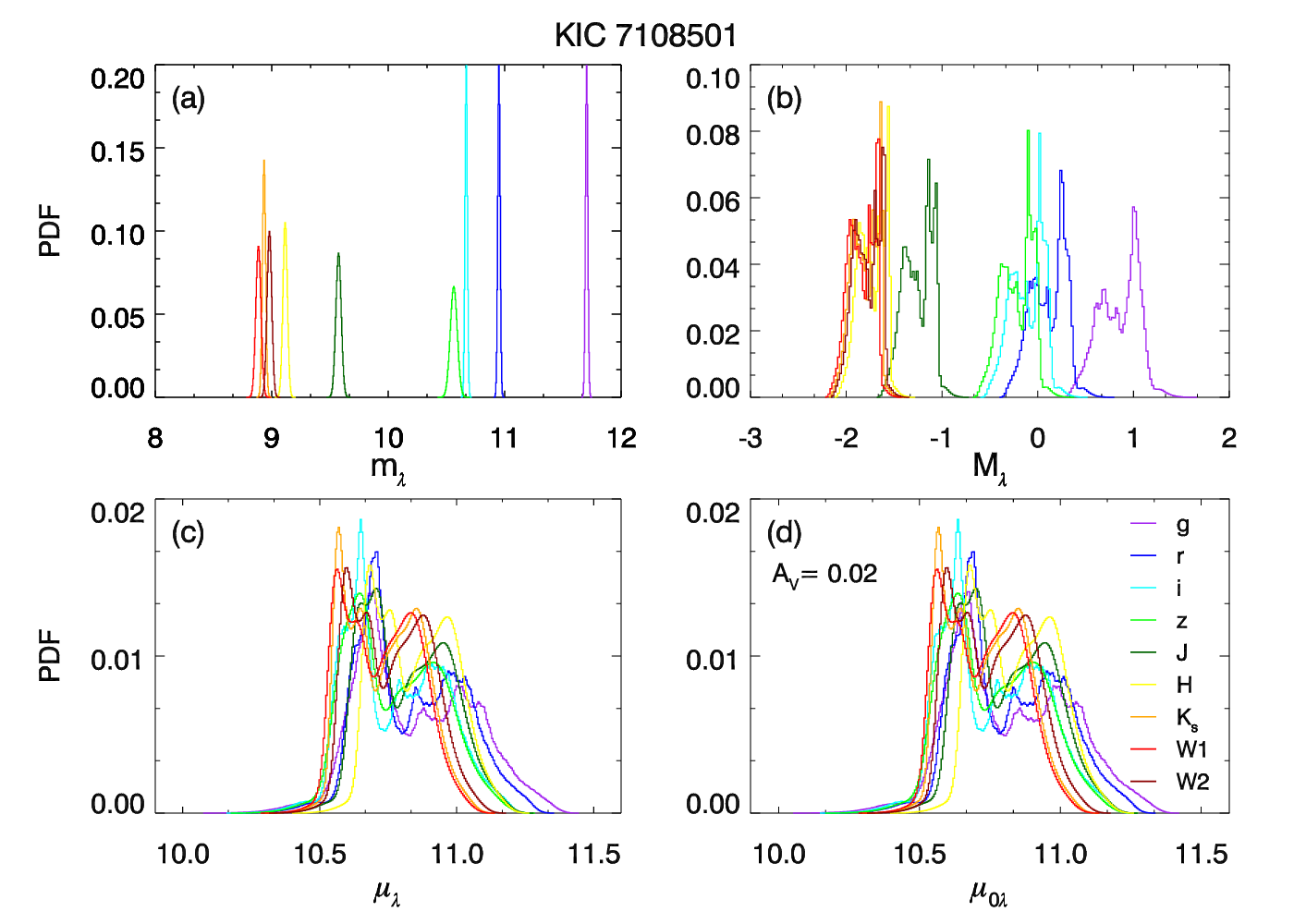}}
      \resizebox{\hsize}{!}{\includegraphics{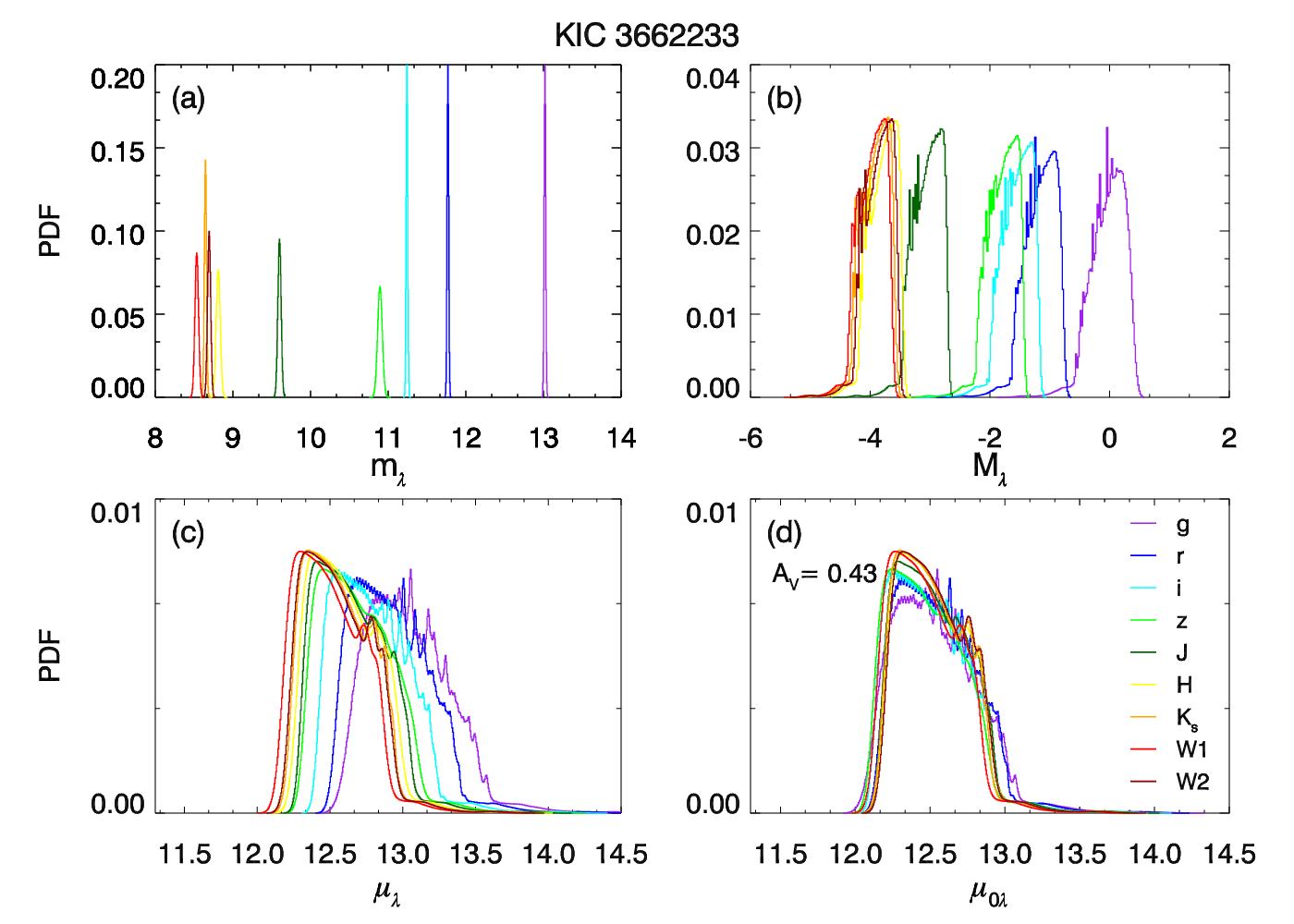}}
    \end{minipage}
    \begin{minipage}{0.49\textwidth}
      \resizebox{\hsize}{!}{\includegraphics{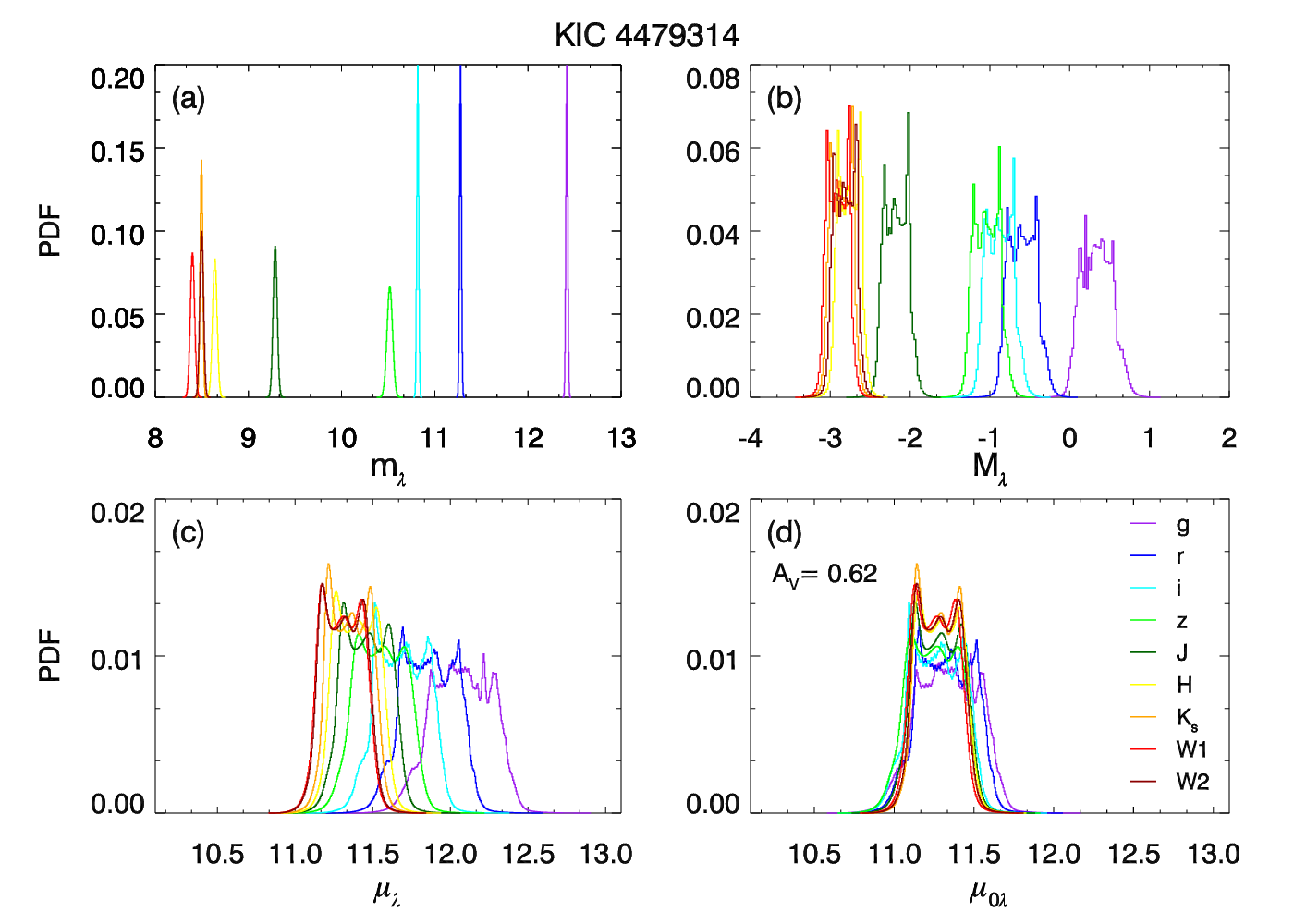}}
      \resizebox{\hsize}{!}{\includegraphics{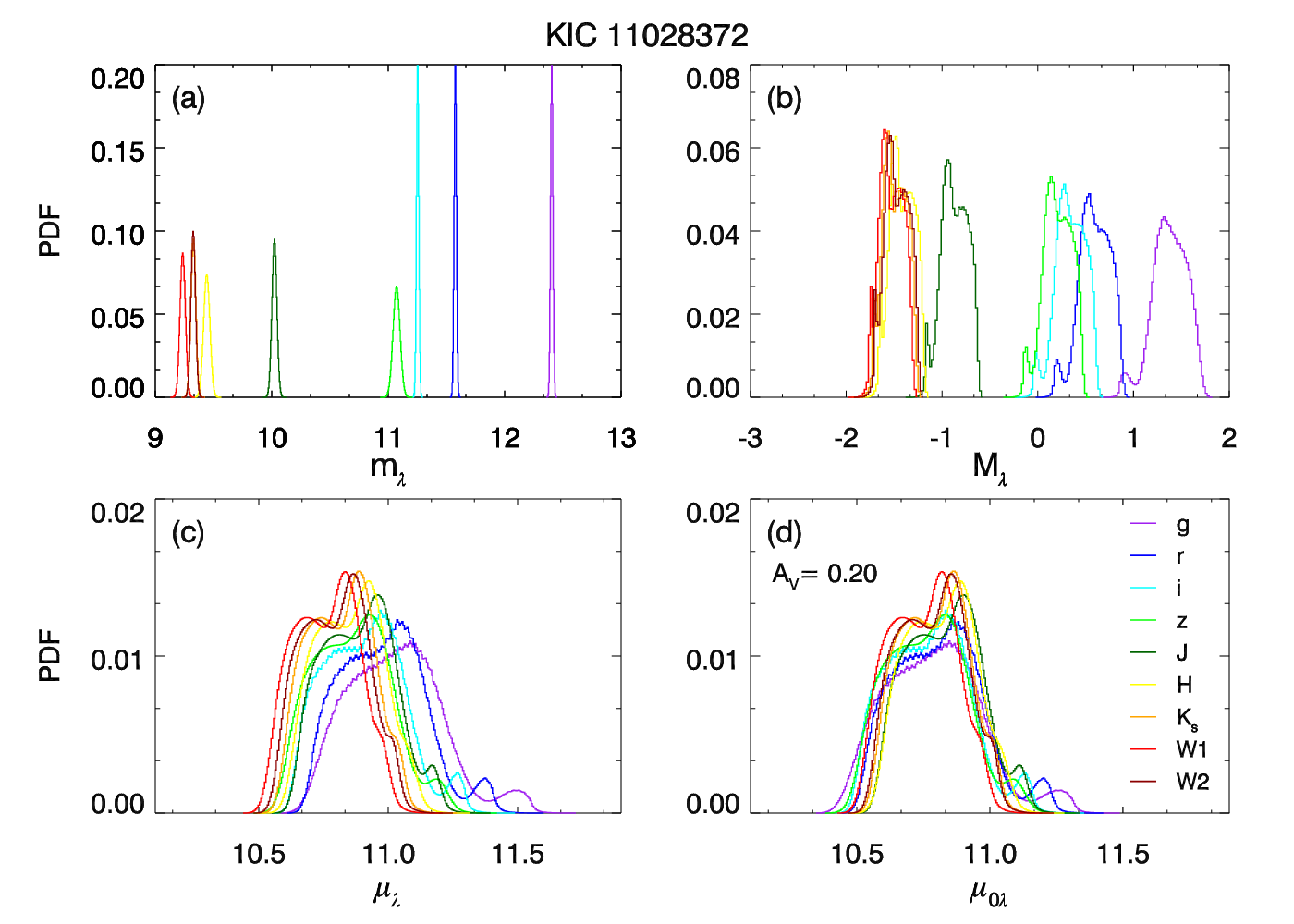}}
    \end{minipage}
  \else
     \begin{minipage}{0.49\textwidth}
      \resizebox{\hsize}{!}{\includegraphics{figs/m_M_mu_mu0/pdf_m_M_mu_mu0_7108501.eps}}
      \resizebox{\hsize}{!}{\includegraphics{figs/m_M_mu_mu0/pdf_m_M_mu_mu0_3662233.eps}}
    \end{minipage}
    \begin{minipage}{0.49\textwidth}
      \resizebox{\hsize}{!}{\includegraphics{figs/m_M_mu_mu0/pdf_m_M_mu_mu0_4479314.eps}}
      \resizebox{\hsize}{!}{\includegraphics{figs/m_M_mu_mu0/pdf_m_M_mu_mu0_11028372.eps}}
    \end{minipage}
  \fi 
 \caption{The same as in Fig.~\ref{fig:m_M_mu_mu0}, but for the stars in Fig.~\ref{fig:m_r_a_g_weird}.}
 \label{fig:m_M_mu_mu0_weird}
\end{figure*}

The entire procedure is illustrated in Figs.~\ref{fig:m_M_mu_mu0} and \ref{fig:m_M_mu_mu0_weird}, which present the PDFs of the apparent and absolute magnitudes, apparent distance modulus, and distance modulus for the same stars as in Figs.~\ref{fig:m_r_a_g} and \ref{fig:m_r_a_g_weird}, and for all available passbands as detailed in the legend. The value of \av\ that provides the best agreement between all curves is indicated in panel (d), for each star.

\begin{figure*}
  \ifpngfig
    \begin{minipage}{0.49\textwidth}
      \resizebox{\hsize}{!}{\includegraphics{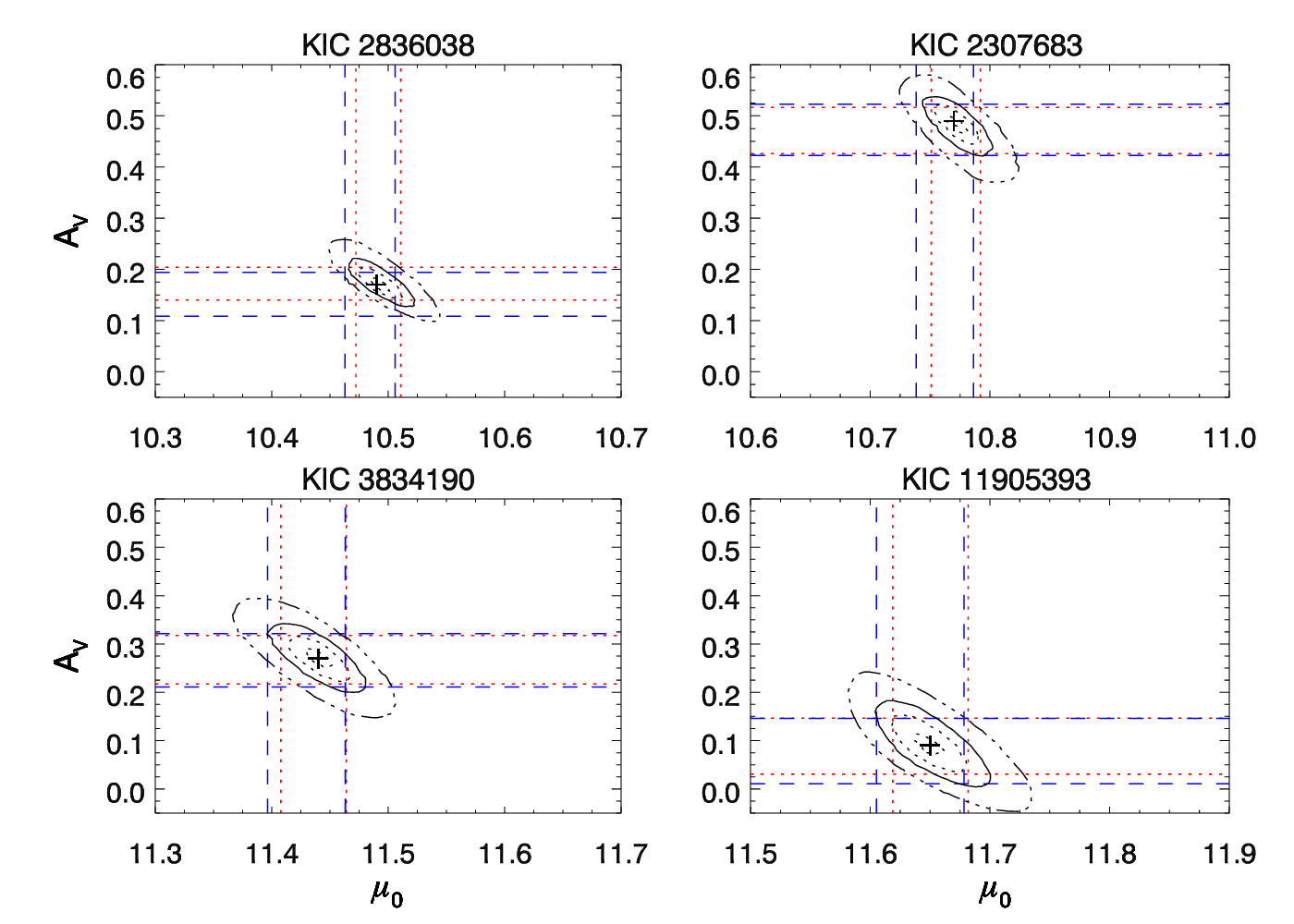}}
    \end{minipage}
    \begin{minipage}{0.49\textwidth}
      \resizebox{\hsize}{!}{\includegraphics{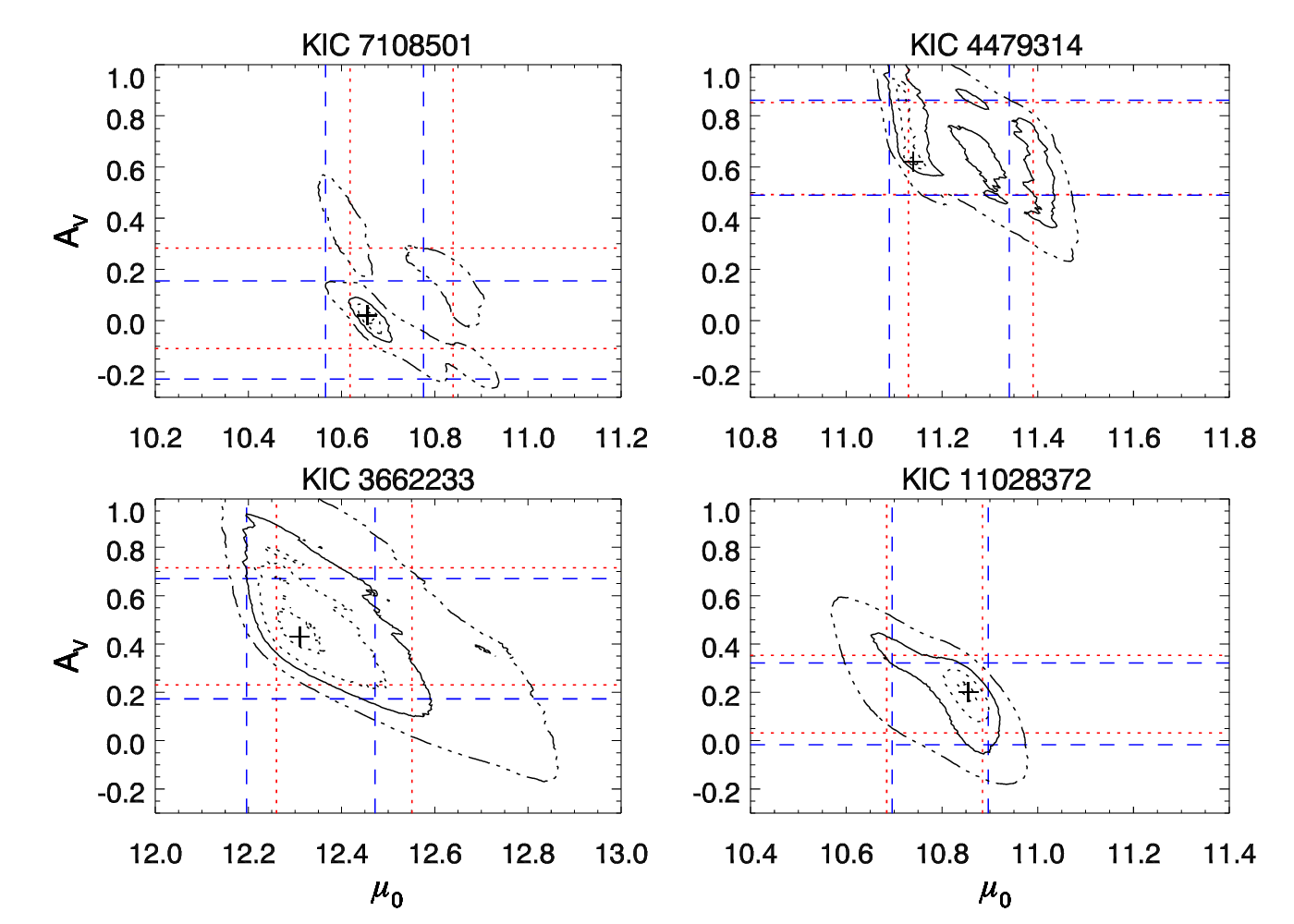}}
    \end{minipage}
  \else
    \begin{minipage}{0.49\textwidth}
      \resizebox{\hsize}{!}{\includegraphics{figs/contour/cont_1.eps}}
    \end{minipage}
    \begin{minipage}{0.49\textwidth}
      \resizebox{\hsize}{!}{\includegraphics{figs/contour/cont_2.eps}}
    \end{minipage}
  \fi
 \caption{Contour levels of the distance modulus and extinction probability space, for the same single-peaked PDF stars as in Fig.~\ref{fig:m_r_a_g} (four left panels) and the broad/multiple-peaked PDF stars as in Fig.~\ref{fig:m_r_a_g_weird} (four right panels). The solid and triple-dot-dashed contours represent the 68 and 95 per cent credible regions. The dashed blue and dotted red lines represent the 68 per cent credible interval for the mode and median of both \av\ and \muo. The plus symbol is the maximum of the joint probability.}
 \label{fig:cont}
\end{figure*}

\begin{figure*}
  \ifpngfig
    \begin{minipage}{0.35\textwidth}
      \resizebox{\hsize}{!}{\includegraphics{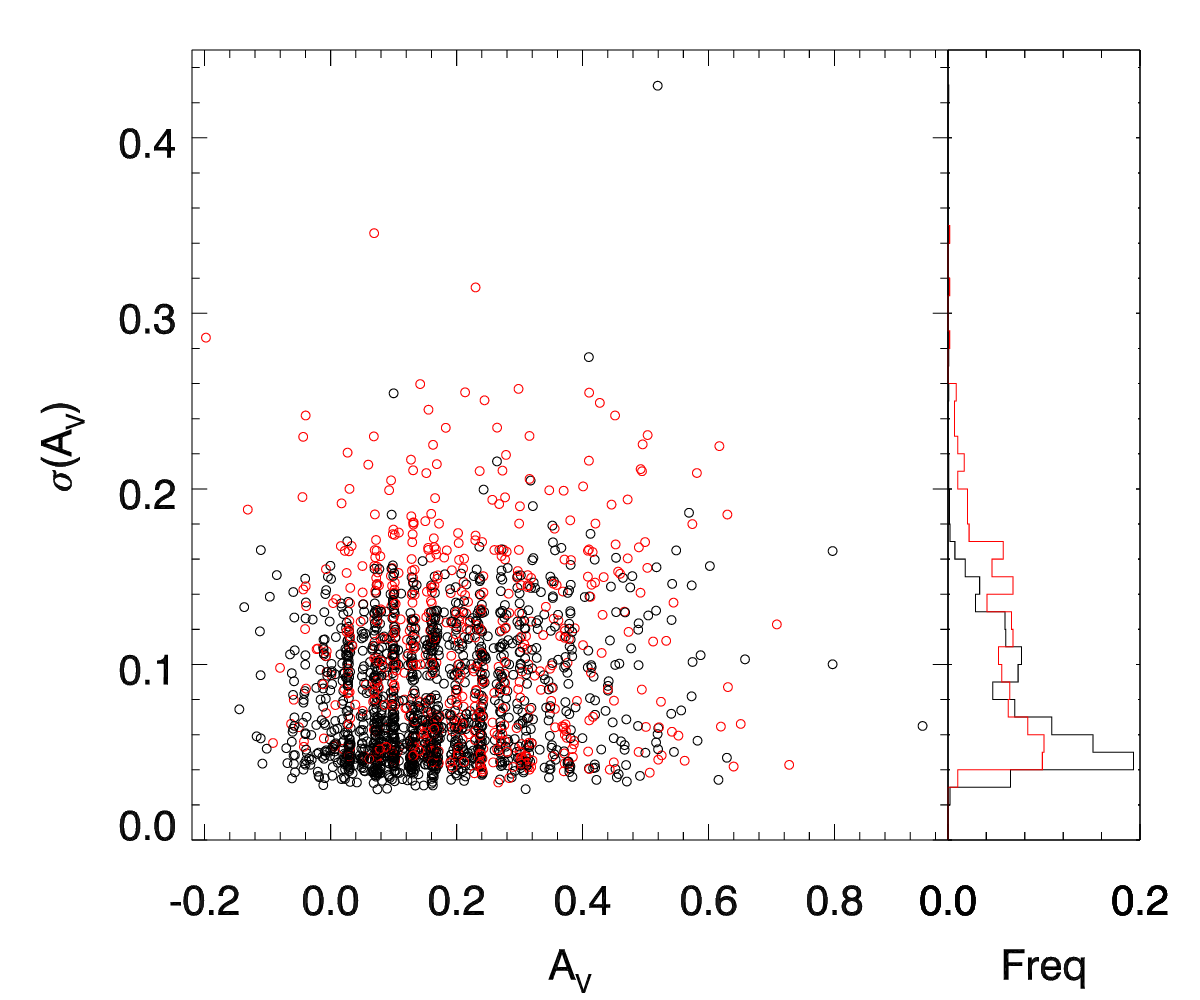}}
    \end{minipage}
    \begin{minipage}{0.35\textwidth}
      \resizebox{\hsize}{!}{\includegraphics{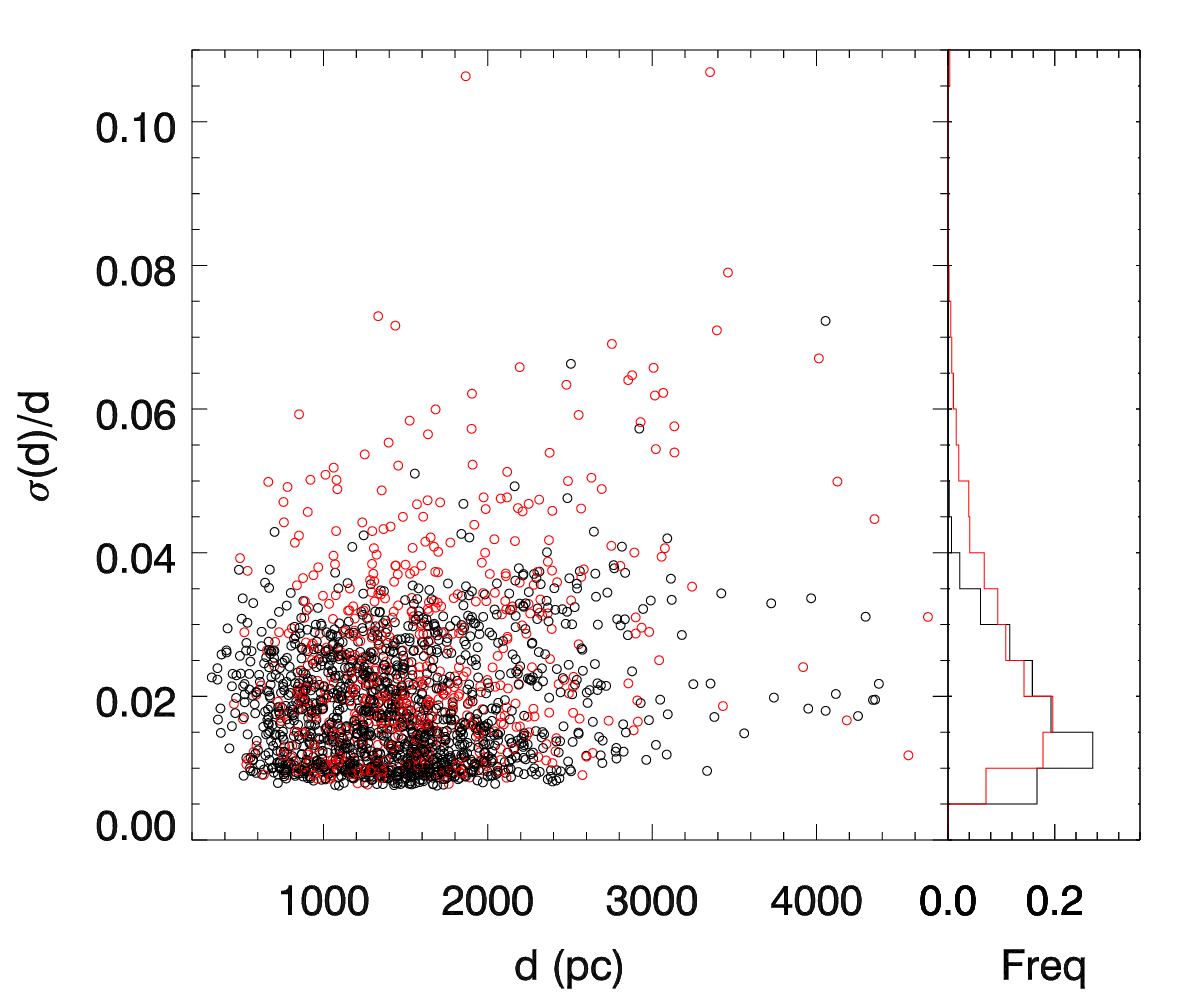}}
    \end{minipage}
  \else
    \begin{minipage}{0.35\textwidth}
       \resizebox{\hsize}{!}{\includegraphics{figs/uncert/av_er_hist.eps}}
    \end{minipage}
    \begin{minipage}{0.35\textwidth}
      \resizebox{\hsize}{!}{\includegraphics{figs/uncert/dist_er_hist.eps}}
    \end{minipage}
  \fi
 \caption{The same as in Fig.~\ref{fig:relaterrors1}, but for the quantities derived in Step~2 (Sec.~\ref{sec:step2}) -- namely \av\ and $d$.}
 \label{fig:relaterrors2}
\end{figure*}

The four left panels in Fig. \ref{fig:cont} present the contour levels of distance modulus and extinction probability space, for the same stars with single-peaked $p(\mu_{0\lambda})$ of Fig.~\ref{fig:m_M_mu_mu0}. The solid and triple-dot-dashed contours represent the 68 and 95 per cent credible regions. The dashed-blue and dotted-red lines represent the same credible interval calculated from the marginal PDF of each parameter, for the mode and median, respectively. The plus symbol is the maximum of the joint probability. What is remarkable in this plot is the excellent precision in determining the distances and \av, with typical (median) uncertainties of $\sigma(d)/d=0.018$ and $\sigma(\av)=0.077$~mag. The uncertainties in extinction and relative uncertainties in distance for the full sample are presented in Fig.~\ref{fig:relaterrors2}. 

\begin{figure*}
  \ifpngfig
    \resizebox{0.52\hsize}{!}{\includegraphics{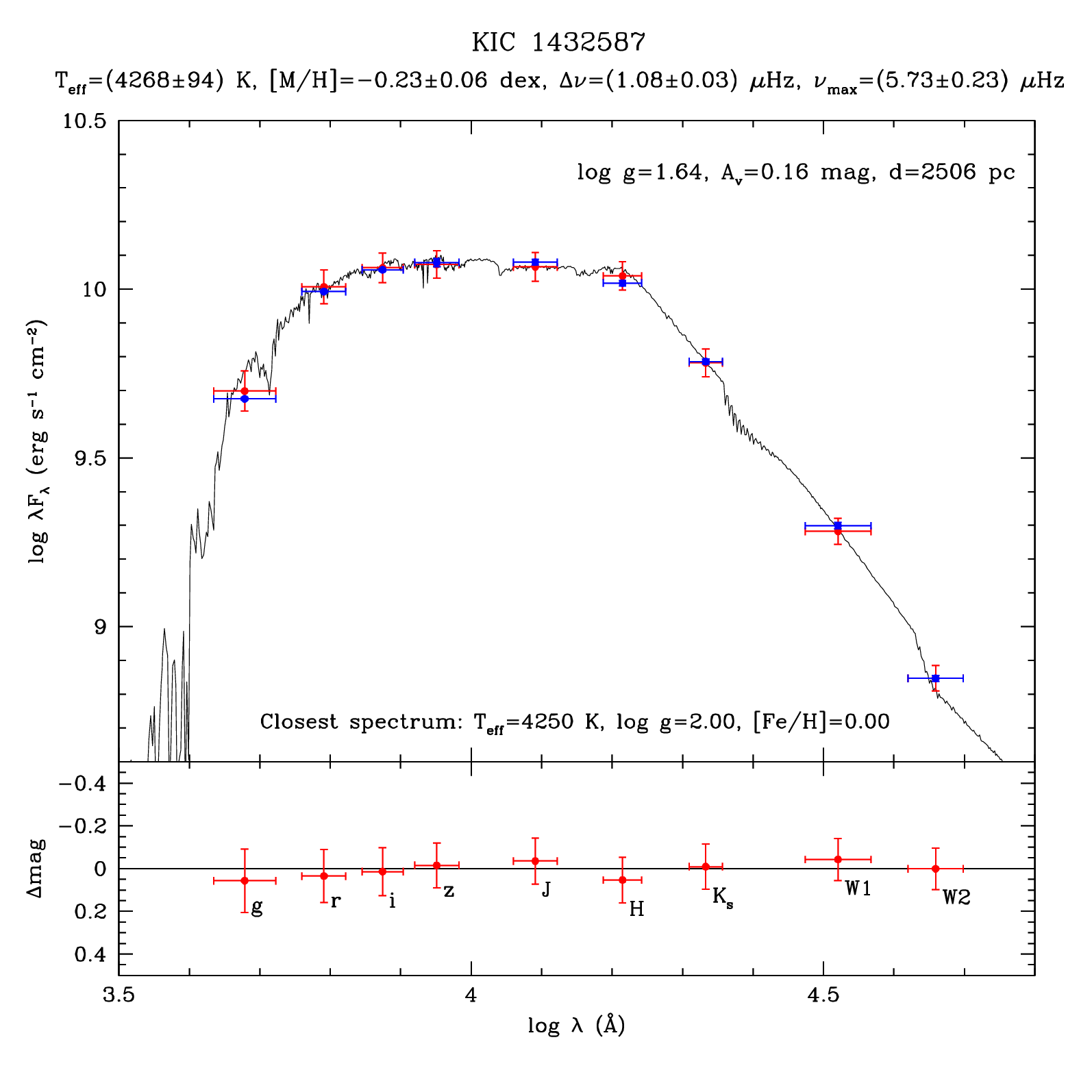}}
  \else
    \resizebox{0.52\hsize}{!}{\includegraphics{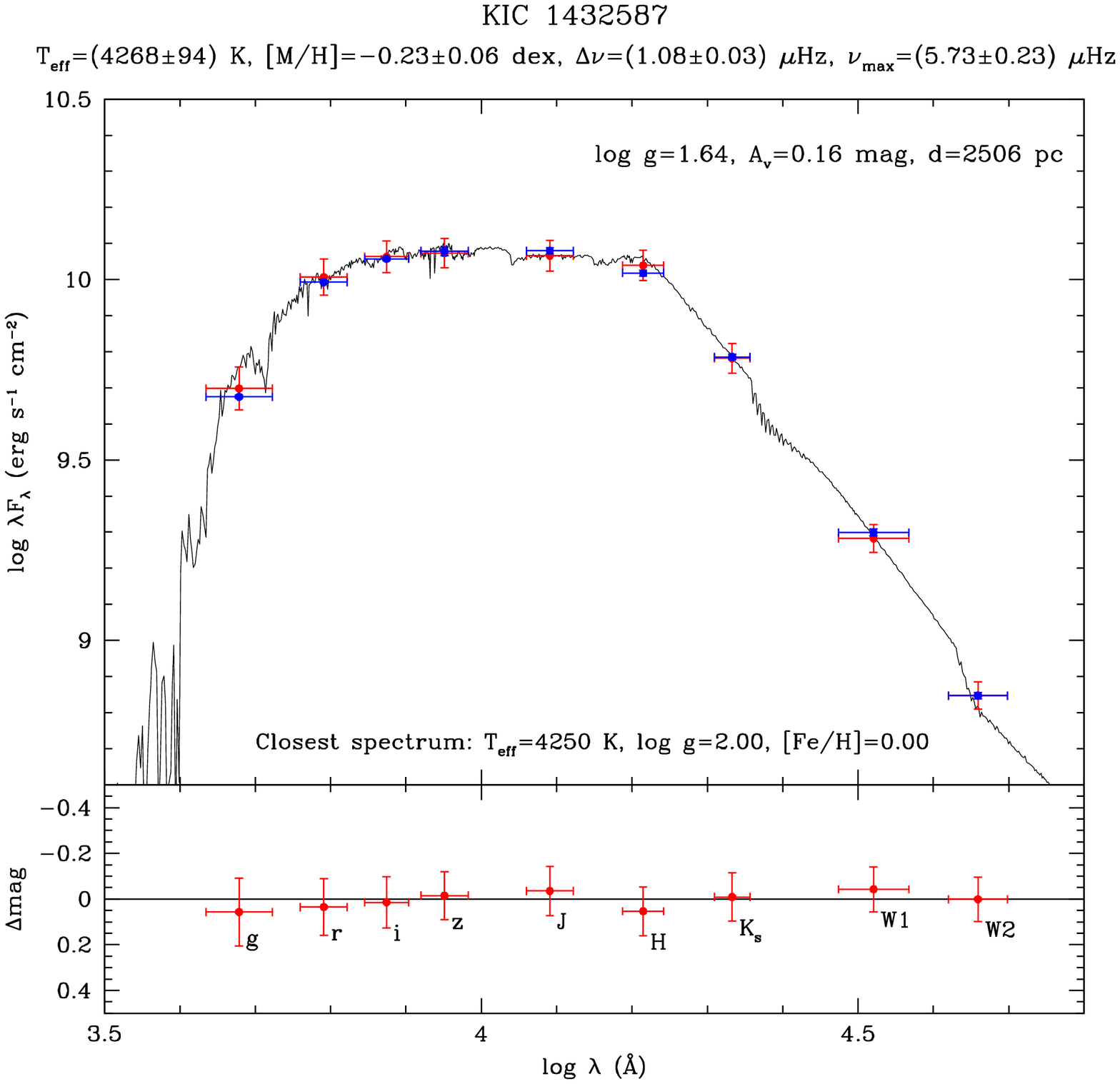}}
  \fi
  \caption{Example of the `SED fitting' being performed by our method. Top panel: The plotted spectrum represents the flux (or, better, $\lambda F_\lambda$) at the stellar surface, for the entry in the ATLAS9 data base with the closest value in \Teff, \logg, and \mh, as the observed star; it is shown for illustrative purposes only. The red dots represent the absolute magnitudes of the star, as inferred from the asteroseismic plus spectroscopic constraints, and converted to the same flux scale as the spectrum. Vertical error bars represent the 68 per cent CI interval, while the horizontal error bars are simply indicating the approximate spectral range of each filter. The blue dots with error bars are the same but for the observed magnitudes, after corrected by the inferred (mode) distance and extinction. Bottom panel: The difference between the inferred magnitudes and the observed ones, as a function of wavelength.}
  \label{fig:sedfitting}
\end{figure*}

The four right panels in Fig.~\ref{fig:cont} present the contour levels of $p(\mu_0,\av)$ for the stars with multimodal PDFs of the Fig.~\ref{fig:m_r_a_g_weird}. The effect of broad/multiple-peaked PDFs is evident: the uncertainty in distance and extinction is much larger, with the presence of secondary peaks (or `extended islands') which represent alternative values for distance and extinction. We treat these cases exactly as before. These stars will appear with larger uncertainties in our final catalogue.

Finally, Fig.~\ref{fig:sedfitting} illustrates that our method is in a way similar to a classical `spectral energy distribution (SED) fitting'. We find the combination of extinction and distance that fits the overall spectrum of the star (as sampled by the photometric points), but in addition, \textit{we consider the tight constraints imposed by the asteroseismic plus spectroscopic data}. In such plots, the bulk of our stars are well-described by a single SED from the $g$ to $W2$ passbands. There are a few cases of stars for which there appears to be a slight excess flux either in the blue or infrared portions of the spectrum, which might indicate the presence of stellar companions. The stars KIC~9479404 and KIC~10157507 present excess flux in the middle of the spectrum, which more likely indicates a problem with the photometry. Such cases will be examined in detail before the next release of the APOKASC catalogue.

\subsection{Why two separate steps?}
\label{sec:whytwo}

It is important to note that the entire procedure of Step~2, as summarized in Eq.~\ref{eq:distance}, apparently does not involve anything else from Step~1 than the PDFs for the absolute magnitudes (Sec.~\ref{sec:step1}), and hence it can be kept separated from the derivation of the other stellar properties performed in Step~1. In other words, we have chosen to approximate $p(\Mact,R,\logg,\tau,M_{\lambda_i},\av,\mu_{0\lambda_i})\sim p(\Mact,R,\logg,\tau,M_{\lambda_i})\,p(\av,\mu_{0\lambda_i})$. This approximation is not perfectly ideal, because Eq.~\ref{eq:distance} involves quantities that depend on the spectral shape -- and hence on \Teff, \logg, and \mh\ -- namely, the set of $M_{\lambda_i}$, and the set of $A_{\lambda_i}/A_V$. Therefore, the most correct procedure would have been a simultaneous derivation of the PDF of all stellar parameters in Steps 1 and 2, using every possible point of the parameter space $(\Mact,\tau,\mh,\Teff,\deltanu,\numax,m_{\lambda_i})$ in the derivation of a posterior probability for $(\Mact,R,\logg,\tau,M_{\lambda_i},\av,\mu_{0\lambda_i})$. The reasons why we do not follow this procedure here are: (1) to keep the required computing resources within reasonable limits; and, mainly, (2) because both effects have a limited impact in our distance estimates, as quantified below. 

The full set of $M_{\lambda}$ varies primarily as a function of \Teff, which is the origin of the well-known \Teff--color relations. So stars with different \Teff\ ranges will result in systematically different sets of intrinsic colors (in Step~1). This may be mistaken by different values of reddening and hence \av, which impacts the distances. This is likely the mechanism that, for stars with multiple-peaked PDFs, result in alternative peaks in the $(A_V,\muo)$ plane in Fig.~\ref{fig:cont}. These cases comprise of less than 30 per cent of our sample. For stars with single-peaked PDFs, we have investigated the effect of a 100~K {\em systematic change} in the \Teff\ scale in Sec.~\ref{sec:systshifts}, which in turn produces a small, although non-negligible effect, in the extinction value. Since the typical uncertainties in our \Teff\ are $\sim$86~K (2 per cent), it is unlikely that the variations of $M_{\lambda_i}$  with \Teff\ (internally to the Bayesian method) can have such a large impact on the final results.

The extinction coefficients $A_\lambda/A_V$ are also a function of the spectral shape, and hence of \Teff, \logg, and \mh\ \citep[see e.g.,][]{grebel95, girardi08}. However, as already mentioned, the changes of $A_\lambda/A_V$ with $(\Teff, \logg, \mh)$, inside the intervals considered in this work, are actually very small, and much less than those caused by spatial variations in the interstellar extinction law \citep{zasowski09}.
 
\subsection{Comparison with a direct method}
\label{sec:compdirmet}

Distances and extinctions can also be derived in a more direct way, starting from the stellar parameters provided by the direct method with Eq.~\ref{eq:dnu_numax}. Essentially, we enter the $R$ and \Teff\ in Eq.~\ref{eq:lum} to derive $L$, which is then transformed into a bolometric absolute magnitude, and into the absolute magnitude in several filters using the bolometric corrections ($BC_\lambda$) inferred from our library of synthetic stellar SEDs. These are then processed through Step 2 of our method, which allows us to identify the distance and extinction, $d_\text{Dir}$ and $A_\text{V,Dir}$, that best fit the set of observed apparent magnitudes. Error bars are obtained by simply propagating the uncertainties in the quantities $R$, \Teff, and $BC_\lambda$, into the absolute magnitudes. The final uncertainties in $d_\text{Dir}$ and $A_\text{V,Dir}$ are derived given by the joint PDF, exactly as in the Bayesian method.
The median uncertainties turn out to be $\sigma(d_\text{Dir})/d_\text{Dir}=0.038$ and $\sigma(A_\text{V,Dir})=0.15$~mag, respectively.

\begin{figure*}
  \ifpngfig
    \begin{minipage}{0.68\columnwidth}
      \resizebox{\hsize}{!}{\includegraphics{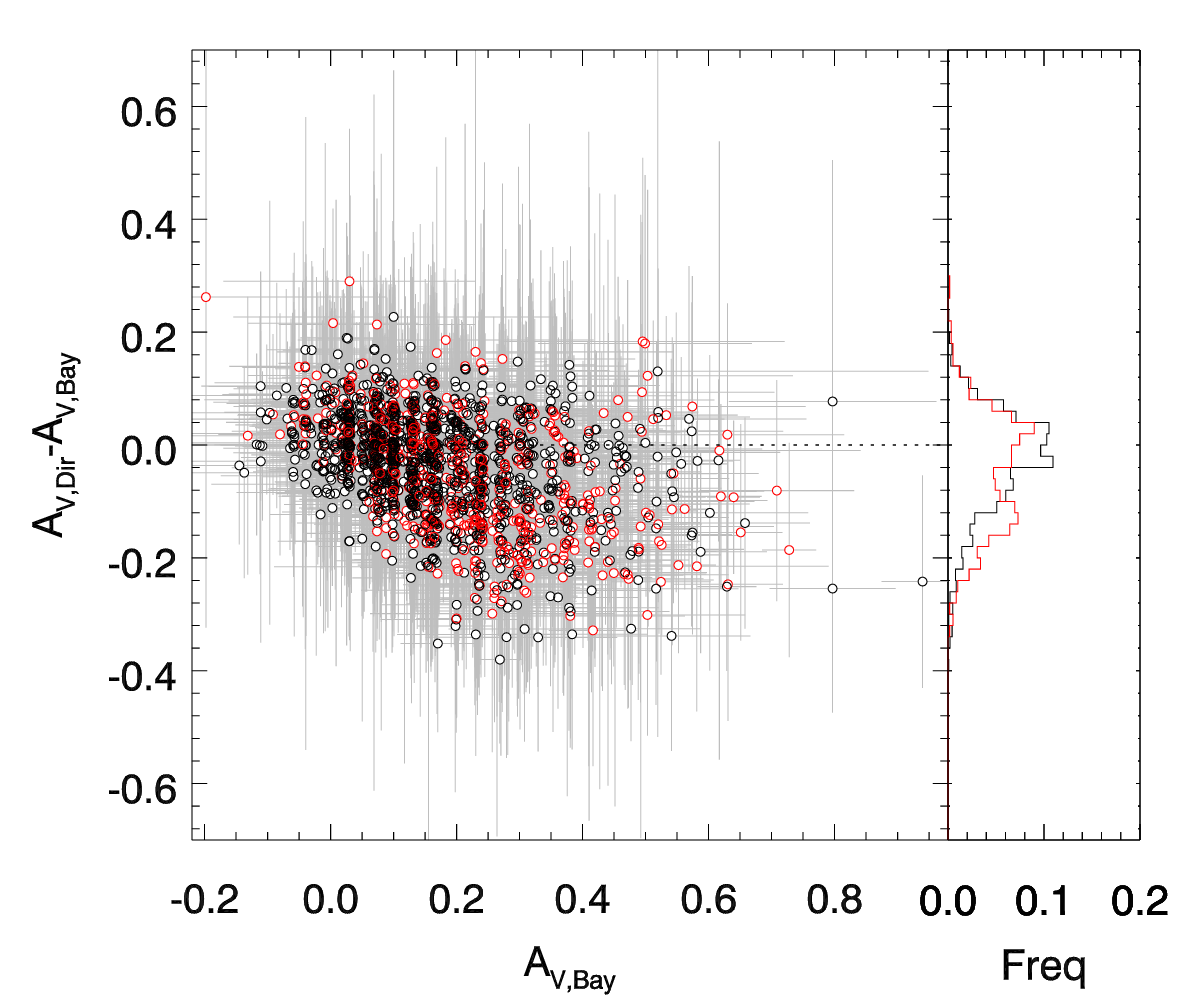}}
    \end{minipage}
    \begin{minipage}{0.68\columnwidth}
      \resizebox{\hsize}{!}{\includegraphics{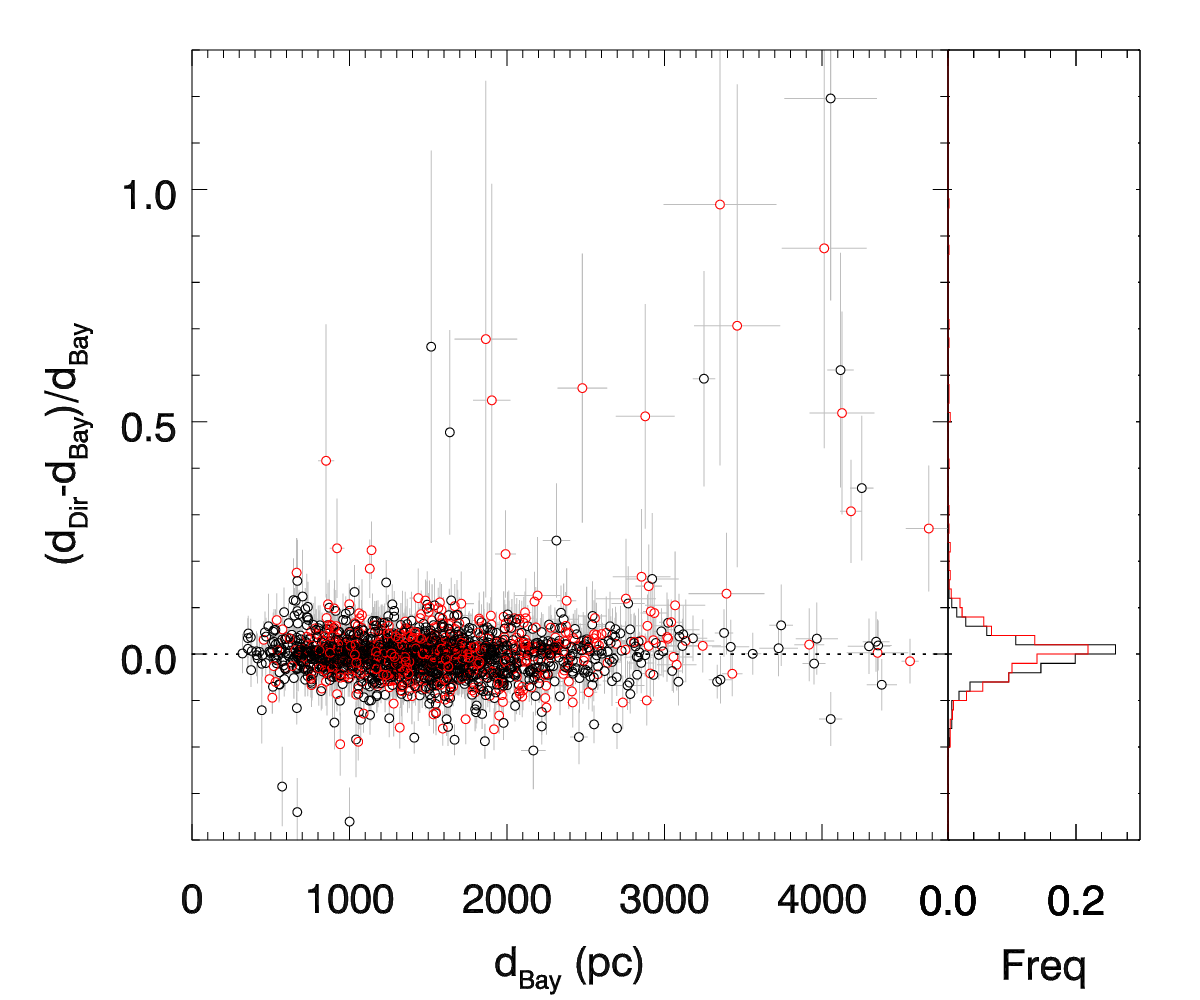}}
    \end{minipage}
  \else
    \begin{minipage}{0.68\columnwidth}
      \resizebox{\hsize}{!}{\includegraphics{figs/metdir/av_bay_dir.eps}}
    \end{minipage}
    \begin{minipage}{0.68\columnwidth}
      \resizebox{\hsize}{!}{\includegraphics{figs/metdir/dist_bay_dir.eps}}
    \end{minipage}
  \fi
  \caption{Relative (left panel) and absolute (right) differences between extinctions and distances for the stars in the sample, derived with the Bayesian and the direct methods.  The dashed line is the identity line. Black dots are stars with single-peaked PDFs, red dots are with broad/multiple-peaked ones. The left sub-panels show histograms of the differences.}
 \label{fig:compdir2}
\end{figure*}

Fig.~\ref{fig:compdir2} shows a comparison between these distances and extinctions with those obtained with the Bayesian method. It is readily evident that they compare well with mean differences $(d_\text{Dir}\!-\!d_\text{Bay})/d_\text{Bay} = -0.009\pm0.001$ and $(A_\text{V,Dir}\!-\!A_\text{V,Bay}) = -0.032\pm0.004$~mag. The large dispersion in extinctions is due the broad PDFs provided by the direct method, which allow a high-probability matching to a wide range of extinctions.

\subsection{Impact of knowing the evolutionary stage}

Stellar parameters derived via the Bayesian method are `forced' to be consistent with the grid of evolutionary tracks being used. This results in smaller uncertainties, which are, typically, a factor of $\sim$2.1 smaller for the Bayesian method than for the direct method.  There are, however, situations in which the Bayesian method produces distance uncertainties similar to the direct method. This happens, in general, for stars with broad and multipeaked PDFs, which often arise from the star being compatible with either a RC or a RGB star.

\begin{figure*}
  \ifpngfig
    \begin{minipage}{0.45\textwidth}
      \resizebox{\hsize}{!}{\includegraphics{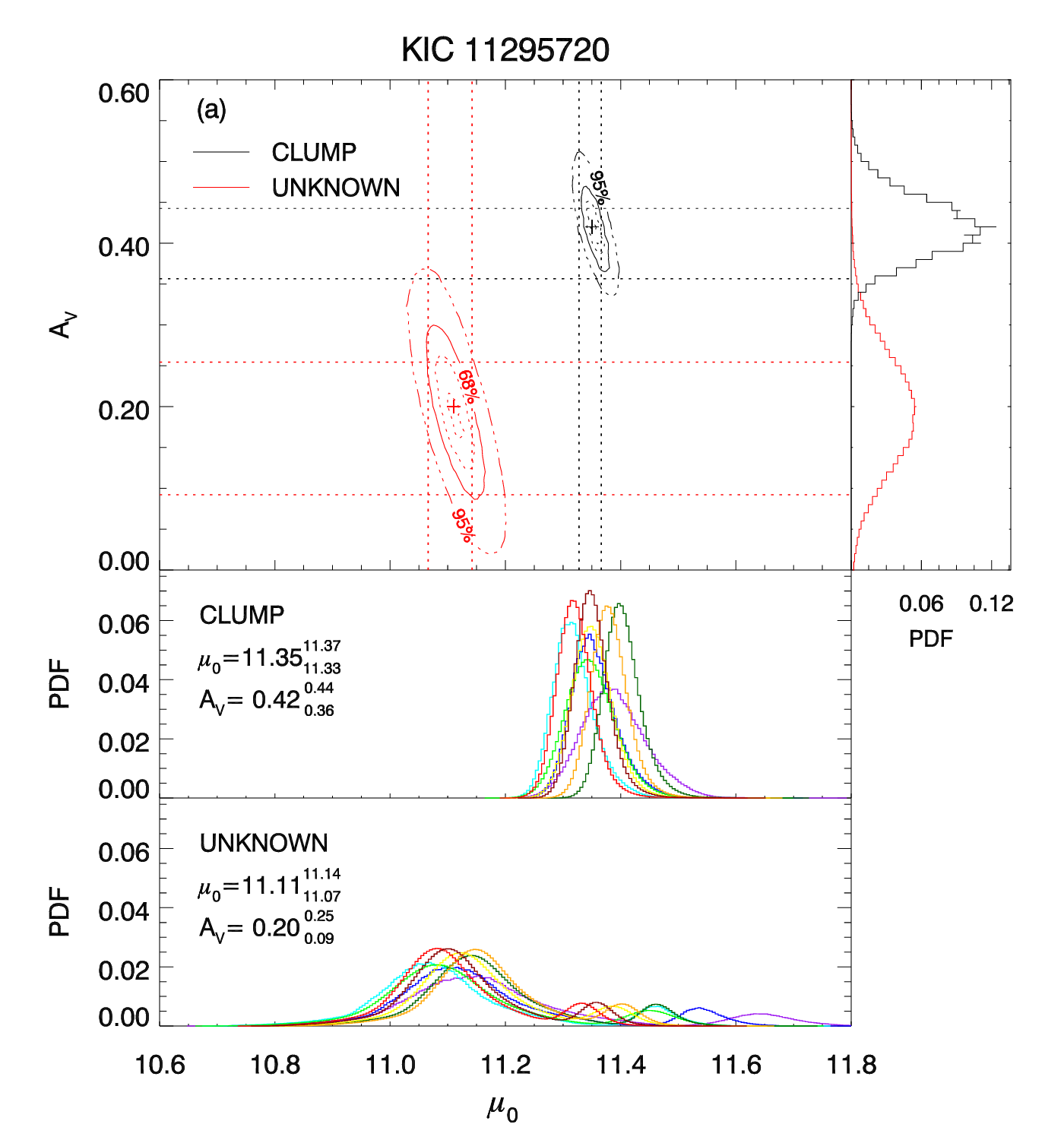}}
    \end{minipage}
    \begin{minipage}{0.45\textwidth}
      \resizebox{\hsize}{!}{\includegraphics{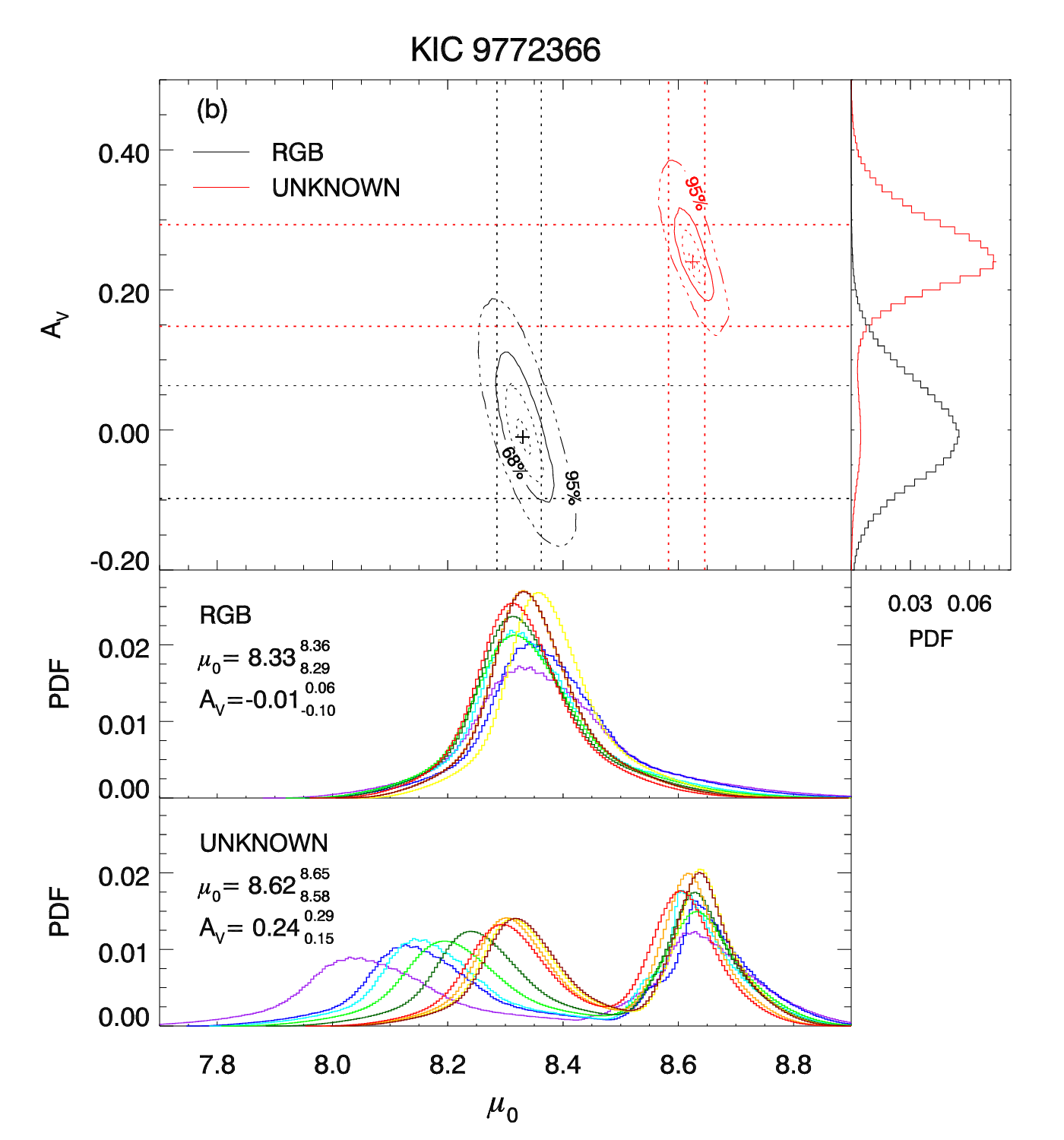}}
    \end{minipage}
  \else
    \begin{minipage}{0.45\textwidth}
      \resizebox{\hsize}{!}{\includegraphics{figs/evol_st/cont_11295720.eps}}
    \end{minipage}
    \begin{minipage}{0.45\textwidth}
      \resizebox{\hsize}{!}{\includegraphics{figs/evol_st/cont_9772366.eps}}
    \end{minipage}
  \fi
\caption{Examples of the joint (\muo,\av) and marginal \av\ PDFs (top panels) of two stars for which we know the evolutionary stage classification, either using this information (black contours), or not using it (red contours). The solid and triple-dot-dashed contours represent the 68 and 95 per cent credible regions. The dotted lines represent the 68 per cent credible interval for the mode of both \av\ and \muo. Their $\mu_{0\lambda}$ PDFs with `unknown' and `known' evolutionary stage classification are shown in the bottom and middle panels, respectively.
}
 \label{fig:classifica}
\end{figure*}

Fig.~\ref{fig:classifica} illustrates the impact of knowing the evolutionary stage on the joint \muo--\av\ PDFs of two stars, classified as CLUMP and RGB in the APOKASC catalogue, respectively. As can be seen, if we assume these stars have an `unknown' classification (red contours), their \muo~PDFs (bottom panels) are clearly bimodal. When we adopt the correct CLUMP classification (black contour) for the RC star KIC~11295720, the peak with the larger distance and extinction is favoured (middle-left panel). If the classification was not available, the peak corresponding to RGB models would have been favoured (bottom-left panel), producing distances $\sim$10 per cent smaller. Curiously, the direct method would have indicated a distance intermediate between those two (the 68 per cent CI being between $\muo=11.06$ and 11.21~mag), although more similar to the `wrong' solution. A similar situation -- but working in the opposite sense -- occurs for the RGB star KIC~9772366, which has its derived distance increased by $\sim$15 per cent when assigned an unknown evolutionary stage. Such significant changes in the distances and extinctions were found for 2 per cent (6 out of 291) of the stars classified as RC, and for 3 per cent (5 out of 199) of those classified as RGB.

Since the initial release of the APOKASC catalogue contains a large number of stars without $\deltaP$ measurements in the interval of $\logg<2.5$, for which confusion between RC and RGB stages can easily occur, it is possible that similar situations are actually present in the catalogue, leading to an increased scatter in our derived distances. Such scatter is likely to be reduced in future versions of the catalogue, when more asteroseismic classification information becomes available.

\subsection{Effect of distance priors}

Since the basic stellar properties derived from \numax, \deltanu, \Teff, and $\mh$ are independent of distance, we have not applied any distance prior in our method. We can, however, estimate the maximum effect that different distance priors would have had, were the distances fully incorporated into the Bayesian part of the method. For this, we have multiplied the distance PDFs by functions of the form  
\beq
p(d)~\propto~\exp{(-\Ract/\Ract_s)}, \quad p(d)~\propto~\exp{(-z/z_s)} \nonumber
\eeq
or a combination of both, where \Ract\ and $z$ are the Galactocentric radius and height above the plane, respectively. These represent the spatial distribution of stars expected in the MW's stellar disk. We adopted as scale factor $\Ract_s = 2600$~pc, and two extreme values of $z_s$, namely $z_s = 100$~pc and  $z_s = 900$~pc. These span the possible range of $z_s$ in going from a young thin disk \citep[e.g.,][]{Maiz01} to the thick disk \citep{Juric08}.

The effect of these multiplicative functions on the PDF medians and modes is very modest, namely: less than 1 per cent changes in the distances for the bulk of the stars, increasing to maximum values of $\sim\!4$ per cent for stars with broad and/or multiple-peaked PDFs. Since these changes are typically smaller than the 68 per cent CI, we can conclude that including prior information on the distances is not worthwhile at this stage. 

\subsection{Effect of systematic shifts in \Teff\ and \mh}
\label{sec:systshifts}

We simply assume that observed stars are well-described by current evolutionary tracks of single stars, which is reasonable as a first approximation. However, it is well known that evolutionary tracks frequently present systematic offsets in the H--R diagram, and especially in the \Teff\ scale of the red giants. This happens primarily because of the approximations used to model the energy transport by convection, such as mixing length theory. In our case, we use evolutionary tracks in which the mixing length parameter is calibrated on a solar model and then applied to all stars \citep[see][]{bressan12}. This approach could cause systematic offsets in the \Teff\ scale of the models.

We explore the possible effect of such offsets by applying the same methods with a grid of stellar models shifted by $\Delta\Teff=+100$~K. The main effect of this shift is that the Bayesian method compares the observed stellar parameters with older/metal-poorer isochrones, causing a mismatch between the derived and the observed stellar SED, which is compensated by an additional extinction. On average, we obtain a change of $A_{V,\Delta T_{\rm eff}}-\av=0.062$~mag. This also slightly impacts the derived distances, which are decreased by $(d_{\Delta T_{\rm eff}}\!-\!d)/d=-0.017$, corresponding to $\sim\!0.9\sigma(d)/d$.

Systematic offsets between the model and data metallicity scales are also possible. We have tested the method applying a systematic shift of $\Delta\mh=+0.1$~dex to the models, which makes the Bayesian method match the observed stellar parameters with younger/metal-richer isochrones. The effect in this case is to produce smaller extinctions with a mean value of $A_{V,\Delta{\rm [M/H]}} \!-\! A_{V} = -0.062$~mag, and to slightly increase the distances by $(d_{\Delta{\rm [M/H]}} \!-\! d)/d = 0.004$. Moreover, in this case we found that a large number of stars have $\mu_{0\lambda}$ PDFs that are better matched with negative $A_V$: $\sim$8 per cent have $A_{V,\Delta{\rm[M/H]}}\leq -0.05$~mag, and $\sim$12 per cent have $-0.05\,{\rm mag}<A_{V,\Delta{\rm[M/H]}}\leq 0.0$.

Since systematic offsets of this order of magnitude are perfectly possible, they can be taken as a rough indication of the possible systematic errors in our distance and extinction estimates.

\begin{figure*}
  \ifpngfig
    \resizebox{0.75\hsize}{!}{\includegraphics{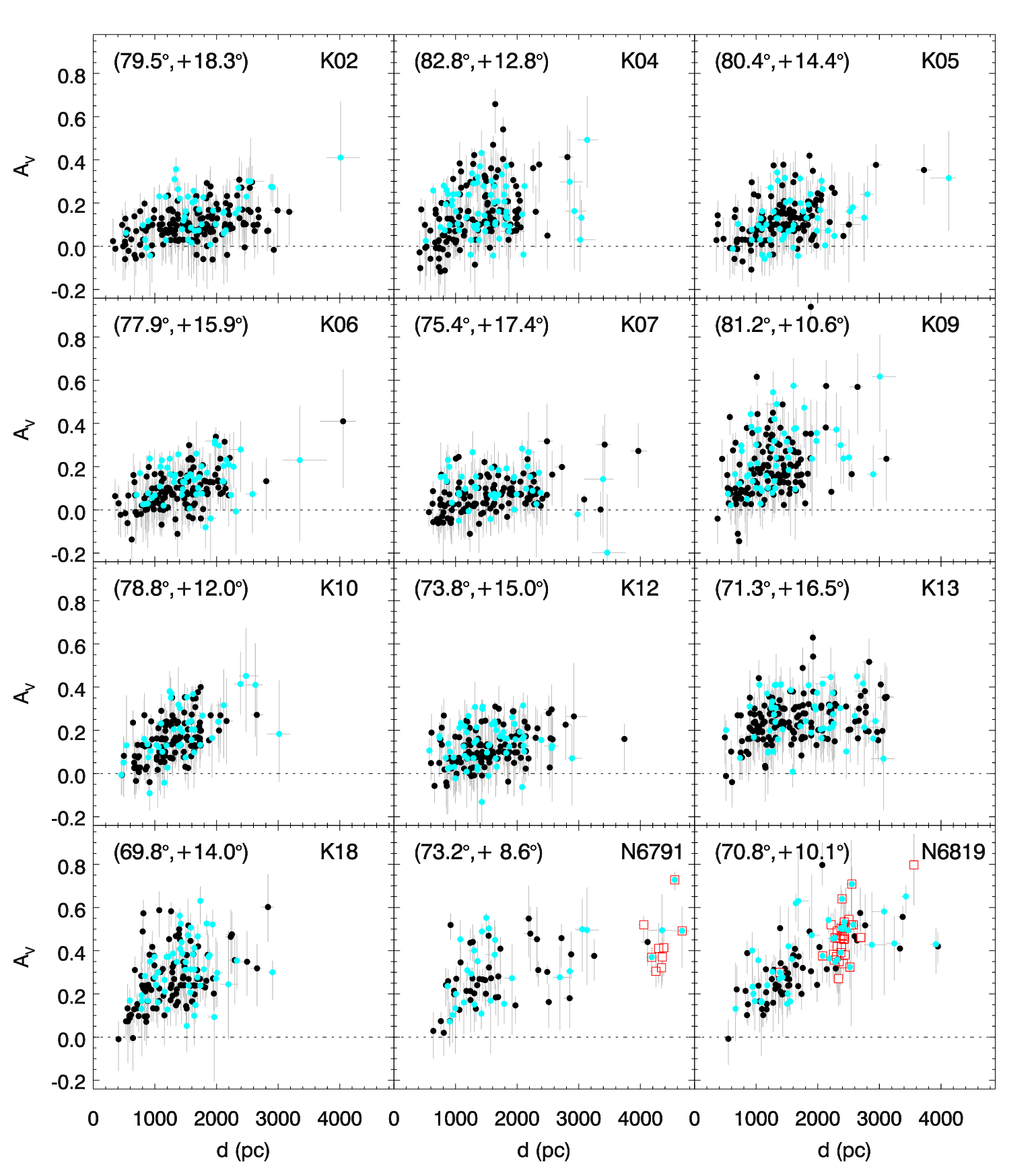}}
  \else
    \resizebox{0.75\hsize}{!}{\includegraphics{figs/all_d_fov/av_dist_fov.eps}}
  \fi
 \caption{$A_{V}$ versus distance ($d$) for the APOKASC fields indicated in Fig.~\ref{fig:l_b_kfov}. The Galactic coordinates of field centers are indicated in the top-left of each panel. The cyan dots are stars whose $\mu_{0\lambda}$ PDFs are broad or multiple-peaked. The small red squares are stars that likely belong to the star clusters NGC~6791 and NGC~6819.}
 \label{fig:av_dist}
\end{figure*} 

As the typical uncertainties in spectroscopic ASPCAP metallicity in the APOKASC sample are of $\sigma(\mh)\simeq0.06$~dex, the experiment of applying $\Delta\mh=+0.1$~dex also gives us an indication about the maximum changes we would have in our distance and extinction estimates, if we had adopted a metallicity prior in the Bayesian method. Indeed, higher metallicities are much more likely in the sample, and could have been more weighted by applying a suitable prior. It is very unlikely, however, that the method would have favoured models more than $2\sigma$ (0.12~dex) away from the measured \mh, which is about the size of the $0.1$~dex shift explored here.

\section{Discussion}
\label{sec:discussion}

\subsection{Typical distances and extinction maps} 

\begin{figure}
  \ifpngfig
    \resizebox{\hsize}{!}{\includegraphics{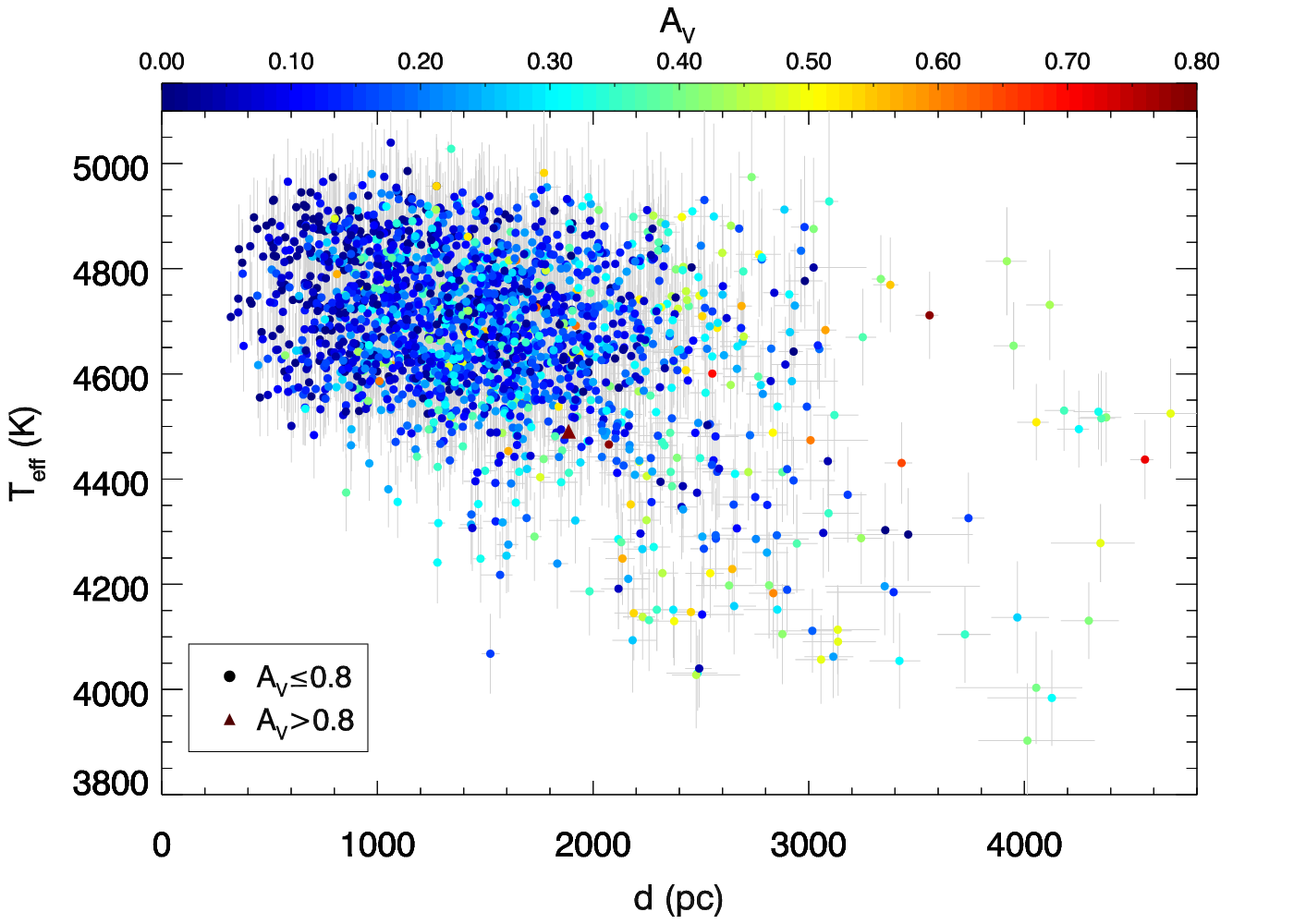}}
  \else
    \resizebox{\hsize}{!}{\includegraphics{figs/all_d_fov/teff_dist_av.eps}}
  \fi
 \caption{Correlation between \Teff, distance, and $A_{V}$.}
 \label{fig:teff_dist_av}
\end{figure}

Fig.~\ref{fig:av_dist} shows $A_{V}$ versus distance ($d$) for all fields showed in Fig.~\ref{fig:l_b_kfov}. This figure indicates that most of the observed stars are located within 2~kpc, whereas almost all stars are within 4~kpc. The cyan dots are stars with broad/multiple-peaked $\mu_{0\lambda}$ PDFs, and hence with more uncertain locations. Note that some stars ($\sim$6 per cent) have $\mu_{0\lambda}$ PDFs that are better matched with slightly negative $A_V$. As shown in Fig.~\ref{fig:teff_dist_av}, there is a trend for stars at larger distances to be cooler than the nearest ones, which is consistent with them being more luminous. Also, high extinction stars (with, say, $A_{V}>0.4$~mag) are observed at larger distances (Figs.~\ref{fig:av_dist} and \ref{fig:teff_dist_av}).  These plots indicate the potential of APOKASC data to provide improved 3D dust extinction maps in the {\it Kepler} fields \citep[see also][]{zasowski14b}.

The distance distribution of APOKASC stars results from a series of factors, comprising the many criteria used to select {\it Kepler} targets, the actual determination of their asteroseismic parameters, and the target prioritization by APOGEE \citep{zasowski13, pinsonneault14}. Discussion of this distribution is postponed to future papers. We note that a large fraction of the targets are RC stars, which in the {\it Kepler} field are preferentially observed within distances of 6~kpc \citep{bovy14}. 
 
\begin{figure*}
  \ifpngfig
    \begin{minipage}{0.49\textwidth}
      \resizebox{\hsize}{!}{\includegraphics{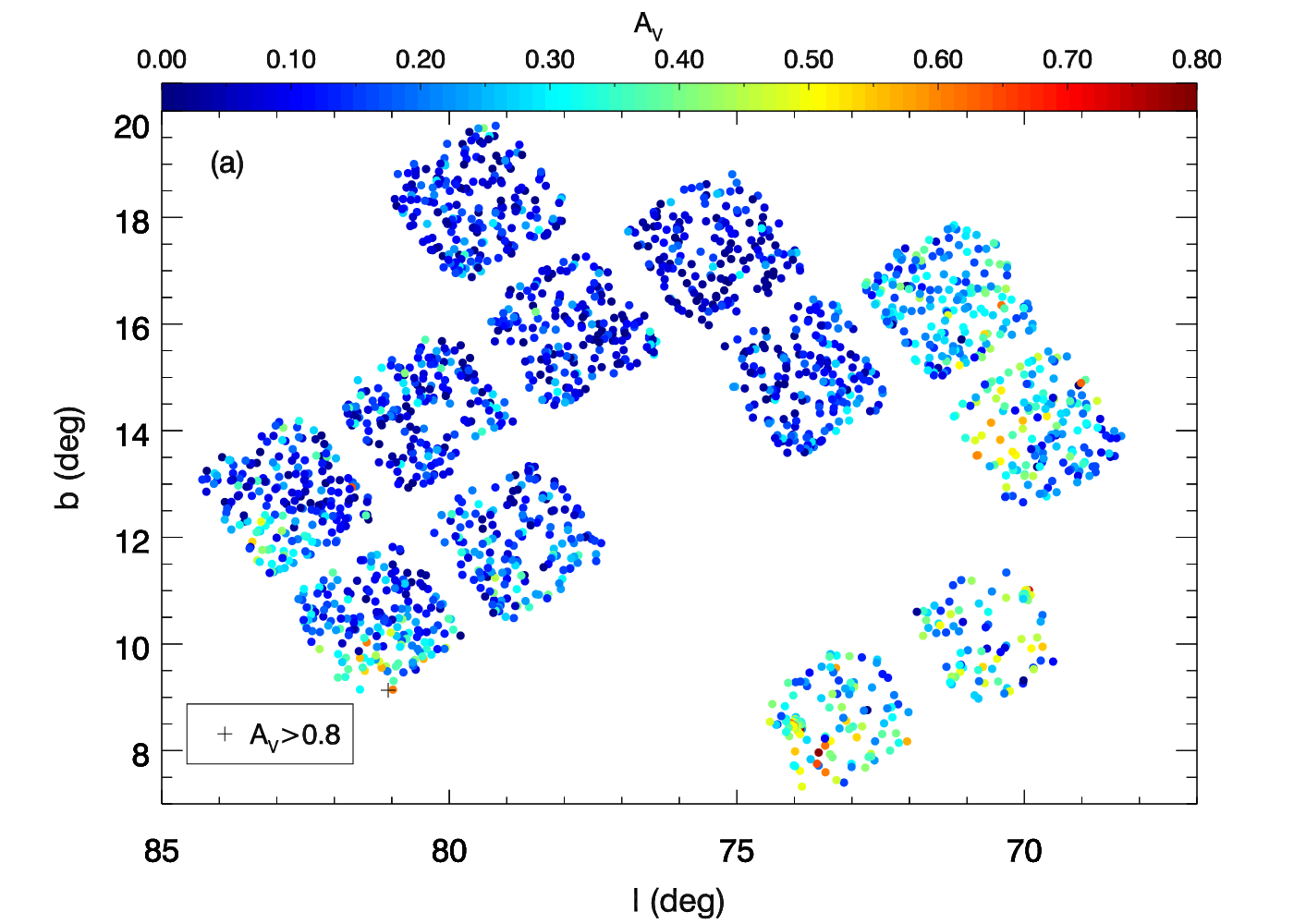}}
      \resizebox{\hsize}{!}{\includegraphics{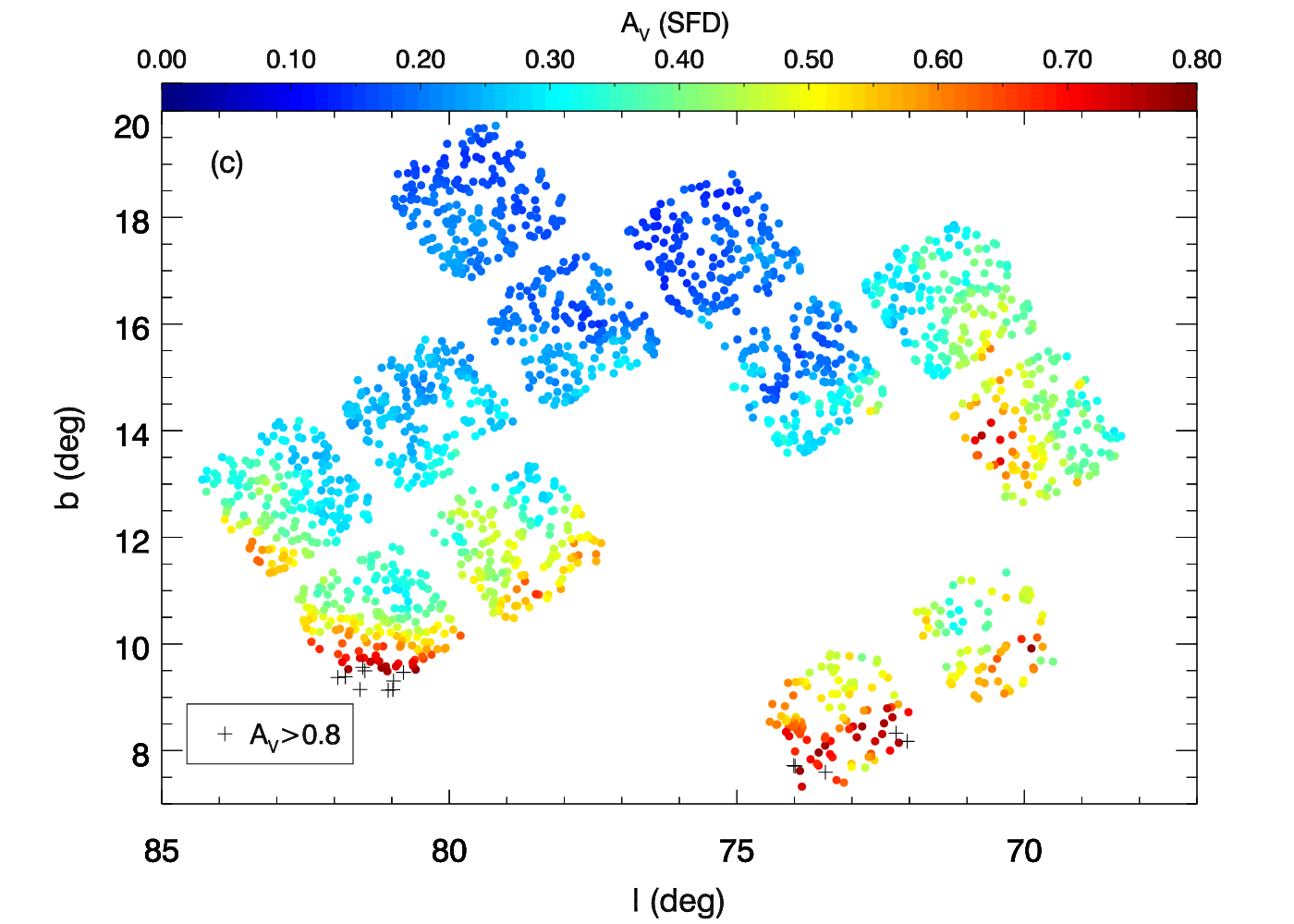}}
    \end{minipage}
    \begin{minipage}{0.49\textwidth}
      \resizebox{\hsize}{!}{\includegraphics{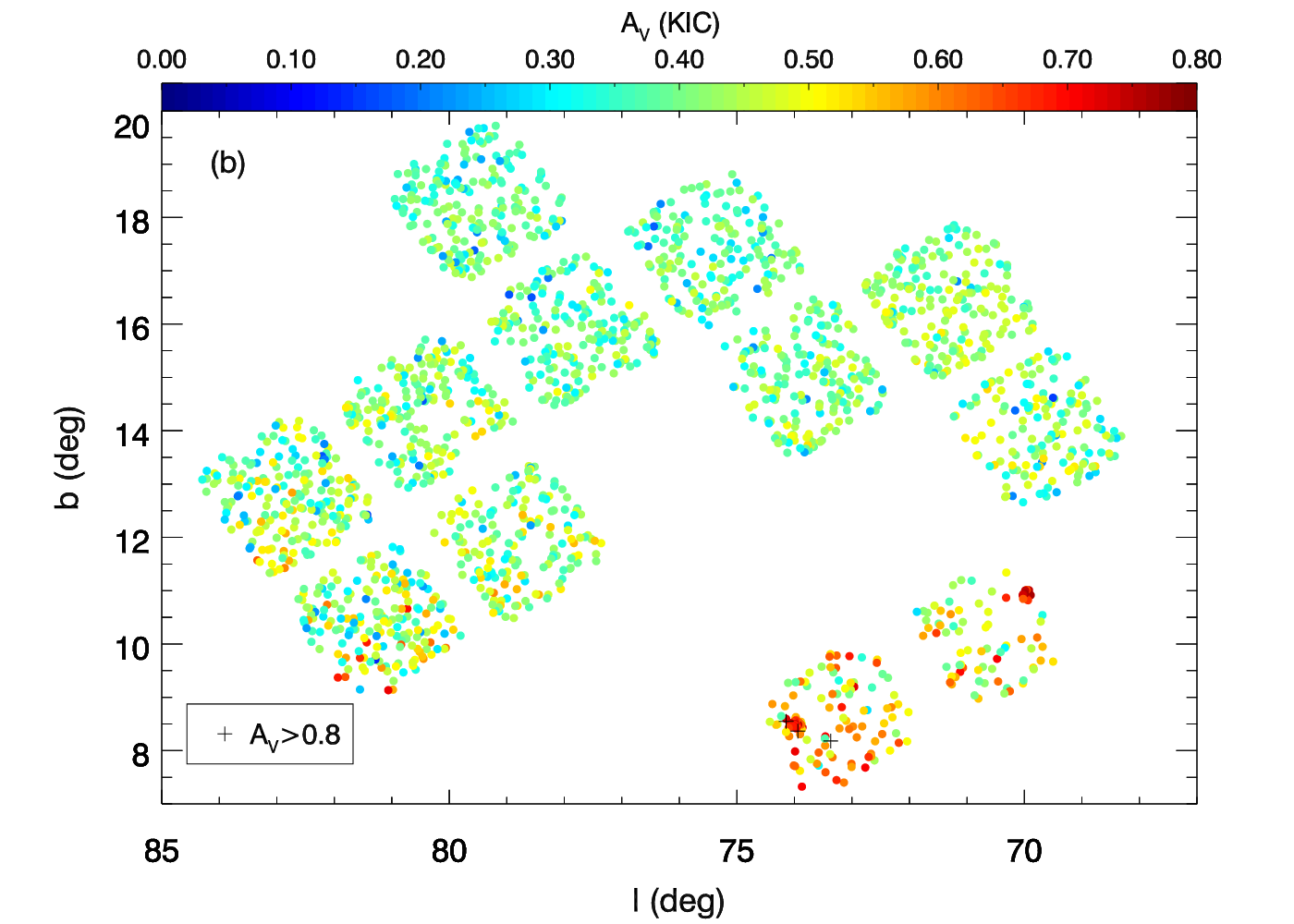}}
      \resizebox{\hsize}{!}{\includegraphics{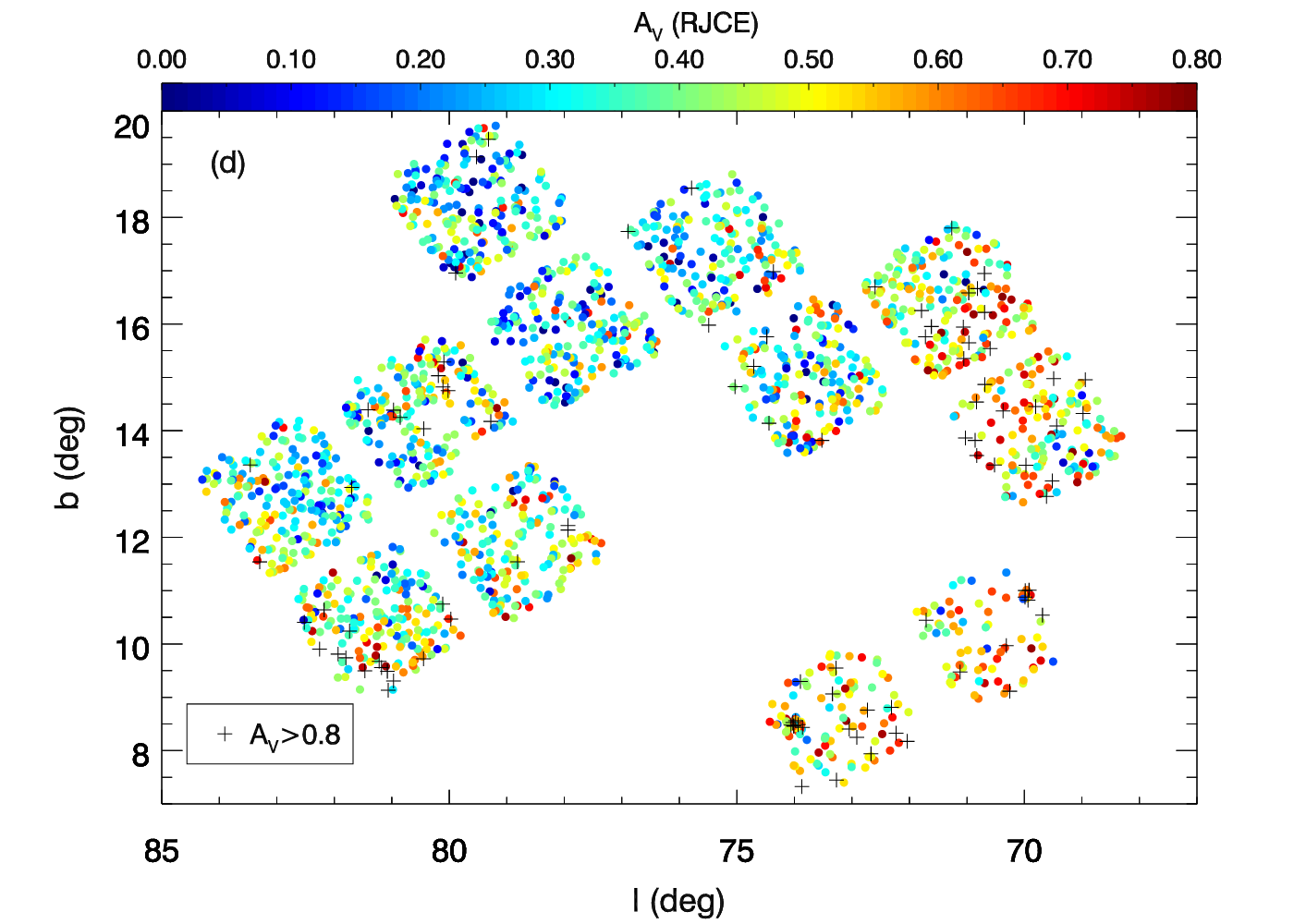}}
    \end{minipage}
  \else
    \begin{minipage}{0.49\textwidth}
      \resizebox{\hsize}{!}{\includegraphics{figs/l_b_av/l_b_avapo.eps}}
      \resizebox{\hsize}{!}{\includegraphics{figs/l_b_av/l_b_avsfd.eps}}
    \end{minipage}
    \begin{minipage}{0.49\textwidth}
      \resizebox{\hsize}{!}{\includegraphics{figs/l_b_av/l_b_avkic.eps}}
      \resizebox{\hsize}{!}{\includegraphics{figs/l_b_av/l_b_avrjce.eps}}
    \end{minipage}
  \fi
 \caption{(a) Our extinction map as compared to (b) KIC, (c) SFD, and (d) RJCE. Extinctions greater than 0.8~mag are represented by a plus symbol. See text for more details.}
 \label{fig:avmaps}
\end{figure*}

Fig.~\ref{fig:avmaps} compares the extinction maps from this work with those derived from the KIC \citep{brown11}, from \citet[][hereafter SFD]{schlegel98}, and with the Rayleigh-Jeans Color Excess \citep[RJCE,][]{majewski11} method. The comparison with the KIC extinction map will be commented further down in this section. The comparison with SFD shows some evident similarities in the position of the highly-extincted regions; the SFD extinctions tend to be much larger than our values, especially in low-latitude fields. This is expected since SFD gives the extinction at infinity, and not along the line-of-sight to every star. Indeed, the excess in the SFD extinction is larger along lines-of-sight close to the Galactic plane, where substantial interstellar material exists between our target stars and infinity. In addition, there are claims that SFD maps overestimate the extinction for regions with $A_V>0.5$~mag anyway \citep{arce99}. 

\begin{figure*}
  \ifpngfig
    \begin{minipage}{0.33\textwidth}
      \resizebox{\hsize}{!}{\includegraphics{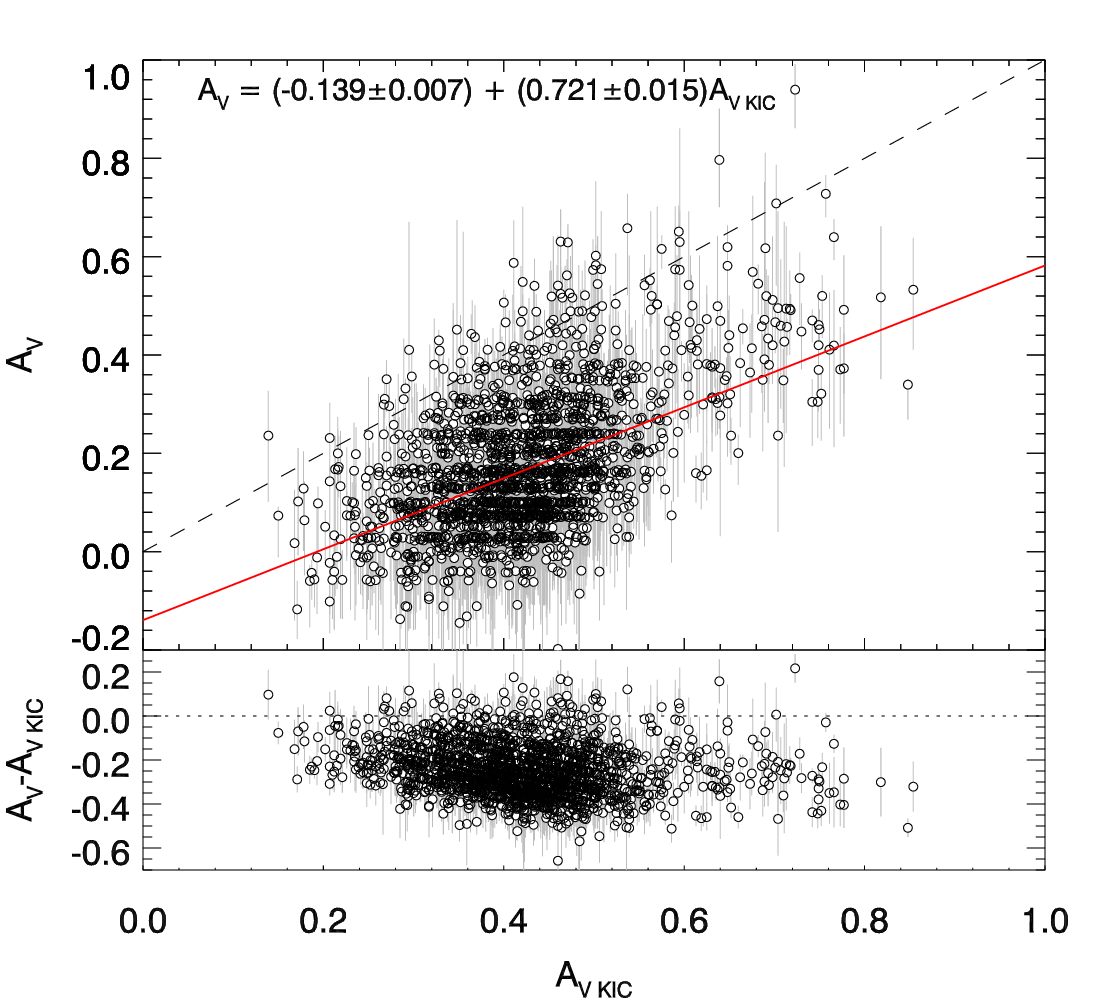}}
    \end{minipage}
    \begin{minipage}{0.33\textwidth}
      \resizebox{\hsize}{!}{\includegraphics{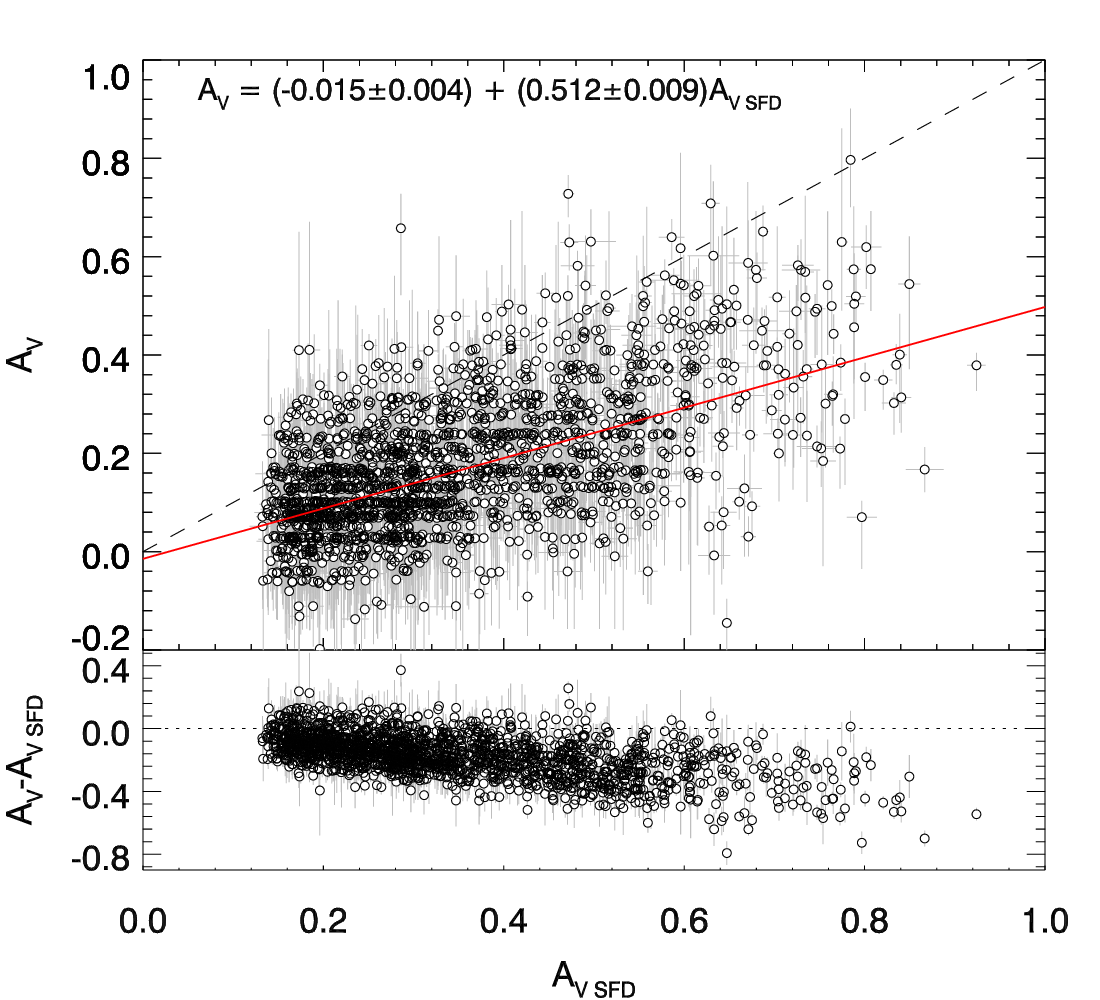}}
    \end{minipage}
    \begin{minipage}{0.33\textwidth}
      \resizebox{\hsize}{!}{\includegraphics{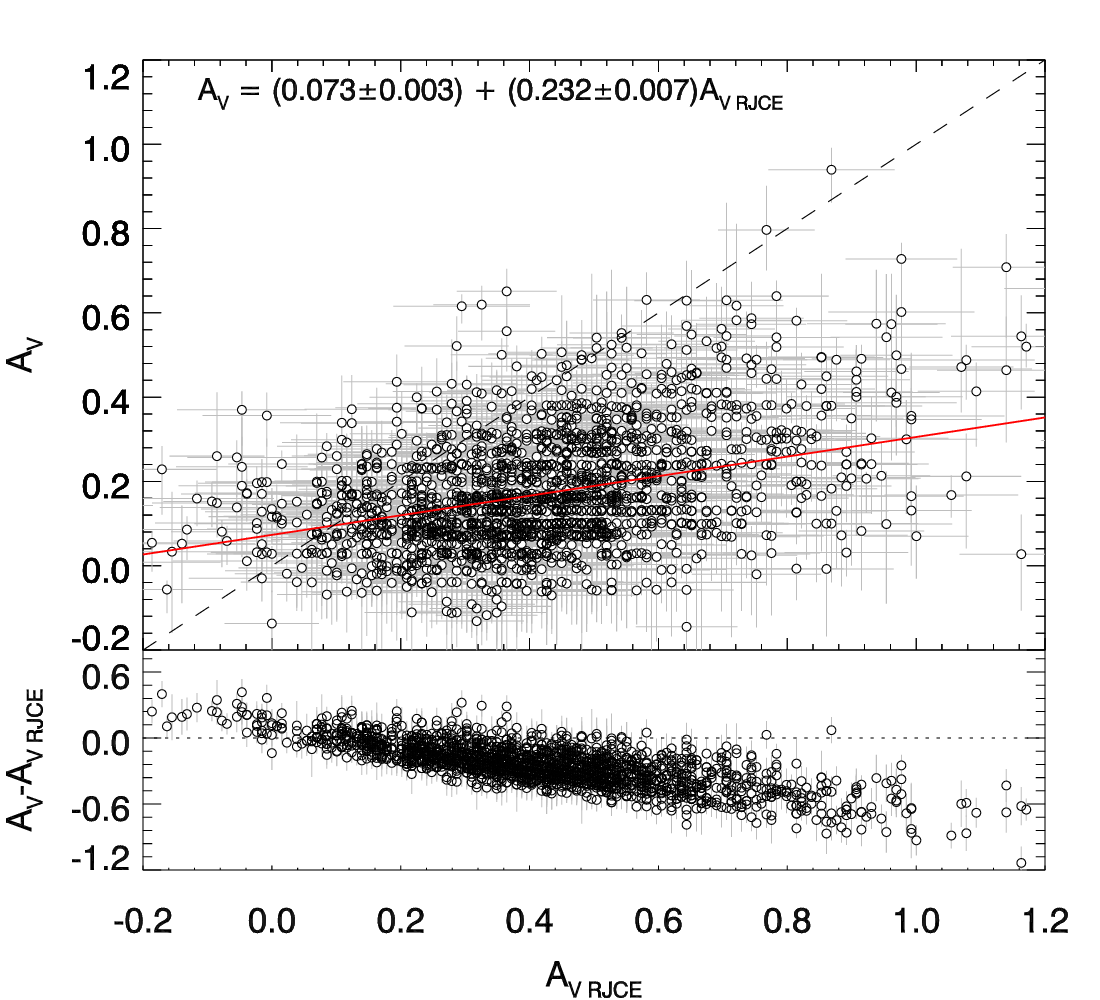}}
    \end{minipage}
  \else
    \begin{minipage}{0.33\textwidth}
      \resizebox{\hsize}{!}{\includegraphics{figs/avs/av_apo_kic_all.eps}}
    \end{minipage}
    \begin{minipage}{0.33\textwidth}
      \resizebox{\hsize}{!}{\includegraphics{figs/avs/av_apo_sfd_all.eps}}
    \end{minipage}
    \begin{minipage}{0.33\textwidth}
      \resizebox{\hsize}{!}{\includegraphics{figs/avs/av_apo_rjce_all.eps}}
    \end{minipage}
  \fi  
 \caption{Comparisons between our \av\ values, with KIC (left panel), SFD (middle), and RJCE (right) extinctions for the sample. The dashed black and solid red lines represent the identity line and the weighted least-squares fit, respectively. The bottom sub-panels show the absolute differences.}
 \label{fig:avs}
\end{figure*} 

The comparison with RJCE deserves some additional explanation. RJCE uses the color excess in $H-W2$ to derive the extinction in the $K_{\rm s}$ band, $A_{K_{\rm s}}$. Since it uses only infrared measurements, it is especially useful to derive extinction values in the regimes of high extinction (say, for $A_V\gtrsim1$~mag) that frequently happen towards the Galactic bulge and across the Galactic mid-plane \citep[see also][]{schultheis14}. In the case of the APOKASC stars, extinction values are never that large, and the typical values of $A_{K_{\rm s}}$ derived with RJCE are of just $\sim\!0.025$~mag. These low values of $A_{K_{\rm s}}$ are then multiplied by 8.45 to convert them to $A_V$. These two facts -- the use of only two photometric measurements, plus the amplification of uncertainties when converting $A_{K_{\rm s}}$ to $A_V$ -- is likely to cause a significant dispersion in the RJCE-derived $A_V$ values of slightly-reddened stars, exactly what we observe in Fig.~\ref{fig:avmaps}. Our extinction maps turn out to be somewhat smoother than the RJCE's; in addition, they also tend to present smaller $A_V$s. It is interesting to note that RJCE produces larger extinction values towards the high-latitude fields of {\it Kepler}, where both our and the SFD maps exhibit relatively low extinction values. The origin of this discrepancy will be investigated in a future paper.

Fig.~\ref{fig:avmaps} also indicates that our $A_{V}$ are valid solutions for the extinctions, since they show close-to-null mean values in the top half of the {\it Kepler} field, at higher latitudes, in rough agreement with the near-absence of dust shown by the SFD maps. 

Fig.~\ref{fig:avs} shows comparisons between our \av\ and KIC, SFD, and RJCE extinctions for the full sample, together with the fitted linear relations between them. Such linear fits could be used to infer the expected extinctions for other {\it Kepler} stars, still not observed by APOKASC. The zero-points in the derived linear relations between $A_{V}$ and $A_{V,{\rm SFD}}$, and between $A_{V}$ and $A_{V,{\rm RJCE}}$ are close to null. Null zero-points can be interpreted as a first evidence that systematic errors, although possible, are probably smaller than those explored in Sec.~\ref{sec:systshifts}.

It is clear that the KIC extinctions appear overestimated with respect to our values. The rms deviation around the linear fit presented in Fig.~\ref{fig:avs} (left panel) is 0.12~mag. If we assume that both are simply proportional to each other (i.e., with no zero-point offset), we obtain that 
\begin{equation}
A_{V} = (0.409\pm 0.003) \, A_{V,\text{KIC}}.
\end{equation}
with a mean rms deviation of 0.12~mag around this relation.

KIC extinctions are derived from a simple geometrical model for the distribution of the dust \citep{brown11}, which is a useful first-order approach for many applications. In this model, the dust density is assumed to decrease exponentially with height $|z|$ above the Galactic plane, with a scaleheight of $h_{z,{\rm dust}}=150$~pc, and a local extinction density of $\kappa_V=1$~mag\,kpc$^{-1}$. Thus, we use our distance and extinction values to recalibrate the values of $h_{z,{\rm dust}}$ and $\kappa_V$. For each pair of values, we integrate the KIC extinction model from the Sun up to every observed star, thus obtaining new estimates of $A_{\text{V,KIC}}$. For the $h_{z,{\rm dust}}$ and $\kappa_V$ adopted by KIC, we obtain a mean difference between our \av\ and the new $A_{\text{V,KIC}}$ of $-0.39$~mag, and an rms dispersion of 0.12~mag around this mean. We then identify the pair of $h_{z,{\rm dust}}$ and $\kappa_V$ values that minimizes the residuals between our \av\ and $A_{\text{V,KIC}}$. These values are $h_{z,{\rm dust}}=234$~pc and $\kappa_V=0.25$~mag\,kpc$^{-1}$. This modified model for KIC extinctions presents a null mean difference  with respect to our estimates, but still a r.m.s. dispersion of 0.12~mag.  This significant dispersion probably reflects the fact that dust extinction is much more patchy along the Galactic disk than assumed in these simple models.
 
Finally, our revised \av\ values allow us to reevaluate the consistency between the different \Teff\ scales included in the APOKASC catalogue. As discussed  thoroughly in \citet{meszaros13}, the zero-point of the ASPCAP \Teff\ was calibrated so as to coincide, on average, with \Teff\ determinations based on the IRFM calibration by \citet{gonzalezhernandez09}, using $\jks$ colours. We can derive the IRFM \Teff\ for all APOKASC targets using the same relations, but the results will be slightly dependent on the extinction values used in the de-reddening of the observed \jks\ colors. Using our \av\ values, we find IRFM \Teff\ estimates that are systematically cooler by $-74$ K, on average, if compared to the ASPCAP-corrected values. \citet{pinsonneault14}, using the KIC extinction maps, find a value of $-193$~K for this offset. Therefore, our smaller extinction values help to reduce, but do not completely eliminate, this systematic difference between the different \Teff\ scales.

\subsection{Results for the star clusters}
\label{sec:clusters}

The last two panels of Fig.~\ref{fig:av_dist} show clear concentrations of stars at distances of $\sim$2.4 and 4.4 kpc, which are obviously caused by the star clusters NGC~6819 and NGC~6791, respectively (red squares). Stars in these clusters provide a useful check of the uncertainties in our methods, since they are expected to be located within a distance interval much smaller than the expected uncertainties. As for the extinction, both clusters spread over tens of arcmin across the {\it Kepler} fields, so that the star-to-star extinction may vary significantly, as indeed indicated by the vertical spread in Fig.~\ref{fig:av_dist}.

\begin{figure*}
  \ifpngfig
    \begin{minipage}{0.45\textwidth}
      \resizebox{\hsize}{!}{\includegraphics{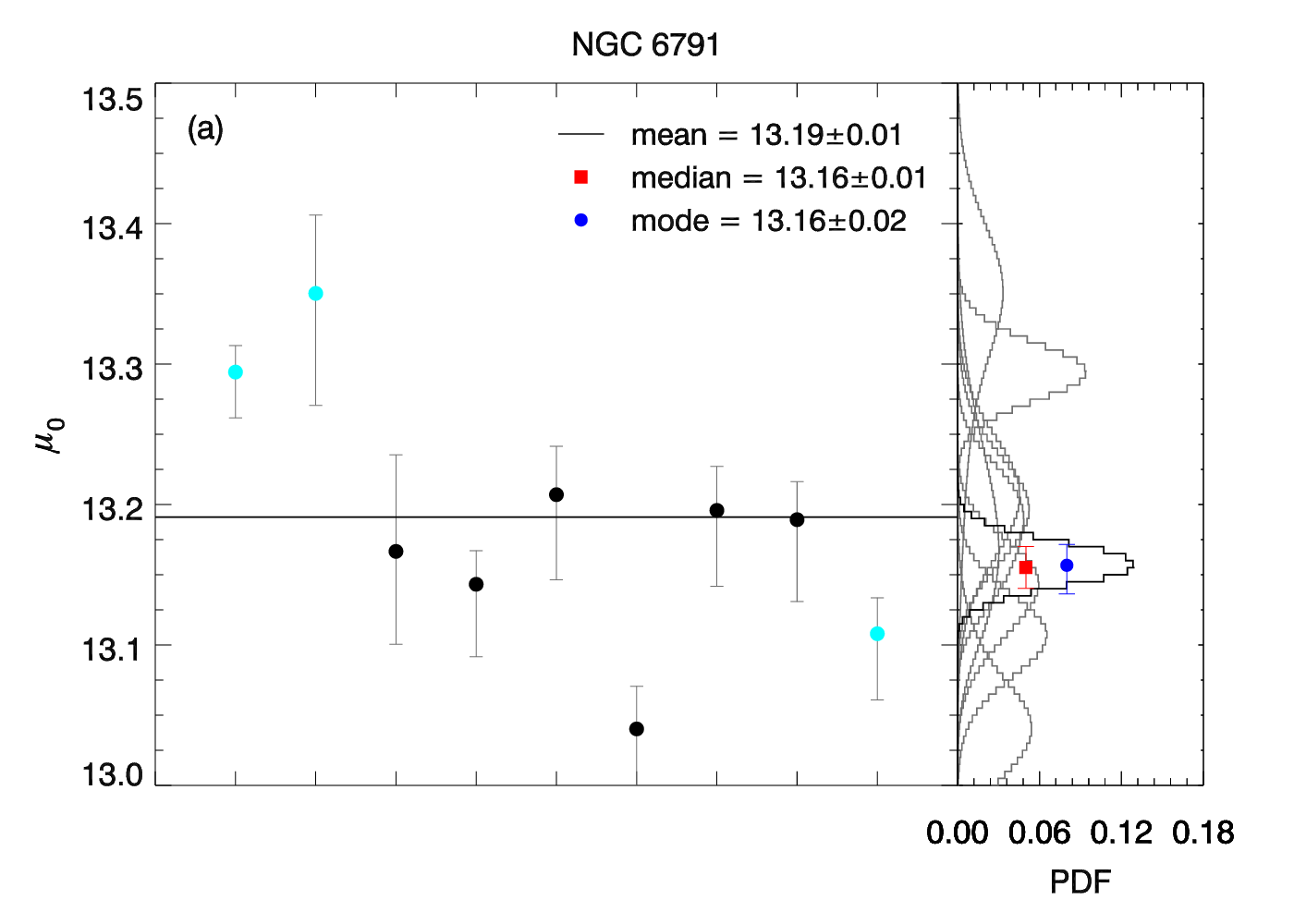}}
    \end{minipage}
    \begin{minipage}{0.45\textwidth}
      \resizebox{\hsize}{!}{\includegraphics{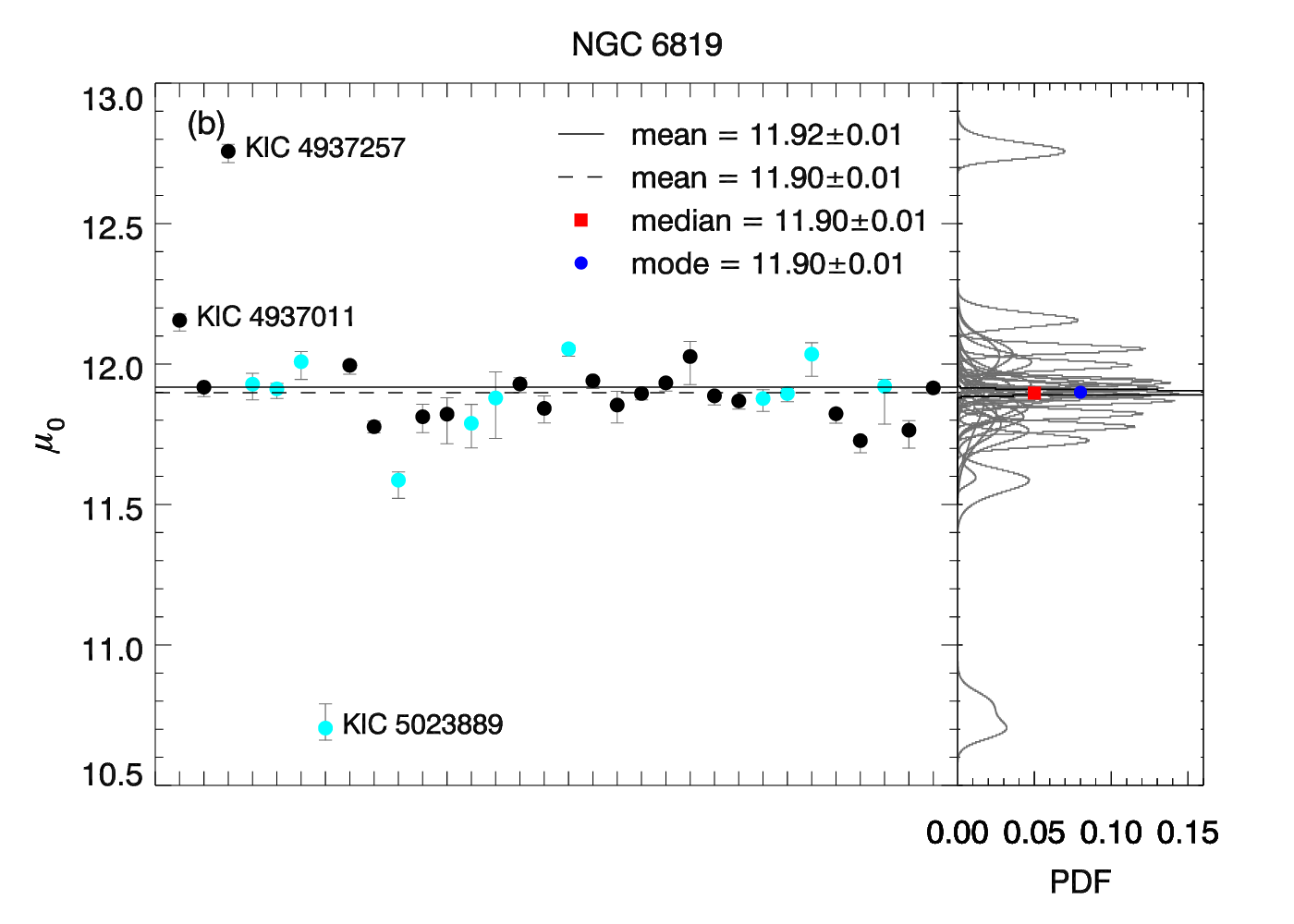}}
    \end{minipage}
  \else
    \begin{minipage}{0.45\textwidth}
      \resizebox{\hsize}{!}{\includegraphics{figs/clusters/pdfs_mu0_all_NGC6791.eps}}
    \end{minipage}
    \begin{minipage}{0.45\textwidth}
      \resizebox{\hsize}{!}{\includegraphics{figs/clusters/pdfs_mu0_all_NGC6819.eps}}
    \end{minipage}
  \fi
 \caption{Distances derived for stars in (a) NGC~6791 and (b) NGC~6819. In both cases, the main panel shows the mode \muo\ and 68 per cent CI for all cluster members. The solid and dashed lines represent the mean weighted values for all stars, and excluding the outliers denoted by their KIC numbers, respectively (see text for more details). The cyan dots are stars whose $\mu_{0\lambda}$ PDFs are broad or multiple-peaked. The smaller sub-panels to the right show the \muo\ PDFs for all cluster members (grey lines). The black line show the results of deriving the distance modulus PDF of the cluster using the product of all individual PDFs, whose mode and median (with their 68 per cent CI) are shown by the blue and red symbols, respectively.}
 \label{fig:clusters}
\end{figure*}

Fig.~\ref{fig:clusters} shows the distance modulus for all cluster members selected by \citet[][same stars as red squares in Fig.~\ref{fig:av_dist}]{stello11_1}, based on photometric membership by \citet{stetson03} for NGC~6791, and with at least 80 per cent membership probability from the radial velocity survey of \citet{hole09} for NGC~6819.
For NGC~6791, nine stars in APOKASC are classified as seismic members by \citet[][]{stello11_1}, and indeed there is a good overlap between their distance modulus PDFs (grey lines in panel a).  
The stars KIC~2435987, 2436688, and 2570214 present broad or double-peaked PDFs (cyan dots in panel a). The second star is classified as potentially affected by blending by \citet{stello11_1}. Also particular is the case of the star KIC~2569055 (sixth in the plot), whose W1 and W2 magnitudes favour a slightly smaller distance.

For NGC~6819, 32 stars selected by \citet[][]{stello11_1} are in the APOKASC sample, out of which 29 were classified as seismic members. The non-seismic members are KIC~4937011, KIC~4937257, and KIC~5023889. Fig.~\ref{fig:clusters}(b) shows the distance modulus PDFs for the 32 stars (grey lines); it is clear that the PDF of these three non-seismic members do not overlap with the others. We have verified that KIC~5023889 presents multiple-peaked PDFs, while the other two stars appear with normal PDFs. Among the other members, we find stars which are somewhat problematic, such as: KIC~4937576, 4937770, 5023732, 5024043, 5024476, 5024851, 5111940, 5112734, 5112744, 5112880, and 5113041 have broad or double-peaked PDFs (cyan dots in panel b); KIC~5024476 is listed as binary likely member by \citet{hole09}; KIC~4937770 could be considered a binary star as argued by \citet{corsaro12}; KIC~5112734 has a known blended star according to \citet{stello11_1}; KIC~5024240 and 5024851 are binary and likely binary members with somewhat widened PDFs; KIC~5112481 and KIC~4937257 have only 2MASS and WISE photometry and hence a somewhat increased uncertainty in \av; for KIC~5024967 the WISE photometry is probably affected by the diffraction spikes of a bright nearby star.

The black lines in the right sub-panels show the results of deriving the distance modulus PDF of the cluster using the product of all individual PDFs. The mode and its 68 per cent CI are represented by the blue symbols and the median and its 68 per cent CI, by red symbols. The mode in the $\mu_{0}$ PDFs -- $13.16\pm0.02$~mag and $11.90\pm0.01$~mag for NGC~6791 and NGC~6819, respectively -- compare well with \citet{basu11} who found $\muo=13.11\pm0.06$~mag ($4.19\pm0.12$~kpc) and $\muo=11.85\pm0.05$~mag ($2.34\pm0.05$~kpc), respectively. They also agree very well with \citet{wu14}, who found $\muo=13.09\pm0.10$~mag for NGC 6791 and $\muo=11.88\pm0.14$~mag for NGC 6819. Eclipsing binaries independently indicate distance moduli of $\muo=13.01\pm0.08$~mag for NGC~6791 \citep{brogaard11} and  $\muo=12.07\pm0.07$~mag for NGC~6819 \citep[][assuming $E(B-V)=0.12$ in this case]{jeffries13}. 

The left main-panels show a summary of the mode and its 68 per cent CI for each star (black and cyan dots). The solid black lines represent the mean weighted values, which is a simpler way to estimate the distance modulus of the cluster without considering the shape of the PDF. The dashed black line in the (b) panel is the mean weighted value for NGC~6819 without the three non-seismic members.

The wide variety of situations we meet in stars belonging to these well-studied clusters -- double-peaked PDFs, binaries, stars with incomplete photometry and/or without evolutionary status -- represents situations we likely have in the entire APOKASC sample.
But overall, it is quite encouraging that we find good agreement in the distances of these stars within their CI. 

\subsection{Distances to stars in the APOGEE-RC catalogue}

\begin{figure}
  \ifpngfig
    \resizebox{\hsize}{!}{\includegraphics{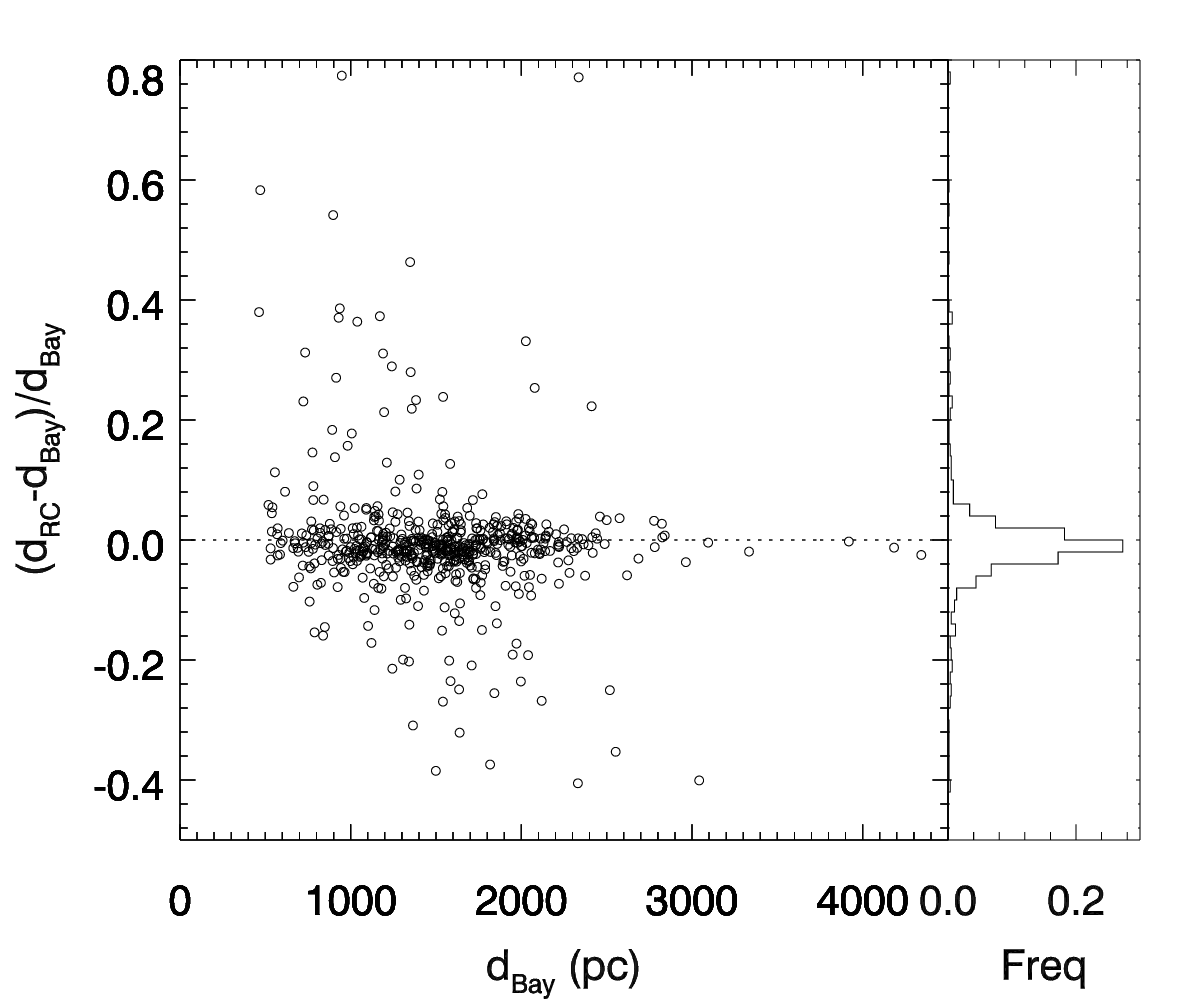}}
  \else
    \resizebox{\hsize}{!}{\includegraphics{figs/rc_dist/dist_bay_rcdist.eps}}
  \fi
 \caption{Relative difference between our Bayesian distances ($d_{\text{Bay}}$; mode) and the RC distances ($d_{\text{RC}}$) derived by \citet{bovy14}. The dashed line is the identity line. The right sub-panel show a histogram of the distribution of this difference.}
 \label{fig:rc_dist}
\end{figure}

\citet{bovy14} have recently released the APOGEE red clump (APOGEE-RC) catalogue, containing stars which, due to their particular values of \Teff, spectroscopic \logg, \mh, and 2MASS $(J\!-\!K_{\rm s})_0$, are very likely RC stars with a well-defined absolute magnitude. Comparison with the {\it Hipparcos}-based absolute magnitude of the RC \citep{laney12} then allows a good determination of their distances. A total of 593 such stars are present in the APOKASC catalogue, and a comparison between our, and the \citet{bovy14} distances is presented in Fig.~\ref{fig:rc_dist}. The mean relative difference between them is only 0.4 per cent. Note that the comparison includes all stars common to both catalogues, which likely includes some misclassified RGB stars in \citet{bovy14}, as well as stars with unknown evolutionary stages in APOKASC; those could easily explain the few outliers in the plot.

Such a tight relation between these two distance scales is remarkable, and very encouraging. It is true that both scales are expected to be somewhat correlated, because they are based on the same set of stellar models from \citet{bressan12} to describe the behaviour of the RC as a function of metallicity and mass. However, the zero-point of the \citet{bovy14} distances does not depend on stellar models. Moreover, the APOGEE-RC catalogue was dereddened using a method quite different from ours, namely the RJCE method by \citet{majewski11}. The comparison between the distances obtained by the direct method (hence independent of stellar models, see Sec.~\ref{sec:compdirmet}) and the \citet{bovy14} distances has also produced excellent agreement, as mentioned in \citet{bovy14}.

\subsection{Distances to stars in the SAGA catalogue}

\begin{figure}
  \ifpngfig
    \resizebox{\hsize}{!}{\includegraphics{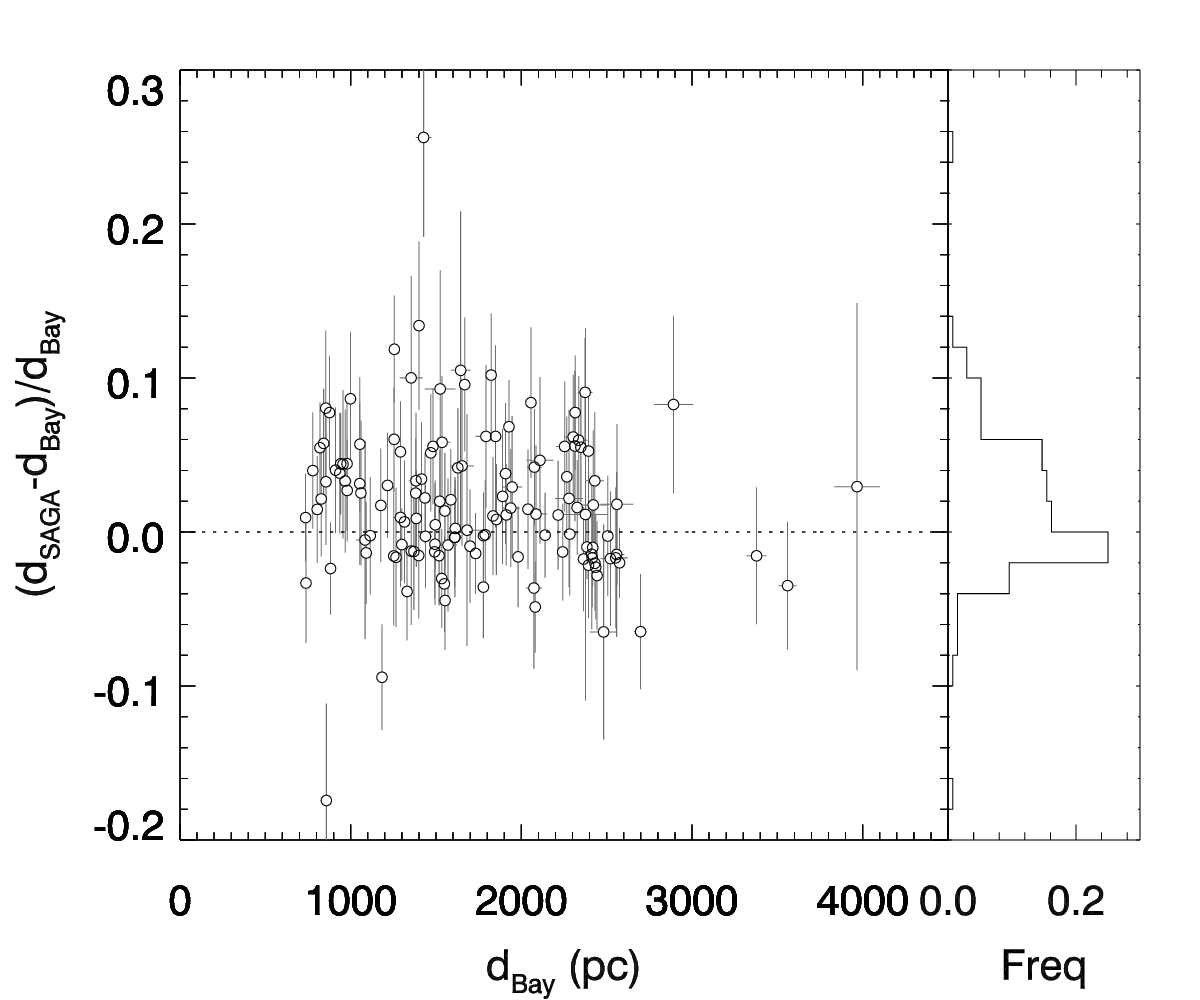}}
  \else
    \resizebox{\hsize}{!}{\includegraphics{figs/SAGAcatalog/dist_bay_saga_dens.eps}}
  \fi
 \caption{Relative difference between our Bayesian distances ($d_{\text{Bay}}$; mode) and the distances ($d_{\text{SAGA}}$) estimated in the SAGA catalog by \citet{casagrande14}. The dashed line is the identity line. The right sub-panel show a histogram of the distribution of this difference.}
 \label{fig:saga_dist}
\end{figure} 

We also compared our distances with those estimated in the SAGA catalogue \citep{casagrande14}, in which the stellar parameters are estimated in a completely independent way, using a combination of Str\"omgren photometry, the IRFM, and several extinction estimates. For the 136 stars in common with the APOKASC catalogue, the mean relative difference in distances is only 1.2 per cent, as shown in Fig.~\ref{fig:saga_dist}. 

Such good agreement is surprising, considering that SAGA \Teff\ are $\approx 90$~K hotter and \mh\ are $\approx 0.14$~dex smaller than the calibrated ASPCAP ones. Indeed, if we apply these zero-point shifts in \Teff\ and \mh\ in our method, the simulations in Sec.~\ref{sec:systshifts} indicate that we should obtain a distance scale 2.1 per cent shorter on average. It is likely that differences in the underlying methods, isochrones (and color-\Teff\ relationships), or in the extinction estimates from SAGA are affecting the results in a way that largely compensates to the offsets between the two \Teff\ and \mh\ scales.

\section{Conclusions}
\label{sec:conclu}

\begin{table*}
\begin{center} 
\caption[]{Derived distances and extinctions with the Bayesian and direct methods for APOKASC stars. LCI and UCI are the lower and upper limits of the 68 per cent CI, respectively. The last column lists the filters for which the photometry is available. A full table is provided in electronic format on the journal website.}
\footnotesize
\setlength{\tabcolsep}{4pt}
\begin{tabular}{ccccccccccccccc}
\hline \hline
& & \multicolumn{6}{c}{Bayesian method} & \multicolumn{6}{c}{Direct method} & \\ \noalign{\smallskip}
KIC ID & 2MASS ID & \multicolumn{3}{c}{$d$ (pc)} & \multicolumn{3}{c}{$\av$ (mag)} & \multicolumn{3}{c}{$d$ (pc)} & \multicolumn{3}{c}{$\av$ (mag)}  & Photometry \\
\cline{3-14}  \noalign{\smallskip}
 & & {\centering mode} & LCI & UCI & mode & LCI & UCI & {\centering mode} & LCI & UCI & mode & LCI & UCI & \\ \hline
1162746	& J19252639+3649116 & 1420 & 1397 & 1435 & 0.11	& 0.04 & 0.14 &	1318 & 1262 & 1368 & 0.10 & -0.08 & 0.23 & griz	JHK$_s$	W$_1$W$_2$ \\
1432587	& J19254985+3701028 & 2514 & 2452 & 2554 & 0.16 & 0.04 & 0.24 &	2418 & 2292 & 2512 & 0.21 & -0.01 & 0.35 & griz	JHK$_s$	W$_1$W$_2$ \\
1433593	& J19264298+3704199 & 1135 & 1105 & 1162 & 0.16 & 0.01 & 0.25 & 1177 & 1128 & 1208 & 0.21 &\ 0.06 & 0.32 & griz	JHK$_s$	W$_1$W$_2$ \\
1433730	& J19265020+3703054 & 1077 & 1043 & 1135 & 0.45 & 0.20 & 0.54 & 1138 & 1087 & 1170 & 0.38 &\ 0.20 & 0.48 & griz	JHK$_s$	W$_1$W$_2$ \\
1435573	& J19282646+3705369 & 1330 & 1308 & 1340 & 0.13 & 0.07 & 0.17 &	1216 & 1159 & 1283 & 0.16 & -0.07 & 0.32 & griz	JHK$_s$	W$_1$W$_2$ \\
\hline
\end{tabular}
\label{tab:dist_ext} 
\end{center} 
\end{table*}

The APOKASC collaboration is providing high-resolution spectroscopy for a large sample of {\it Kepler} asteroseismic targets, using the SDSS-III/APOGEE spectrograph. The sample discussed in this work represents only a small fraction of the final APOKASC sample, that will include over 10,000 giants and dwarfs in the {\it Kepler} field. Moreover, there are advanced plans for further expanding the sample as part of the SDSS-IV/APOGEE-2 survey. The APOKASC collaboration adds accurate and homogeneous determinations of effective temperatures and surface chemical abundances to a large sample of stars for which we have precious information from the oscillation spectra. 

In this paper we have employed a Bayesian method to determine basic stellar parameters (mass, surface gravity, radius), distances, and extinctions for 1989 giants present in the first version of the APOKASC catalogue \citep{pinsonneault14}. The results are very encouraging: distances and extinctions are derived with very small formal uncertainties, for stars located as far away as 5~kpc. The vast majority of the stars produced results that are internally consistent, well-behaved, and considered reliable. The final diagnostics of the correctness of our distances and extinctions come essentially from three sources: (1) for regions of the {\em Kepler} field for which very small extinction is expected from the SFD dust maps, our own extinction maps do not show evidence of systematic offsets. (2) Stars in the NGC~6791 and NGC~6819 star clusters present essentially the same distances, with small variations in their extinctions. (3) Stars in the APOGEE-RC catalogue \citep{bovy14} -- which are very likely RC stars and hence can be assigned precise {\it Hipparcos}-calibrated spectrophotometric distances -- correlate well with our independent distances.

The present results are also very encouraging for the application of these distances and extinctions in studies of Galactic structure and evolution (as will be done in following papers). Nevertheless, we have identified difficulties that will have to be dealt with in future studies. The most serious one appears to be the lack of evolutionary status for a large fraction of APOKASC stars, which causes some degeneracy in the determination of their stellar parameters, and increased uncertainties. This problem will probably be much reduced in future releases of the APOKASC catalogue, after careful revision of the asteroseismic parameters from the individual oscillation spectra, and the measurement of \deltaP\ for more stars. Also, the parameters \numax, \deltanu, and \deltaP\ themselves will be progressively replaced by a more detailed star-by-star analysis of the observed oscillation modes.

All the distance and extinction values are made available for download alongside this paper on the journal website. A sample table is reported in Table~\ref{tab:dist_ext}.


\section*{Acknowledgments}
We warmly thank David Valls-Gabaud and Pascal Bord\'e for the useful discussions. T.S.R.\ acknowledges support from CNPq-Brazil. A.M., W.J.C. and Y.E acknowledge support from the UK Science and Technology Facilities Council. D.B. acknowledges support from the University of Birmingham. Funding for the Stellar Astrophysics Centre is provided by The Danish National Research Foundation (grant agreement 
No.: DNRF106). J.B.\ was supported by NASA through Hubble Fellowship grant HST-HF-51285.01 from the Space Telescope Science Institute, which is operated by the Association of Universities for Research in Astronomy, Incorporated, under NASA contract NAS5-26555. C.R.E., J.A.J., and M.P.\ acknowledge support from National Science Foundation grant AST-1211673. The research leading to the presented results has received funding from the European Research Council under the European Community's Seventh Framework Programme (FP7/2007-2013)/ERC grant agreement no 338251 (StellarAges). T.C.B.\ acknowledges partial support for this work by grant PHY 08-22648: Physics Frontiers Center/Joint Institute for Nuclear Astrophysics (JINA), awarded by the U.S.\ National Science Foundation. A.S.\ is supported by the MICINN grant AYA2011-24704 and by the ESF EUROCORES Program EuroGENESIS (MICINN grant EUI2009-04170). G.Z. is supported by an NSF Astronomy \& Astrophysics Postdoctoral Fellowship under Award No. AST-1203017.
 
We thank the entire {\it Kepler} team. Funding for this Discovery mission is provided by NASA's Science Mission Directorate.

Funding for SDSS-III has been provided by the Alfred P. Sloan Foundation, the Participating Institutions, the National Science Foundation, and the U.S. Department of Energy Office of Science. The SDSS-III web site is http://www.sdss3.org/.

SDSS-III is managed by the Astrophysical Research Consortium for the Participating Institutions of the SDSS-III Collaboration including the University of Arizona, the Brazilian Participation Group, Brookhaven National Laboratory, Carnegie Mellon University, University of Florida, the French Participation Group, the German Participation Group, Harvard University, the Instituto de Astrofisica de Canarias, the Michigan State/Notre Dame/JINA Participation Group, Johns Hopkins University, Lawrence Berkeley National Laboratory, Max Planck Institute for Astrophysics, Max Planck Institute for Extraterrestrial Physics, New Mexico State University, New York University, Ohio State University, Pennsylvania State University, University of Portsmouth, Princeton University, the Spanish Participation Group, University of Tokyo, University of Utah, Vanderbilt University, University of Virginia, University of Washington, and Yale University.

Funding for the Brazilian Participation Group has been provided by the Minist\'erio de Ci\^{e}ncia e Tecnologia (MCT), Funda\c{c}\~{a}o Carlos Chagas Filho de Amparo \`{a} Pesquisa do Estado do Rio de Janeiro (FAPERJ), Conselho Nacional de Desenvolvimento Cient\'ifico e Tecnol\'ogico (CNPq), and Financiadora de Estudos e Projetos (FINEP).

\bibliographystyle{mn2e/mn2e_new}

\bibliography{apokasc} 
%


\label{lastpage}

\end{document}